\documentclass[a4paper,fleqn,usenatbib]{mnras}
\usepackage[T1]{fontenc}
\usepackage{ae,aecompl}
\usepackage{graphicx}	
\usepackage{amsmath}	
\usepackage{amssymb}	
\usepackage{rotating}
\newcommand{\ie}{$i.e.,\;$}
\newcommand{\eg}{$e.g.,\;$}

\title[Investigating kpc-scale radio properties of NLS1s]{Investigating kpc-scale radio emission properties of narrow-line Seyfert 1 galaxies}
\author[Singh et al.]{Veeresh Singh$^{1}$\thanks{E-mail: veeresh@prl.res.in} and Hum Chand$^{2}$ 
\\
$^{1}$Astronomy and Astrophysics Division, Physical Research Laboratory, Ahmedabad 380009, India \\
$^{2}$Aryabhatta Research Institute of Observational Sciences (ARIES), Manora Peak, Nainital 263002 India \\
}
\date{Accepted XXX. Received YYY; in original form ZZZ}
\pubyear{2018}
\begin{document}
\label{firstpage}
\pagerange{\pageref{firstpage}--\pageref{lastpage}}
\maketitle
%
\begin{abstract}
%
In recent years, several Radio-Loud Narrow-Line Seyfert 1 galaxies (RL-NLS1) possessing relativistic jets have come into attention 
with their detections in Very Large Baseline Array (VLBA) and in $\gamma$-ray observations. 
In this paper we attempt to understand the nature of radio-jets in NLS1s by examining the kpc-scale radio 
properties of, hitherto, the largest sample of 11101 optically-selected NLS1s. 
Using 1.4 GHz FIRST, 1.4 GHz NVSS, 327 MHz WENNS, and 150 MHz TGSS catalogues 
we find the radio-detection of merely $\sim$ 4.5 per cent (498/11101) NLS1s, with majority 
(407/498 $\sim$ 81.7 per cent) of them being RL-NLS1s. Our study yields the highest number of RL-NLS1s and it can only be a lower limit. 
We find that the most of our radio-detected NLS1s are compact ($<$ 30 kpc), exhibit both flat as well as steep radio spectra, 
and are distributed across a wide range of 1.4 GHz radio luminosities (10$^{22}$ $-$ 10$^{27}$ W Hz$^{-1}$). 
At the high end of radio luminosity our NLS1s often tend to show blazar-like properties such as compact radio-size, 
flat/inverted radio spectrum, radio variability and polarization. 
The diagnostic plots based on the mid-IR colours suggest that the radio emission in NLS1s is mostly powered by AGN, while 
nuclear star-formation may have a significant contribution in NLS1s of low radio luminosities.   
The radio luminosity versus radio-size plot infers that the radio-jets in NLS1s are either 
in the early evolutionary phase or possibly remain confined within the nuclear region due to low-power or intermittent AGN activity.  
\end{abstract}
%
%
%
%
\section{Introduction}
\label{sec:Intro}
Narrow-line Seyfert 1 galaxies (NLS1s) are a subclass of Active Galactic Nuclei (AGN) that show 
broad permitted Balmer emission lines of widths relatively narrower (FWHM $<$ 2000 km s$^{-1}$) than that are 
usually seen in normal AGN {\ie}FWHM $\sim$ 5000$-$10000 km s$^{-1}$ \citep{Osterbrock85,Goodrich89}.
Also, unlike Broad-Line AGN (BL-AGN), NLS1s exhibit relatively weaker forbidden emission lines 
({\eg}[O III]$\lambda$5007{\AA}/H${\beta}$ $<$ 3), strong permitted Fe II emission lines, steep soft X-ray spectra, 
rapid X-ray variability \citep{Panessa11,Zhang11}. 
NLS1s are believed to possess smaller Super-Massive Black Holes (SMBHs; $\sim$ 10$^{5}$~$-$10$^{8}$ M$\odot$) 
with relatively higher accretion rates \citep{Komossa08,Foschini15}. 
Based on the radio-loudness parameter (R) which is defined as the ratio of 5 GHz radio flux density 
to the optical flux at $\lambda$4400{\AA}, AGN are divided into radio-quiet (R $<$ 10) and radio-loud (R $\geq$ 10) 
categories \citep{Kellermann89}. 
Radio studies suggest that, in general, NLS1s tend to be radio-quiet \citep{Zhou06}, 
and hence, the nature of radio emission in NLS1s is poorly understood. 
Also, unlike BL-AGN, a much smaller fraction of NLS1s is believed to be radio-loud. 
For instance, only 7 per cent of NLS1s are found to be radio-loud \citep{Zhou06,Komossa06}, 
while 10$-$20 per cent BL-AGN are reported to be radio-loud \citep[see][]{Falcke96,Kellermann16}. 
In fact, the fraction of strong radio-loud NLS1s (R $>$ 100) is even much lower ($\sim$ 2.5 per cent; \citealt{Komossa08}). 
Therefore, the scarcity of RL-NLS1s is intriguing, and the nature of radio emission in NLS1s needs 
to be investigated in a substantially large sample. 
\par
Targeted radio observations of the small samples of NLS1s show that RL-NLS1s often display compact radio emission but of diverse spectral properties 
{\eg}steep, flat, and inverted radio spectra \citep{Komossa06,Gu15}. 
Although, a handful cases of RL-NLS1s with extended kpc-scale jet-lobe morphology have also been found 
\citep[see][]{Doi12,Richards15,Berton18}. 
The kpc-scale jet-lobe structures detected in a few RL-NLS1s resemble with the classical radio galaxies. 
While, the compact steep-spectrum radio emission in RL-NLS1s suggests their similarity with the young 
radio galaxies, and the NLS1s with flat or inverted radio spectra may be blazar-like sources.    
In fact, strong radio-loud NLS1s (R $>$ 100) display several properties 
({\eg}compact radio cores, very high brightness temperatures, large-amplitude radio flux and spectral variability, 
enhanced optical continuum emission, flat X-ray spectra, and $\gamma$-ray detection) similar to blazars 
\citep{Yuan08,Abdo09a,Abdo09b,Foschini12,Foschini15}. 
Also, parsec-scale high resolution Very Large Baseline Array (VLBA) radio observations of 
few RL-NLS1s reveal presence of relativistic jets in them \citep{Doi07,Gu15}. 
The detection of $\gamma$-ray emission in some RL-NLS1s further corroborates the presence of relativistic jets 
\citep{Abdo09a,Abdo09b,Foschini11,DAmmando12,DAmmando15}. 
Moreover, it is unclear if these NLS1s are intrinsically radio-loud or apparently radio-loud due to jet beaming 
effects \citep{Gu10,Doi11}. Also, due to the limited number of NLS1s detected in VLBA and $\gamma$-ray observations, it is 
unclear if the presence of relativistic jets in NLS1s is a common feature. 
Therefore, it does seem that despite the increasing studies on NLS1s, 
the mechanisms driving the diverse radio properties in NLS1s, are still poorly understood. 
\par 
In order to understand the nature of radio-jets in NLS1s, we investigate radio properties of, hitherto, the largest sample of 11101 NLS1s. 
The details of our NLS1s sample are outlined in Sect.~\ref{sec:sample}. 
To detect radio emission in our NLS1s we use large-area radio surveys such as the 
1.4 GHz Faint Images of the Radio Sky at Twenty-cm (FIRST, \citealt{Becker95}), 
1.4 GHz NRAO VLA Sky Survey (NVSS, \citealt{Condon98}), 327 MHz Westerbork Northern Sky Survey (WENSS; \citealt{Rengelink97}), 
and 150 MHz TIFR GMRT Sky Survey (TGSS; \citealt{Intema17}). 
The details of these radio surveys are given in Sect.~\ref{sec:radiodata}. 
The radio detection of our NLS1s in different surveys is discussed in Sect.~\ref{sec:radiodet}. 
The radio properties ({\ie}detection rates, flux densities, luminosities, sizes, radio-loudness parameters, 
variability and polarization) of our NLS1s are presented in Sect.~\ref{sec:RadioProp}. 
In Sect.~\ref{sec:SF} we discuss the possibility of star-formation contribution to the radio emission in NLS1s. 
Sect.~\ref{sec:Discussion} is devoted to a discussion on the evolutionary stage of radio-jets in NLS1s.
The conclusions of our study are outlined in Sect.~\ref{sec:Conclusions}.
\\     
In this paper, we use cosmological parameters H$_{\rm 0}$ = 71 km s$^{-1}$ Mpc$^{-1}$, 
${\Omega}_{\rm M}$ $=$ 0.27, ${\Omega}_{\Lambda}$ $=$ 0.73. 
\section{The Sample}
\label{sec:sample}
Our sample of NLS1s is taken from \cite{Rakshit17} who present an optically-selected sample of 11101 NLS1s ($z$ $\leq$ 0.8) 
derived from the Sloan Digital Sky Survey Data Release 12 (SDSS DR12). 
This is the most recent and the largest sample of NLS1s till date. It is nearly five times larger than the previous largest NLS1s sample reported by 
\cite{Zhou06}. The NLS1s in this sample are selected as the AGN with 
(i) FWHM (H${\beta}$) $\leq$ 2200 km s$^{-1}$, and (ii) the flux ratio of [O III] to H$\beta$ ([O III]/H${\beta}$) $<$ 3. 
We note that these selection criteria are conventionally used in the literature \citep{Osterbrock85,Goodrich89,Zhou06}.
The cutoff on FWMH (H${\beta}$), considered in this sample, is 2200 km s$^{-1}$ instead of the originally proposed value 
of 2000 km s$^{-1}$ \citep{Goodrich89}. This is mainly due to the fact that the distribution of FWHM (H${\beta}$) is rather smooth 
and does not show a cutoff at 2000 km s$^{-1}$ \citep[see][]{Zhou06}. 
In fact, the cutoff value on FWMH (H$\beta$) is somewhat arbitrary, and, there are suggestions to adopt a 
luminosity-dependent cutoff value of FWHM (H$\beta$) \citep[see][]{Laor2000,Veron-Cetty01,Shemmer04} or Eddington ratio 
(L/L$_{\rm Edd}$ $>$ 0.25; \citealt{Netzer07}) to classify NLS1s. 
We caution that the inclusion of sources with relatively larger FWHM (H$\beta$) can introduce a bias. 
More details about our NLS1s sample can be found in \cite{Rakshit17}. 
\section{Radio Data}
\label{sec:radiodata}
\begin{figure}
\includegraphics[angle=0,width=9cm,trim={0.5cm 0.5cm 0.5cm 0.5cm},clip]{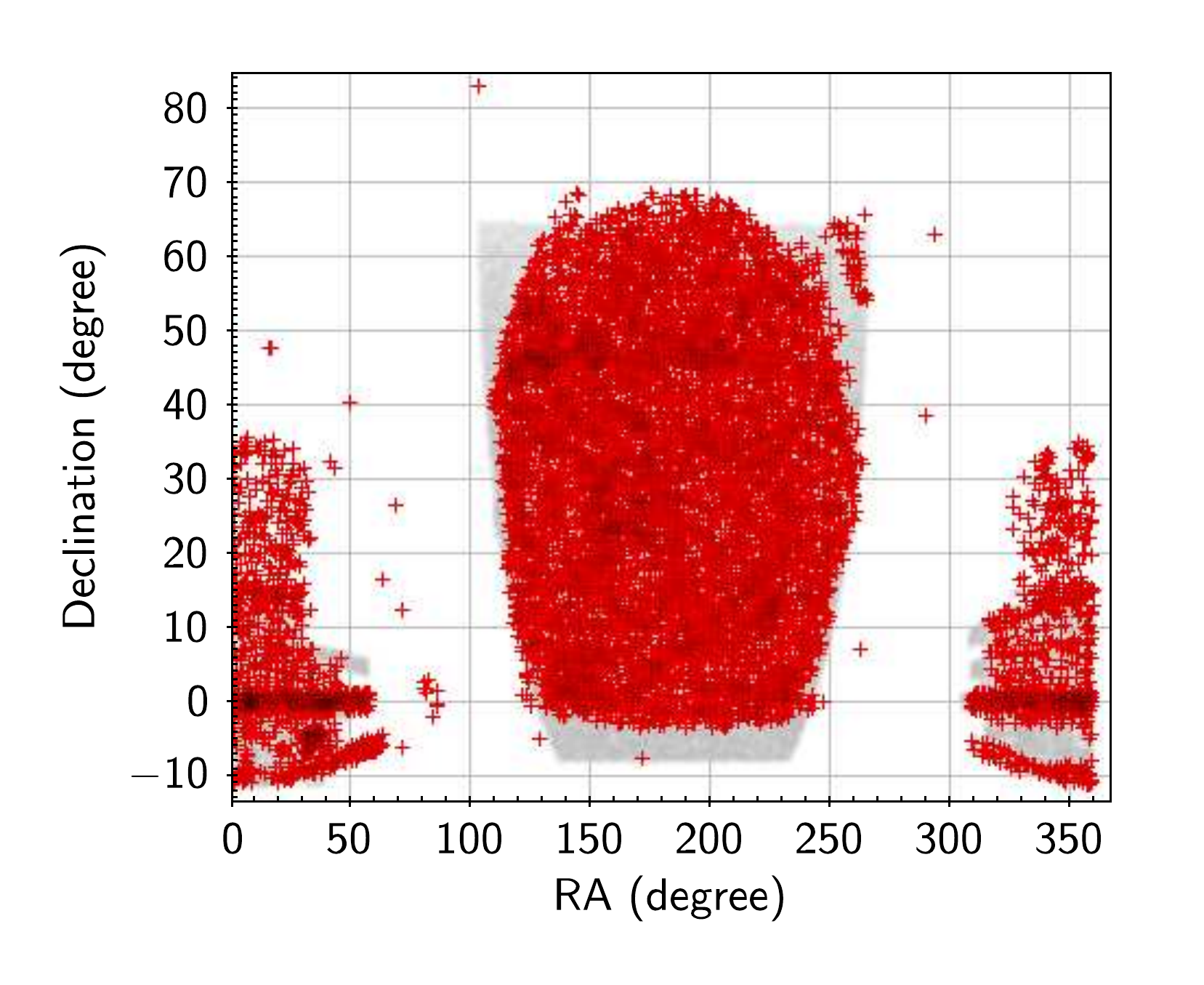}
\caption{NLS1s sky distribution. NLS1s are represented by `+' symbol in red colour, and 
FIRST coverage is shown in grey shades. 
The NVSS, TGSS and WENSS cover all the sky above declination of $-$40, $-$53 and +30 degrees, respectively.}
\label{fig:NLS1FIRSTCoverage} 
\end{figure}

To find the radio counterparts of our NLS1s we primarily use the FIRST radio survey which has substantial overlap with the SDSS. 
In Fig.~\ref{fig:NLS1FIRSTCoverage} we show the sky coverages of our NLS1s and the FIRST survey. 
We also use 1.4 GHz NVSS, 327 MHz WENSS and 150 MHz TGSS. 
The details of these radio surveys are given below.  
\label{sec:RadioData} 
\subsection{The FIRST survey}
The FIRST survey is carried out at 1.4 GHz with the Very Large Array (VLA) in `B' configuration. 
It covers a total sky area of about 10575 deg$^{2}$ {\ie}8,444 deg$^{2}$ in the north Galactic cap 
and 2,131 deg$^{2}$ in the south Galactic cap{\footnote{http://sundog.stsci.edu/}}. 
The FIRST radio images have typical resolution of $\sim$ 5$\arcsec$.0 and the rms noise of $\sim$ 0.15 mJy. 
The FIRST catalogue contains a total of 946,432 sources and lists peak as well as integrated flux densities and radio sizes derived by fitting a two dimensional
Gaussian to a source detected at $\geq$ 5$\sigma$ ($\sim$ 1.0 mJy) flux limit. 
The astrometric reference frame of the FIRST maps is accurate to $\sim$ 0$\arcsec$.05 and individual sources have $\sim$ 90 per cent 
confidence error circles of radius $<$ 0$\arcsec$.5 at 3 mJy level and $\sim$ 1$\arcsec$.0 at 1.0 mJy. 
We use the most recent version of the FIRST catalogue released on 17 December 2014. 
\subsection{The NVSS} 
The NVSS is a 1.4 GHz continuum survey carried out with the VLA in `D' configuration. 
It covers entire sky north of $-$40 degree declination and produces radio images with the resolution 
of $\sim$ 45$\arcsec$ and sensitivity of $\sim$ 2.5 mJy at 5$\sigma$ level \citep{Condon98}. 
We use the NVSS source catalogue that contains nearly 1.8 million sources and lists 
integrated 1.4 GHz flux densities, major axis, minor axis, position angle for all the radio sources detected at $\geq$ 2.5 mJy. 
The astrometric accuracy in the NVSS ranges from $\sim$ 1$\arcsec$.0 for the brightest detections to $\sim$ 7$\arcsec$.0 for the 
faintest detections.
It is important to note that owing to a relatively large beam-size, NVSS is advantageous in detecting 
extended low-surface-brightness radio emission that can be missed in the FIRST. 
Therefore, both FIRST and NVSS, carried out at same frequency, can be used in a complementary fashion. 
\subsection{The WENSS}
327 MHz WENSS is carried out with the Westerbork Synthesis Radio Telescope (WSRT), 
and covers entire sky north of +30 degree declination \citep{Rengelink97}. 
The WENSS is fairly shallow with 5$\sigma$ sensitivity limit of 
$\sim$ 18 mJy and resolution of $\sim$ 54$\arcsec$ $\times$ 54$\arcsec$ cosec($\delta$). 
The WENSS catalogue contains nearly 220,000 sources with positional accuracy of $\sim$ 1$\arcsec$.5 for bright sources and 
$<$ 5$\arcsec$.0 for faint sources. 
\subsection{The TGSS}
150 MHz TGSS is carried out with the Giant Metrewave Radio Telescope (GMRT) during April 2010 to March 2012, 
and covers nearly 36,900 deg$^{2}$ area between $-$53$^{\circ}$ and +90$^{\circ}$ declination. 
\cite{Intema17} present TGSS source catalogue containing 620,000 sources above 7$\sigma$ $\sim$ 24.5 mJy. 
The TGSS images have approximate resolution of 25$\arcsec$ $\times$ 25$\arcsec$ with 
astrometric accuracy of $\sim$ 2$\arcsec$.0. 
\section{Search for radio counterparts} 
\label{sec:radiodet}
In following subsections we provide the details of the search for radio counterparts of our NLS1s in different radio surveys. 
The number of NLS1s detected in different radio surveys are listed in Table~\ref{table:RadioDet}.  
\subsection{1.4 GHz radio counterparts in the FIRST}
\begin{figure*}
\includegraphics[angle=0,width=8.5cm,trim={0.5cm 0.5cm 0.5cm 0.5cm},clip]{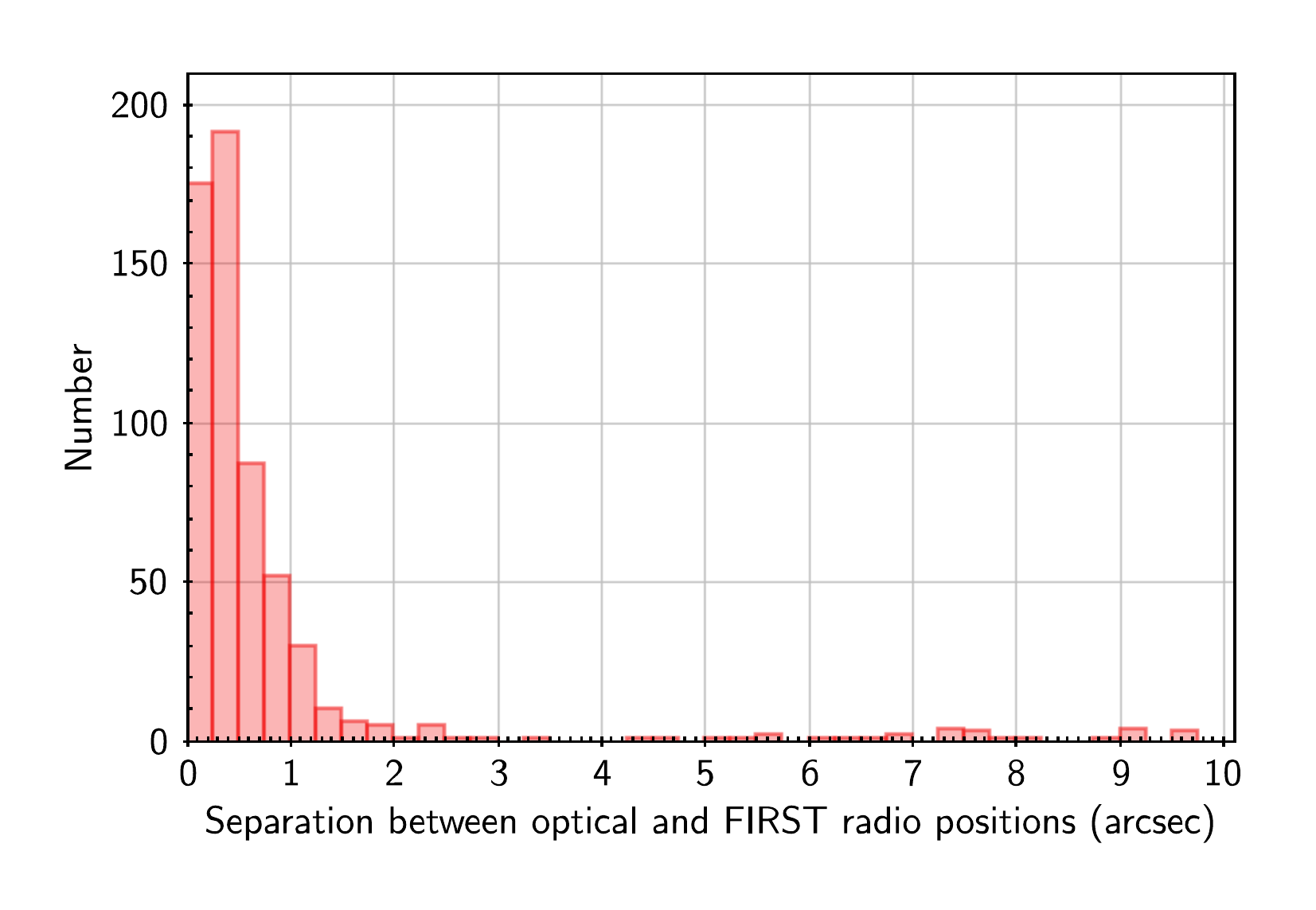}{\includegraphics[angle=0,width=8.5cm,trim={0.5cm 0.5cm 0.5cm 0.5cm},clip]{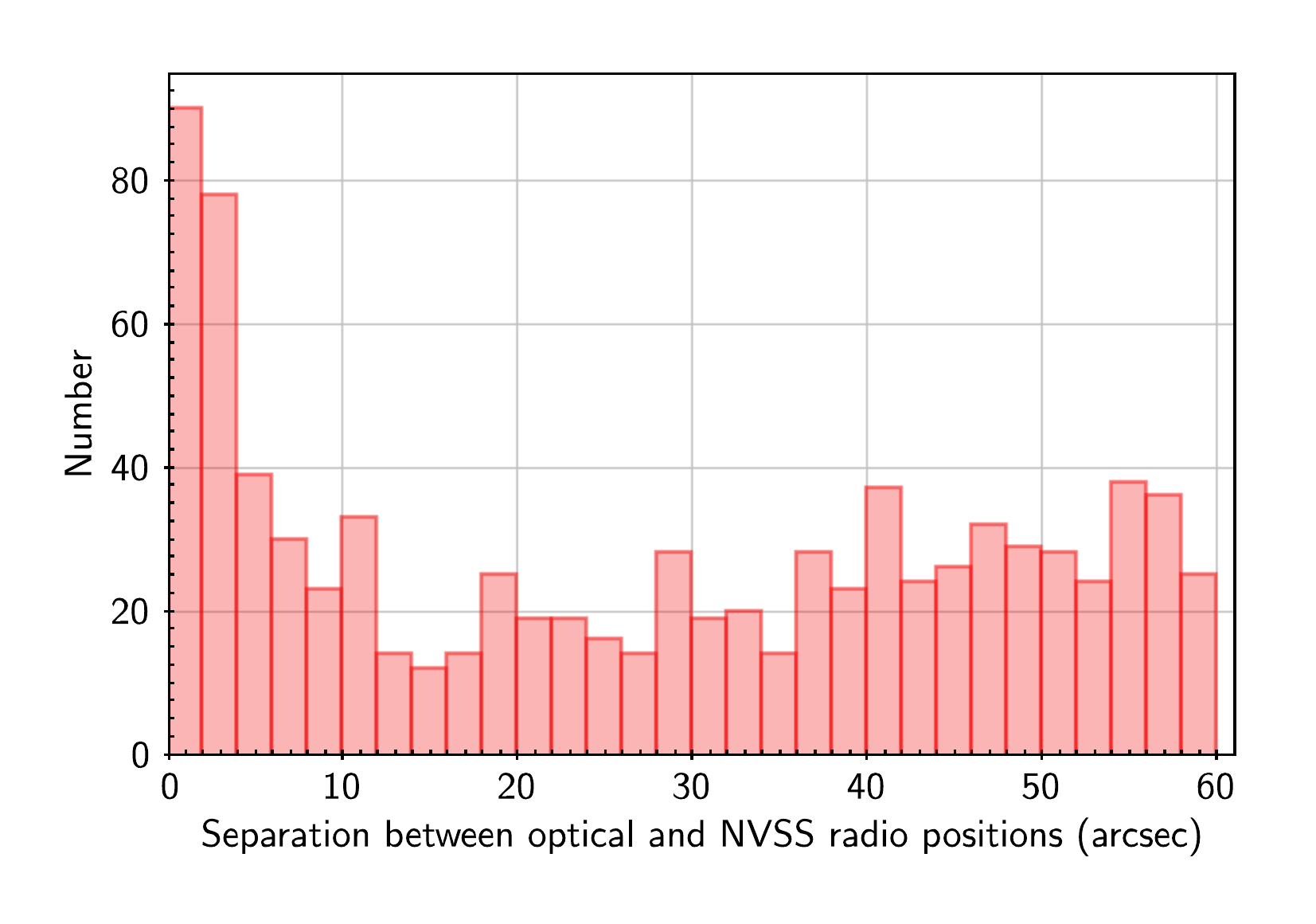}}
\includegraphics[angle=0,width=8.5cm,trim={0.5cm 0.5cm 0.5cm 0.5cm},clip]{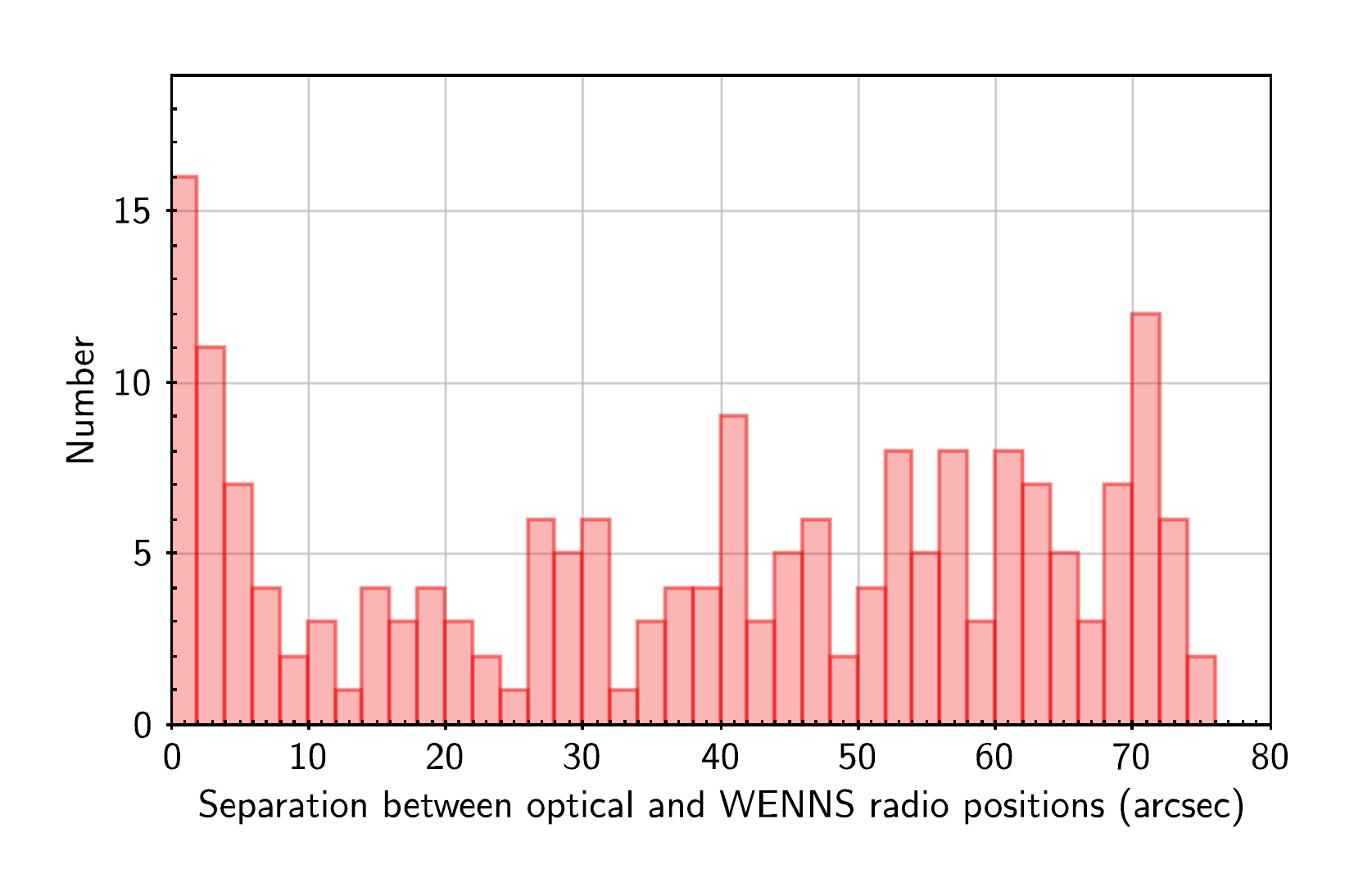}{\includegraphics[angle=0,width=8.5cm,trim={0.5cm 0.5cm 0.5cm 0.5cm},clip]{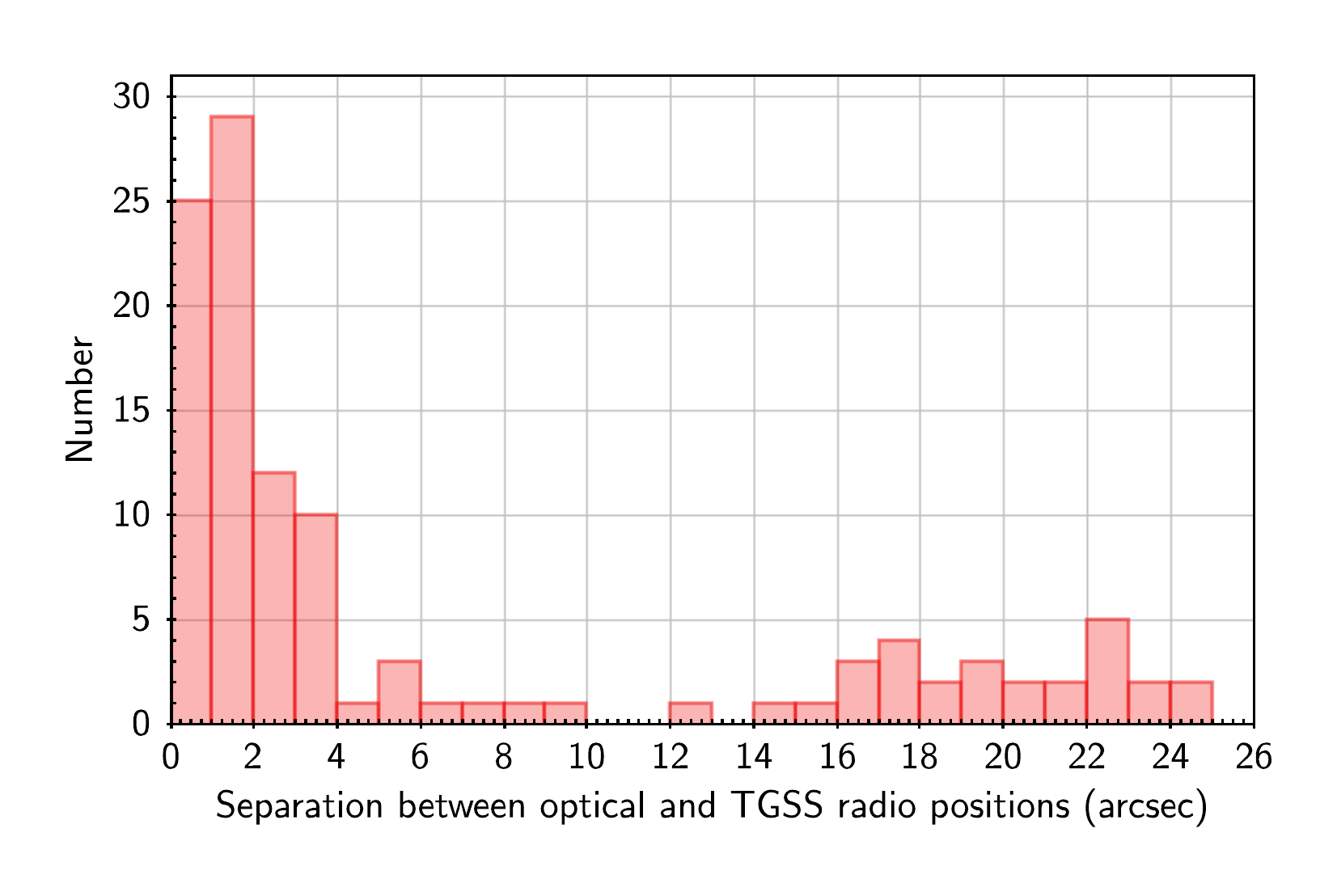}}
\caption{Histograms for the separation between optical and radio positions in different radio surveys. 
The different search radii are used for different surveys due to their differing beam-sizes.}
\label{fig:SepHist} 
\end{figure*}

We find the radio counterparts of our NLS1s in the FIRST survey by performing positional 
cross-matching between the optical catalogue of NLS1s and the FIRST radio catalogue. 
The catalogues are cross-matched using `Tool for OPerations on Catalogues And Tables' (TOPCAT) 
software \citep{Taylor05}. 
We note that it is important to use an optimum cross-matching radius as it decides the degree of completeness and contamination. 
A smaller search radius minimizes the contamination at the expense of completeness, 
while a larger search radius gives higher completeness but with higher contamination.  
To find the optimum value of search radius we first cross-match the FIRST and NLS1s catalogues by using a relatively large search radius of 10$\arcsec$.   
We plot the histogram of the separation between optical and radio positions, 
and find that the histogram follows nearly a Gaussian distribution that tails off beyond 2$\arcsec$.0 (see Fig.~\ref{fig:SepHist}, top left panel). 
Therefore, beyond 2$\arcsec$.0, the number of random matches begins to dominate over the number of real matches.  
Thus, we choose the optimum value of search radius as 2$\arcsec$.0. 
In fact, our search radius of 2$\arcsec$.0 is similar to the one used in some of the previous studies 
that performed the cross-matching of FIRST and SDSS sources \citep[see][]{Kimball08}. 
We also note that there is no systematic offset between the radio and optical positions {\ie}the median value of 
the differences between the optical and radio positions 
($\Delta$RA (RA$_{\rm optical}$ $-$ RA$_{\rm FIRST}$) $=$ -0$\arcsec$.04 and $\Delta$DEC (DEC$_{\rm optical}$ $-$ DEC$_{\rm FIRST}$) 
$=$ 0$\arcsec$.02) are within the astrometric uncertainties. 
\par
The cross-matching of our NLS1s and the FIRST catalogue using a search radius of 2$\arcsec$.0 yields the radio counterparts for 
556 NLS1s. To ensure the reliability of radio counterparts we consider only those FIRST sources that 
have side-lobe probability less than 0.1 {\ie}there is 90 per cent probability that the radio detection is real 
and not a side-lobe artefact. We note that our cut-off limit on the side-lobe probability is fairly conservative 
to ensure that the radio counterparts of our NLS1s are not contaminated by spurious sources. 
After excluding 75 radio counterparts with side-lobe probability more than 0.1, we get 
FIRST radio counterparts for only 481 NLS1s. We note that all the 75 radio sources excluded due to higher side-lobe 
probability are faint. 
We also check the level of contamination due to random chance matches, by shifting the optical positions of NLS1s 
with 60$\arcsec$ in random directions and cross-matching it with the FIRST catalogue. 
With shifted positions of NLS1s we find only 0 to 1 cross-matched source, 
which infers that the level of contamination in our radio-detected NLS1s is $\leq$ (1/481) 0.2 per cent. 
Also, for our redshift limited NLS1s sample ({\it z} $\leq$ 0.8), 
the cross-matching radius of 2$\arcsec$.0 corresponds to the physical scale of 15 kpc at the highest redshift of 
{\it z} $\simeq$ 0.8, which is smaller than the size of a typical galaxy. 
\par
It is important to note that the FIRST source catalogue derived from relatively higher resolution ($\sim$ 5$\arcsec$.4) radio maps can erroneously 
identify a multicomponent extended radio source as more than one individual source. 
Therefore, to find multicomponent extended radio counterparts, we search for the presence of additional radio sources 
within the circle of 120$\arcsec$ radius centred at the optical positions of our all radio-detected NLS1s. 
The opted search radius of 120$\arcsec$ corresponds to the physical scale of $\sim$ 900 kpc at the redshift cut-off (z $<$ 0.8) of our sample. 
Thus, the diameter of circle within which additional radio components are searched, is similar to the size of a typical giant radio galaxy \citep[see][]{Dabhade17}. 
We visually inspected 5$\arcmin$~$\times$~5$\arcmin$ FIRST image cut-outs of all the sources that show one or more additional 
radio sources within the circle of 120$\arcsec$ radius. 
In most cases, the additional radio source(s) is (are) found to be completely unrelated to the central radio source. 
We find only 08/481 sources showing multicomponent radio emission in the FIRST image cut-outs. 
With visual inspection we also find two additional sources that are missed in the cross-matching. 
These two sources exhibit extended lobe dominated radio morphology such that their core components are not identified as the 
individual sources but the centroid positions match with the optical positions. 
We also visually inspected 5$\arcmin$~$\times$~5$\arcmin$ FIRST image cut-outs of all the discarded sources with side-lobe 
probability more than 0.1, and we do not find any source with multicomponent extended radio structure.
Therefore, using FIRST survey we find the radio counterparts for a total of 483 NLS1s in which 10 sources exhibit 
multicomponent extended radio morphology.   
For multicomponent extended radio sources we obtain the total flux density by adding the flux densities of all the individual components.  
Since we do not perform the visual inspection of the FIRST image cut-outs of all 11101 optically-selected NLS1s, 
we can miss relic radio galaxy like sources which do not show a core radio component. 
However, NLS1s with radio structures similar to the core-less relic radio galaxy are not known, 
and would be extremely rare, if they exist.
\subsection{1.4 GHz radio counterparts in the NVSS}
We find 1.4 GHz NVSS radio counterparts of our NLS1s by cross-matching our NLS1s with the NVSS radio catalogue. 
The optimum search radius of 15$\arcsec$ is obtained by using the method described in the previous subsection 
(see Fig.~\ref{fig:SepHist}, top right panel). 
From Fig.~\ref{fig:SepHist} (top right panel), it is apparent that beyond 15$\arcsec$, 
the number of cross-matched sources begins to increase due to the increase in random matches. 
We note that the level of contamination due to random matches depends on search radius, accuracy of astrometry and source 
surface number density in the two surveys \citep[see][]{Wadadekar04}. 
Thus, unlike to FIRST survey, the higher level of contamination due to random matches in cross-matching the NLS1s and NVSS catalogues 
is due to relatively less accurate astrometry in the NVSS and the larger search radius. 
The cross-matching of NLS1s and NVSS source catalogues using 15$\arcsec$ search radius yields a total of 314 sources. 
We note that there is no systematic offset between optical and NVSS radio positions 
{\ie}$\Delta$RA (RA$_{\rm optical}$ $-$ RA$_{\rm NVSS}$) $=$ 0$\arcsec$.12 and $\Delta$DEC 
(DEC$_{\rm optical}$ $-$ DEC$_{\rm NVSS}$) $=$ 0$\arcsec$.08. 
To check the level contamination by random chance matches we shift the optical positions of our 
NLS1s by 1$\arcmin$~$-$2$\arcmin$ in random directions, and cross-match it 
with the NVSS catalogue using 15$\arcsec$ search radius. The cross-matching with the shifted optical positions gives 
nearly 25$-$32 sources and therefore, the level of contamination due to chance matches is limited only to (25$-$32)/314 $\sim$ 8$-$10 per cent. 
The relatively higher contamination, in compared to the FIRST, can be attributed to larger positional uncertainties in the NVSS. 
We note that, in general, relatively bright NVSS radio sources (S$_{\rm peak}$ $\geq$ 10 mJy) are matched within few arcsec, 
while faint sources tend to have larger matching radii.  
\par
Given the higher contamination in the NVSS detections we examine 
all the NVSS detections by checking their FIRST counterparts, whenever available. 
The FIRST survey with better resolution and sensitivity allows us to identify radio counterparts more reliably. 
We note that among 314 NVSS detections, only 304 sources lie within the FIRST coverage. 
And, we find that only 253/304 sources show FIRST counterparts within 2$\arcsec$.0. 
Indeed, it is intriguing to have a large number of NVSS detections with no FIRST counterparts as 
the FIRST survey (5$\sigma$ $\sim$ 1.0 mJy) is more sensitive than the NVSS (5$\sigma$ $\sim$ 2.5 mJy). 
Therefore, for remaining 51 NVSS detections with no FIRST counterparts within 2$\arcsec$.0, 
we make visual inspection of both NVSS as well as FIRST image cut-outs. 
We also over-lay the NVSS and FIRST radio contours on to the corresponding SDSS optical images, and 
find 44 NVSS detections in which FIRST contours show 
that the radio emission is from a neighbouring source and not from our NLS1. 
Therefore, we discard 44 NVSS detections and include only seven NVSS detections in which 
NVSS radio contours spatially match with the optical positions of NLS1s. 
Importantly, all these seven NLS1s are faint in NVSS with their flux densities nearly $\sim$ 2.5 mJy 
and do not show counterparts in FIRST. 
The lack of FIRST counterparts in these NVSS-detected NLS1s suggests them to be variable.  
We also find that there are six FIRST-detected bright extended sources that have NVSS counterparts with offset $>$ 15$\arcsec$. 
For 10 sources falling outside the FIRST coverage we over-lay NVSS contours on to the corresponding SDSS optical images and 
find only eight sources in which the radio contours spatially match with the optical position. 
Thus, our exercise yields the reliable NVSS detections for only 274 (253 + 7 + 6 + 8) NLS1s. 
We note that among 274 NVSS detections only 259 sources show FIRST counterparts, 
and therefore, the use of NVSS helps us in finding the radio counterparts for additional 15 NLS1s.  
\subsection{325 MHz radio counterparts in the WENNS}
To find 325 MHz radio counterparts in the WENNS we cross-match our NLS1s with the WENSS source catalogue. 
Based on the histogram of the separation between optical and radio positions we choose search radius of 10$\arcsec$. 
The histogram of the separation between optical and radio positions follows a normal distribution up to 10$\arcsec$, 
and random matches begin to dominate beyond 10$\arcsec$ (see Fig.~\ref{fig:SepHist}, lower left panel). 
The cross-matching of our NLS1s and WENSS catalogues yields only 40 sources. 
All 40 WENSS detections fall within the FIRST coverage. 
We examine the reliability of WENSS counterparts by over-laying WENNS and FIRST radio contours on to their corresponding SDSS images. 
We exclude one WENNS detection at 10$\arcsec$ separation as the FIRST contours show that the radio 
emission is from a neighbouring galaxy and not from the NLS1. 
Therefore, we obtain WENNS radio counterparts for only 39 NLS1s. 
All 39 WENNS detections have FIRST as well as NVSS counterparts. 
We note that the relatively less number of counterparts found in the WENSS is primarily due to its 
much lower sensitivity (5$\sigma$ $\sim$ 18 mJy), in compared to the FIRST (5$\sigma$ $\sim$ 1.0 mJy) 
and NVSS (5$\sigma$ $\sim$ 2.5 mJy). 
Only relatively bright FIRST-detected sources will be detected in the WENSS even if they have steep radio spectra. 
For instance, a radio source detected in the WENNS at 5${\sigma}$ level (S$_{\rm 325~MHz}$ = 18 mJy) 
would have 1.4 GHz flux density of 4.18 mJy for ${\alpha}$ ${\sim}$ $-$1.0,  8.67 mJy for ${\alpha}$ ${\sim}$ $-$0.5, 
and 18.0 mJy for ${\alpha}$ ${\sim}$ 0; where ${\alpha}$ is the power law (S$_{\nu}$ ${\propto}$ ${\nu}^{\alpha}$) 
spectral index between 325 MHz and 1.4 GHz. 
\subsection{150 MHz radio counterparts in the TGSS}
We obtain 150 MHz radio counterparts of our NLS1s by cross-matching our NLS1s with 150 MHz TGSS catalogue. 
Based on the histogram of the separation between optical and 150 MHz radio positions we choose the optimum 
search radius of 6$\arcsec$.0, beyond 
which the number of random matches begins to increase (see Fig.~\ref{fig:SepHist}, lower right panel).   
The cross-matching of our NLS1s and 150 MHz radio catalogue yields a total of 79 sources. 
Among the 79 TGSS-detected sources 76 sources have FIRST counterparts, while three sources lie outside the FIRST coverage and have NVSS counterparts. 
\begin{table*}
\begin{minipage}{140mm}
\caption{NLS1s detected in different radio surveys.}
\begin{tabular}{@{}cccccccc@{}}
\hline
NLS1s                                  &    No. of              & Radio        &  ${z}$            & ${z}$               &  ${z}$    &  ${z}$     &  ${z}$      \\  
                                       &    sources             & Det. ($\%$)  &   range           &     median          &   Q1      &   Q3       &    IQR      \\ \hline
Full optically-selected sample         &    11101               &              & 0.009$-$0.8       &       0.47          &  0.26     &  0.65      &  0.39       \\ 
Within FIRST coverage                  &    10508               &              & 0.009$-$0.8       &       0.47          &  0.26     &  0.65      &  0.39       \\        
Within NVSS/TGSS coverage              &    11101               &              & 0.009$-$0.8       &       0.47          &  0.26     &  0.65      &  0.39       \\        
Within WENSS coverage                  &     4842               &              & 0.009$-$0.8       &       0.48          &  0.27     &  0.65      &  0.38       \\        
Detected in 1.4 GHz FIRST              &    483/10508           &  4.6         & 0.0098$-$0.8      &       0.36          &  0.14     &  0.54      &  0.40       \\        
Detected in 1.4 GHz NVSS               &    274/11101           &  2.5         & 0.0098$-$0.8      &       0.37          &  0.15     &  0.53      &  0.38       \\           
Detected in both FIRST and NVSS        &    259/10508           &  2.5         & 0.0098$-$0.8      &       0.38          &  0.15     &  0.53      &  0.38       \\         
Detected in 327 MHz WENSS              &    39/4842             &  0.8         & 0.028$-$0.79      &       0.45          &  0.19     &  0.52      &  0.33       \\         
Detected in 150 MHz TGSS               &    79/11101            &  0.7         & 0.028$-$0.8       &       0.47          &  0.30     &  0.59      &  0.29       \\         
Total radio detected (FIRST and NVSS)  &    498/11101           &  4.5         & 0.0098$-$0.8      &       0.36          &  0.14     &  0.54      &  0.40       \\         
\hline   
\end{tabular}
\label{table:RadioDet} 
\\
{\it Notes}. All 11101 optically-selected NLS1s of our sample are falling within the coverage areas of NVSS and and TGSS. 
All our NLS1s detected in WENSS and TGSS are also detected at 1.4 GHz in FIRST or in NVSS. 
Q1, Q3 and IQR = Q3 $-$ Q1 are first quartile, third quartile, and interquartile range of the distributions, respectively. 
\end{minipage}
\end{table*}
\begin{table*}
\begin{minipage}{140mm}
\caption{Radio properties of our NLS1s.}
\begin{tabular}{@{}cccccccc@{}}
\hline
Parameter             & Unit                & Range                 &   Median    &    Q1       &     Q3      &  IQR (IQR$^{-{\delta}}$, IQR$^{+{\delta}}$)  & No.            \\    
                      &                     &                       &             &             &             &                                         &  of sources    \\ \hline
S$_{\rm 1.4~GHz}$     &      (mJy)          & 1.0$-$8360.6          & 2.60        &  1.63       & 6.28        &  4.65 (4.48, 4.85)    &   498         \\
                      &                     &                       &             &             &             &                       &               \\ 
S$_{\rm 1.4~GHz}^{\rm FIRST}$ &  (mJy)      & 1.0$-$8360.6          & 2.56        &  1.63       &  6.31       &  4.68 (4.49, 4.89)    &   483         \\
                      &                     &                       &             &             &             &                       &               \\ 
S$_{\rm 1.4~GHz}^{\rm NVSS}$  &   (mJy)     & 2.2$-$8283.1          &   5.5       &  3.3        &  18.1       &  14.8 (14.5, 15.0)    &   274         \\
                      &                     &                       &             &             &             &                       &               \\ 
S$_{\rm 327~MHz}$     &         (mJy)       &  18$-$30410           &  103        &   49        &  241        &  192.0 (186.0, 198.3) &   39          \\
                      &                     &                       &             &             &             &                       &               \\ 
S$_{\rm 150~MHz}$     &          (mJy)      & 13.9$-$53958          &   100.6     &  45.4       &  246.2      &  200.8 (183.1, 219.0) &   79          \\
                      &                     &                       &             &             &             &                       &               \\ 
logL$_{\rm 1.4~GHz}$  & (W Hz$^{-1}$)       & 21.85$-$27.14         &   24.02     &  23.12      &  24.56      &  1.44 (1.45, 1.44)    &   498         \\
                      &                     &                       &             &             &             &                       &               \\ 
logL$_{\rm 327~MHz}$  & (W Hz$^{-1}$)       & 23.22$-$27.60         &   25.74     &   25.11     &  26.22      &  1.11 (1.17, 1.08)    &   39          \\
                      &                     &                       &             &             &             &                       &               \\ 
logL$_{\rm 150~MHz}$  & (W Hz$^{-1}$)       & 23.27$-$27.69         &    25.79    &   25.29     &  26.32      &  1.03 (1.05, 1.01)    &   79          \\
                      &                     &                       &             &             &             &                       &               \\ 
${\alpha}_{\rm 150~MHz}^{\rm 1.4~GHz}$&     & $-$1.20$-$0.58        &  $-$0.53    &   $-$0.72   &  $-$0.38    &  0.34 (0.34, 0.34)    &   79          \\
                      &                     &                       &             &             &             &                       &               \\ 
${\alpha}_{\rm 327~MHz}^{\rm 1.4~GHz}$&     &  $-$1.32$-$0.33       &  $-$0.59    & $-$0.80     &  $-$0.24    &  0.56 (0.55, 0.56)    &   39          \\
                      &                     &                       &             &             &             &                       &               \\ 
${\alpha}_{\rm 150~MHz}^{\rm 327~MHz}$&     &  $-$1.20$-$1.22       &  $-$0.36    &  $-$0.74    & $-$0.15     &  0.59 (0.57, 0.64)    &   35          \\
                      &                     &                       &             &             &             &                       &               \\ 
Radio-size             & (kpc)              &   1.13$-$1113         &   11.95     &   6.3       &   21.2      &  14.9 (13.4, 16.4)    &   55          \\
                      &                     &                       &             &             &             &                       &               \\ 
logR$_{\rm 1.4~GHz}$   &                    & 0.35$-$5.76           &   1.85      &  1.40       &   2.58      &  1.18 (1.19, 1.18)    &   498         \\
                      &                     &                       &             &             &             &                       &               \\ 
${\sigma}_{\rm var}$$^{a}$&                 &  $-$26.64$-$7.74      &  $-$1.01    & $-$2.63     &  0.22       &  2.85 (2.85, 2.85)    &   259         \\
                      &                     &                       &             &             &             &                       &               \\ 
Fract. Pol.$^{b}$      &                    &   0.10$-$10.15        &   3.29      &    1.89     &   5.33      &  3.44 (3.13, 4.03)    &    24         \\
\hline  
\end{tabular}
\label{table:HistProp} 
\\
{\it Notes}. a : ${\sigma}_{\rm var}$ represents the degree of radio variability estimated from the comparison of FIRST and NVSS flux densities. 
b : Fractional polarization (Fract. Pol.) is defined as `100 $\times$ (S$_{\rm pol}$/S$_{\rm int}$)'. 
Q1, Q3 and IQR = Q3 $-$ Q1 are the first quartile, third quartile and interquartile of a distribution, respectively. 
IQR$^{+{\delta}}$ and IQR$^{-{\delta}}$ are the IQRs of the distributions obtained with the additions and subtraction of errors, respectively.   
\end{minipage}
\end{table*}
\section{Radio properties}
\label{sec:RadioProp}
In this section we investigate radio properties 
({\ie}detection rates, flux densities, luminosities, spectral indices, radio-sizes, radio-loudness, variability and polarization) 
of our NLS1s. A representative sub-sample of our radio-detected NLS1s is presented in Table~\ref{table:Sample}. 
\subsection{Radio detection rates} 
\label{sec:detectionRate}
The uniform large-area radio surveys allow us to obtain the radio detection rates of our optically-selected NLS1s.  
In Table~\ref{table:RadioDet} we list the radio detection rates of our NLS1s found in the FIRST, NVSS, WENSS and TGSS. 
As expected the FIRST survey being the most sensitive survey yields the highest detection rate 
{\ie}$\sim$ (483/10508) 4.6 per cent of our NLS1s. 
The detection rate in NVSS is only $\sim$ (274/11101) 2.47 per cent. 
While, the detection rates of our NLS1s in 325 MHz WENSS and 150 MHz TGSS are even much lower {\ie}(39/4842) $\simeq$ 0.8 per cent and 
(79/11101) $\simeq$ 0.7 per cent, respectively. 
The combination of FIRST and NVSS yields only $\sim$ (498/11101) 4.5 per cent radio detection rate of our NLS1s. 
Interestingly, there is a large fraction (224/483 $\simeq$ 46.4 per cent) of FIRST-detected NLS1s with no 
NVSS counterparts, as these sources are fainter than the NVSS detection limit of 2.5 mJy. 
The presence of a substantially large fraction of only FIRST-detected sources at faint end shows 
that the detection rate increases sharply with the sensitivity of a radio survey. 
Therefore, we can expect a much higher detection rate of NLS1s in deep radio surveys.  
\par
\begin{figure*}
\includegraphics[angle=0,width=9.0cm,trim={0.5cm 0.5cm 0.5cm 0.5cm},clip]{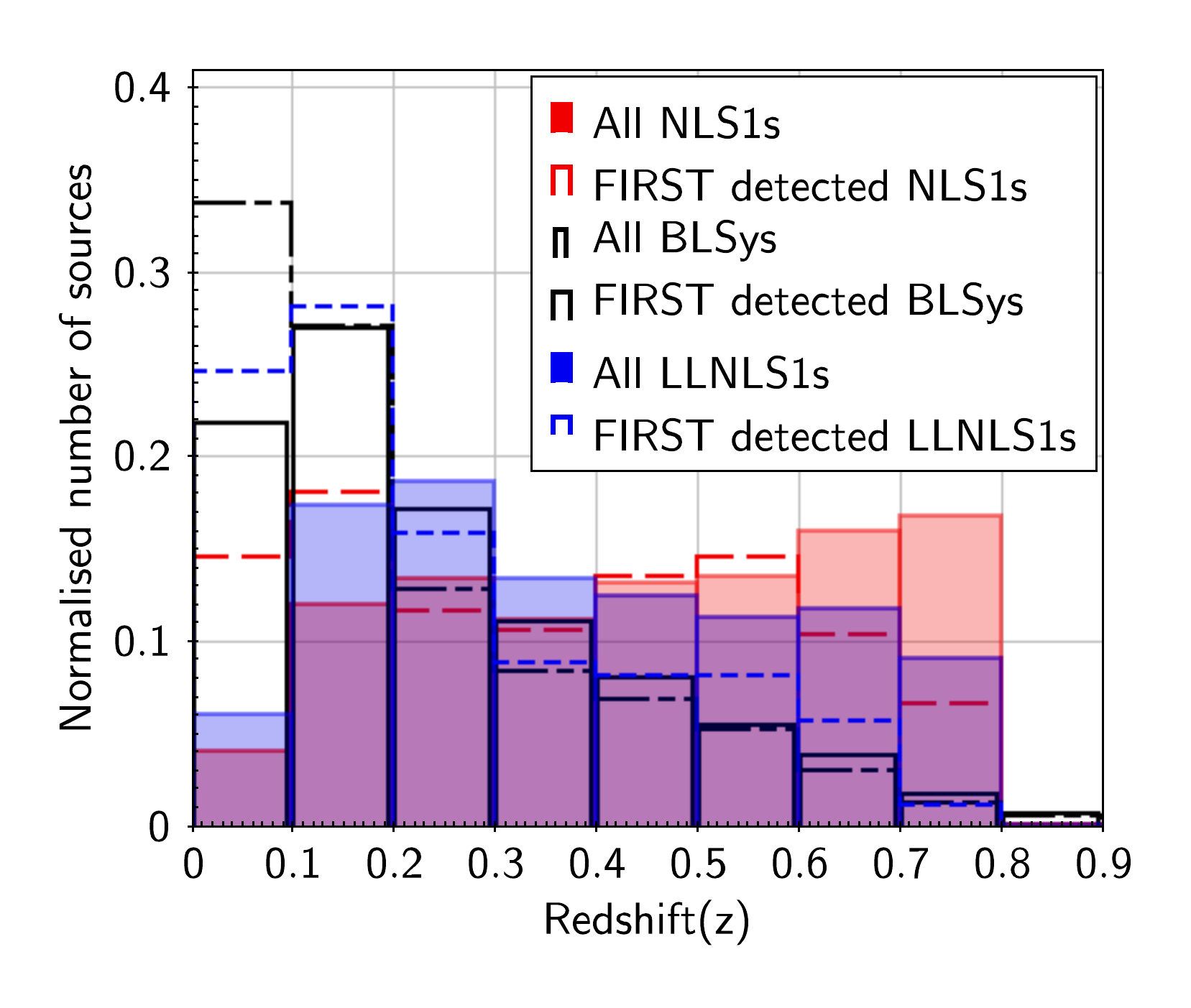}{\includegraphics[angle=0,width=9.0cm,trim={0.5cm 0.5cm 0.5cm 0.5cm},clip]{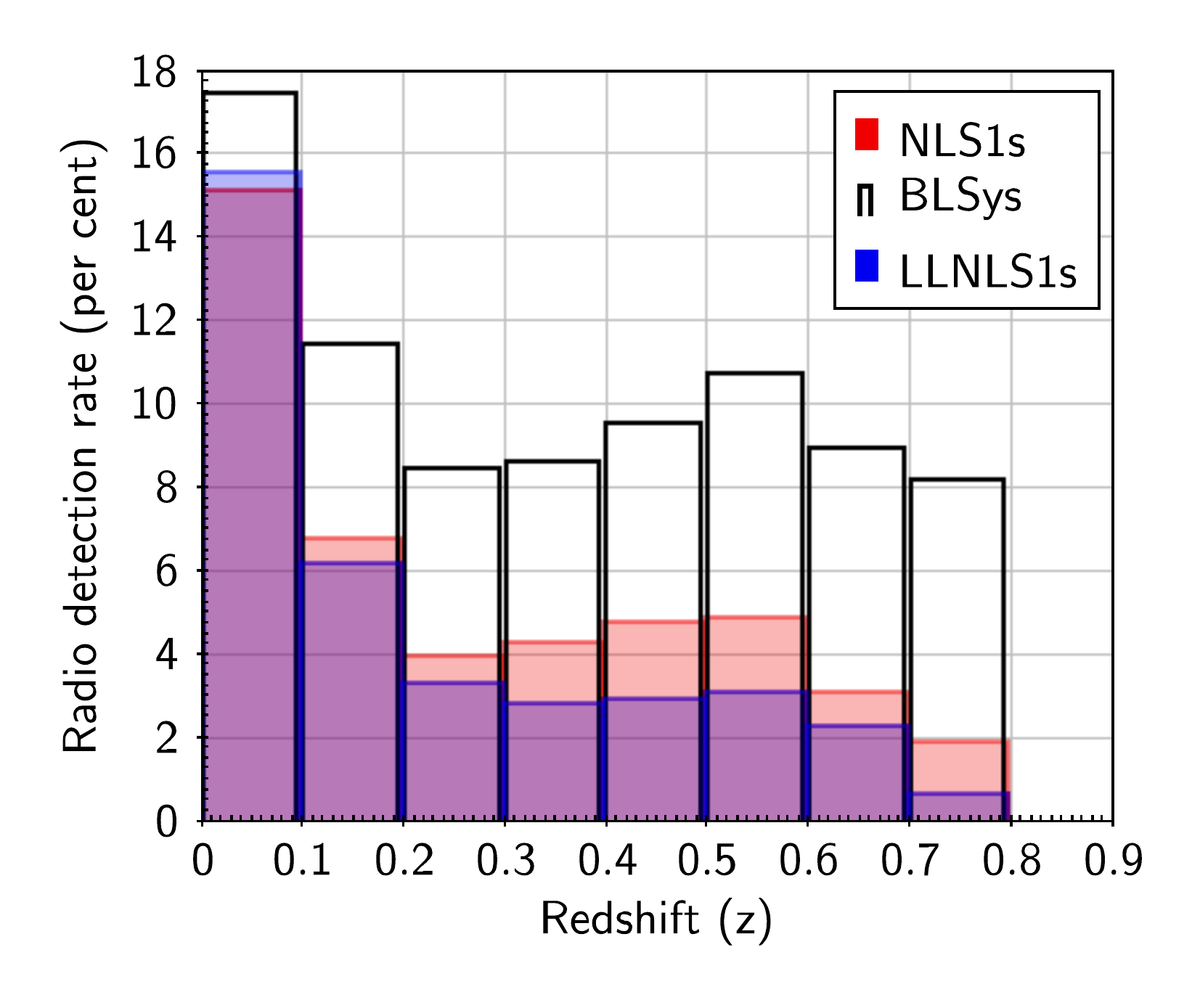}}
\caption{{\it Left panel} : Redshift distributions for our NLS1s (in light red shades), LLNLS1s (in dark blue shades), 
and BL-Seyfert galaxies (in solid black lines). The redshift distributions of our FIRST-detected NLS1s, LLNLS1s and BL-Seyfert galaxies are shown by 
dashed red lines, small-dashed blue lines and broken long-dashed black lines, respectively.
For better visualisation of the comparison of different histograms the normalised number of sources 
(number of sources in each bin normalised with the total number of sources) are shown in y-axis. 
The same convention is followed in other plots as well.
{\it Right panel} : The radio detection rates for our NLS1s (in light red shades), LLNLS1s (in dark blue shades) 
and BL-Seyfert galaxies (in solid black lines) in the FIRST survey. 
The detection rates for BL-Seyfert galaxies are considered up to $z$ = 0.8 as our NLS1s sample is limited to $z$ = 0.8. 
BL-Seyfert galaxies are taken from Singh et al. (2015).}
\label{fig:RedshiftHist} 
\end{figure*}
It appears that the radio detection rate for our NLS1s is lower than that for BL-AGN. 
For instance, using a sample of 2800 BL-Seyfert galaxies \cite{Wadadekar04} found the FIRST counterparts for $\sim$ 27 per cent sources. 
While, in a much larger sample of 23448 BL-Seyfert galaxies \cite{Singh15a} reported the FIRST detection of nearly 11.3 per cent sources. 
In a sample of 8434 low-redshift ($z$ $<$ 0.35) BL-AGN derived from SDSS DR4 \cite{Rafter09} found the FIRST detection of $\sim$ 10 per cent sources. 
It is worth to note that the same radio survey ({\ie}FIRST) is used in all these studies, 
and therefore, the different detection rates in different samples can be attributed to factor(s) other than the sensitivity of 
radio survey. The radio detection rate of AGN can depend on the intrinsic factor(s) such as the jet power, 
jet production efficiency, star-formation rate and/or the observational bias 
introduced by the redshift distribution. 
Fig.~\ref{fig:RedshiftHist} (left panel) shows the redshift distributions for our NLS1s and BL-Seyfert galaxies studied in \cite{Singh15a}.  
We find that the redshift distribution for our 10508 NLS1s falling within the FIRST coverage ranges 
from 0.009 to 0.8 with a median of $\sim$ 0.47, while the FIRST-detected NLS1s are distributed over the same redshift range (0.009$-$0.8) 
but with a lower median redshift of $\sim$ 0.36. 
Furthermore, the redshift distribution of our NLS1s is completely different than that for BL-Seyfert galaxies. 
Contrary to our NLS1s, the redshift distribution for BL-Seyfert galaxies peaks at $z$ $\simeq$ 0.1$-$0.2, 
and declines sharply at $z$ $\geq$ 0.3. 
The other BL-AGN samples also contain relatively low-$z$ sources \citep[see][]{Wadadekar04,Rafter09}. 
Therefore, the lower radio detection rate for our NLS1s as compared to BL-Seyfert galaxies can be understood 
as our NLS1s are distributed at relatively higher redshifts. 
\\
We note that unlike our optically-selected NLS1s, BL-Seyfert galaxies 
in the sample of \cite{Singh15a} have absolute B-band magnitude (M$_{\rm B}$) lower than $-$22.25. 
Therefore, BL-Seyfert galaxies contain relatively less optically-luminous sources that are found preferentially at lower redshifts, 
while our NLS1s sample has a substantial fraction of more optically-luminous sources at higher redshifts. 
To account for this bias we consider a sub-sample of our NLS1s with sources of M$_{\rm B}$ $>$ $-$22.25. 
The B-band magnitudes for our NLS1s are derived from $u$ and $g$ band SDSS magnitudes 
using equation `B = $g$ + 0.17($u$ $-$ $g$) +0.11' \citep{Jester05}. 
As expected the low-luminosity NLS1s (LLNLS1s;  M$_{\rm B}$ $>$ $-$22.25) tend to decrease at 
higher redshifts (see Fig.~\ref{fig:RedshiftHist} left panel). 
The LLNLS1s are distributed over the redshift of 0.009 to 0.8 with a median of 0.33, 
while the redshifts of FIRST-detected LLNLS1s range from 0.028 to 0.8 with a median of 0.19. 
It is evident that the redshift distributions of FIRST-detected LLNLS1s shows a trend similar to that of 
FIRST-detected BL-Seyfert galaxies. 
\par
In Fig.~\ref{fig:RedshiftHist} (right panel) we also show the radio detection rates in different redshift bins 
for NLS1s, LLNLS1s and BL-Seyfert galaxies. 
It is apparent that both NLS1s and LLNLS1s have systematically lower detection rates in compared to BL-Seyfert galaxies 
in all redshift bins. At $z$ $\geq$ 0.1, the detection rate of BL-Seyfert galaxies ($\sim$ 10 per cent) 
is nearly twice of the detection rate of our NLS1s ($\sim$ 5 per cent). 
The radio detection rates for LLNLS1s is even lower than that for NLS1s.
Thus, we conclude that NLS1s tend to show lower radio detection rate in compared to BL-Seyfert galaxies. 
However, in the lowest redshift bin ($z$ $<$ 0.1), the radio detection rates of our NLS1s and BL-Seyfert galaxies are not too different 
({\ie}15 per cent and 17.5 per cent, respectively). 
The higher detection rate in the lowest redshift bin can be due to an observational bias where relatively less luminous 
radio sources are detected in the lowest redshift bin but remain undetected at higher redshifts. 
The lower radio detection rate of NLS1s in compared to BL-Seyfert galaxies can be indicative of the fact 
that, unlike in BL-Seyfert galaxies the radio emission in NLS1s is due to low power jets, and/or star-formation. 
\subsection{Radio flux densities}
\label{sec:fluxdensities} 
\begin{figure*}
\includegraphics[angle=0,width=9.0cm,trim={0.5cm 0.5cm 0.5cm 0.5cm},clip]{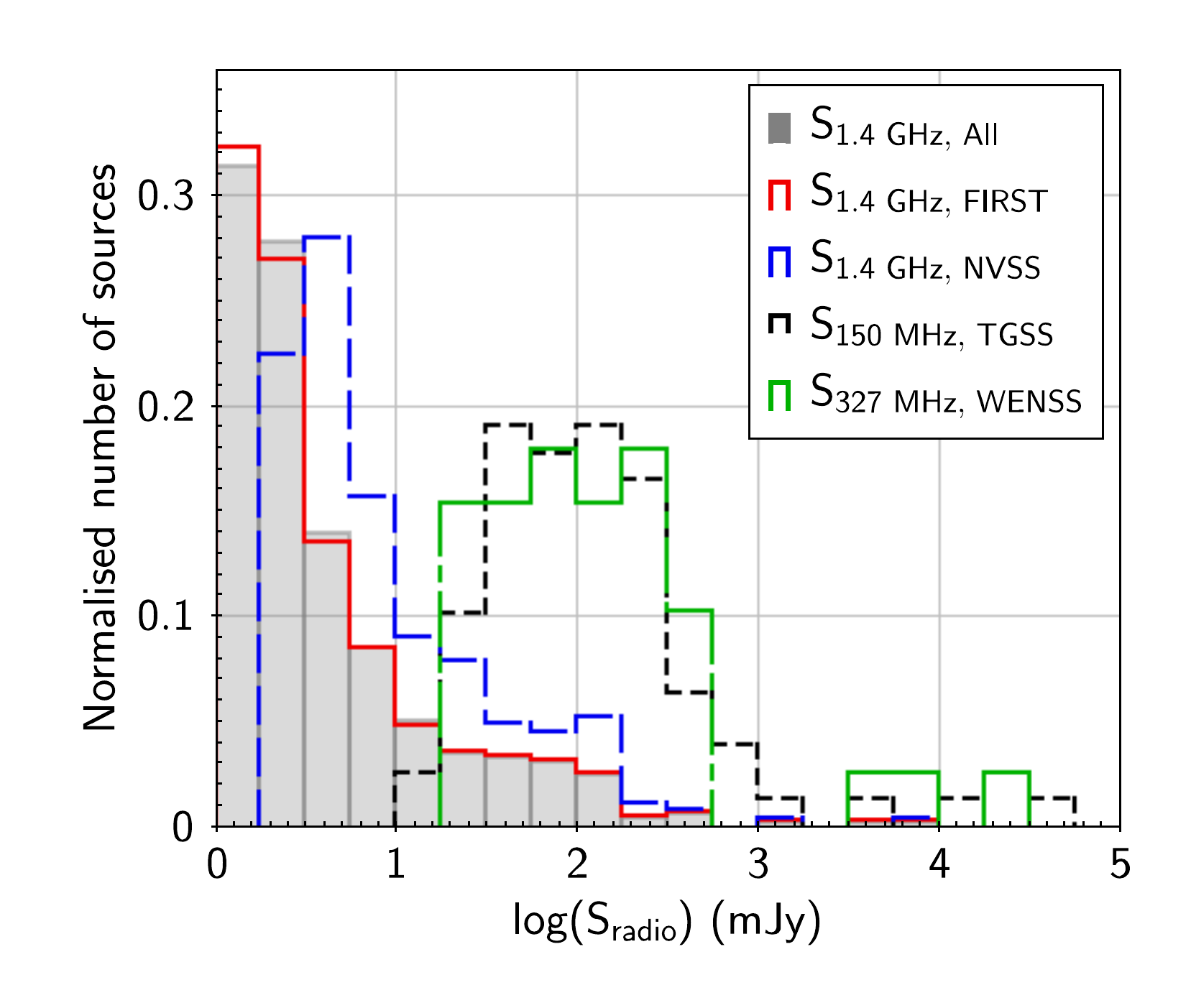}{\includegraphics[angle=0,width=9.0cm,trim={0.5cm 0.5cm 0.5cm 0.5cm},clip]{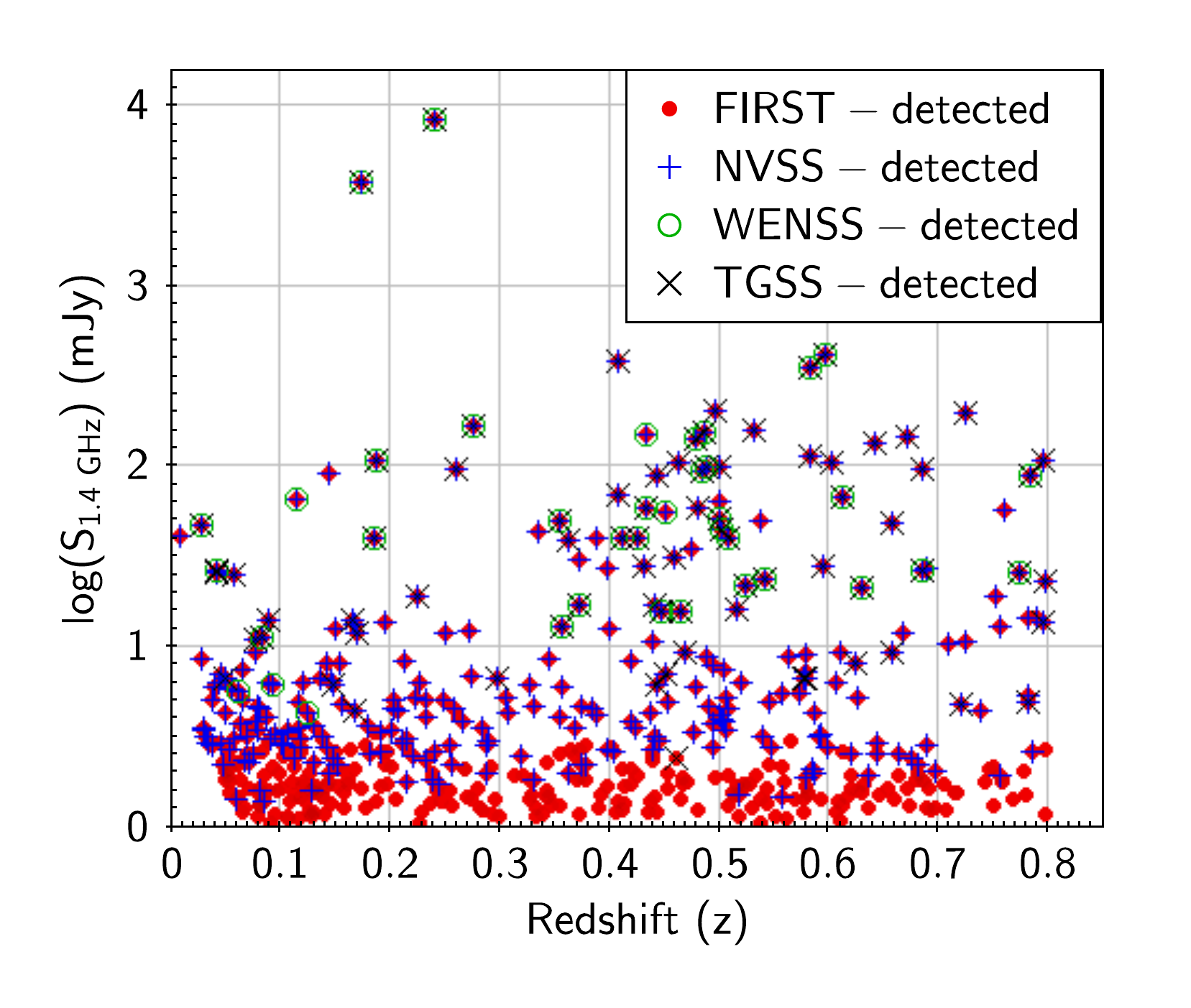}}
\caption{{\it Left panel} : Radio flux density distributions for our all radio-detected NLS1s (in grey shades) and 
NLS1s detected in the FIRST (in red solid lines), NVSS (in dashed blue lines), WENSS (in broken long-dashed green lines) 
and TGSS (small-dashed black lines). 
{\it Right panel} : Redshift versus 1.4 GHz flux density plot for our NLS1s. The 1.4 GHz flux density is from the FIRST survey, 
whenever available, otherwise NVSS flux density is considered. The sizes of error bars are smaller than the sizes of symbols.}
\label{fig:FluxHist} 
\end{figure*}

In Fig.~\ref{fig:FluxHist} (left panel) we show the 1.4 GHz, 327 MHz and 150 MHz radio flux density distributions of our NLS1s. 
The range, median, first quartile (Q1), third quartile (Q3), and interquartile range (IQR) of 
the flux density distributions of our NLS1s are listed in Table~\ref{table:HistProp}.  
The 1.4 GHz flux densities of our 498 radio-detected NLS1s range from 1.0 mJy to 8360.6 mJy with a median of 2.60 mJy. 
For 1.4 GHz flux density distribution of our NLS1s we consider FIRST flux density, whenever available, otherwise 
NVSS flux density is considered. 
The 1.4 GHz flux density distribution is mainly dominated by the FIRST-detections 
as 483/498 detections are from the FIRST and remaining are from the NVSS. 
In fact, flux densities of FIRST-detected NLS1s also range from 1.0 mJy to 8360.6 mJy with a median of 2.56 mJy. 
As expected, the NVSS detections are limited to relatively bright sources. 
The NVSS flux densities of 274 NLS1s range from 2.2 mJy to 8283.1 mJy with a median of 5.5 mJy. 
The TGSS and WENSS detect only radio-bright NLS1s due to their low sensitivities. 
The 1.4 GHz flux densities of 79 TGSS-detected NLS1s span over 4.2 mJy to 8360 mJy with a
median of 47.2 mJy. And, 39 WENNS-detected NLS1s have 1.4 GHz flux densities in the range of 2.4 mJy to 8360 mJy
with a median of 38.1 mJy. 
The 39 WENNS-detected NLS1s have 327 MHz flux densities in the range of 18 mJy to 30410 mJy with a median of 103 mJy. 
While, 150 MHz flux densities of 79 TGSS-detected NLS1s are distributed over 13.9 mJy to 53958 mJy with a median of 100.6 mJy. 
We note that the errors on flux densities are limited to a few per cent. 
NVSS and TGSS catalogues list the errors on total flux densities. 
While, for FIRST sources we estimate the errors in flux densities by considering 5 per cent systematic uncertainty in addition to 
the uncertainty due to local rms noise \citep[see][]{Becker95}. The errors on 327 MHz flux densities from WENSS are computed using the 
formula `${\sigma}_{\rm s}/{\rm S}$ $=$ $[{\rm C}_{\rm 1}^{2} + {\rm C}_{\rm 2}^{2}({\sigma}_{\rm rms}/{\rm S})]^{1/2}$'; where ${\sigma}_{\rm s}$ 
is the error in source flux density, ${\sigma}_{\rm rms}$ is the local rms noise, S is the source flux density, 
C$_{\rm 1}$ and C$_{\rm 2}$ are constants with values 0.04 and 1.3, respectively \citep[see][]{Rengelink97}.       
\par
The flux density distribution of our NLS1s suggests that NLS1s are mostly faint. 
The 1.4 GHz flux density distribution of our NLS1s peaks in the faintest bin of 1.0$-$2.5 mJy, 
and declines sharply above $\sim$ 5.0 mJy (see Fig.~\ref{fig:FluxHist}, left panel).  
In fact 50 per cent of the FIRST-detected NLS1s have flux densities in the range of 1.0 mJy to 2.6 mJy 
{\ie}the median flux density is $\sim$ 2.6 mJy. 
Also, there are 224/483 $\sim$ 47.6 per cent FIRST-detected sources that fall below the NVSS detection limit of 2.5 mJy. 
The faint radio emission in NLS1s is further vindicated by the fact that the radio detection rate of our NLS1s 
in the FIRST is merely 4.6 per cent, and therefore, remaining 95.4 per cent NLS1s have 1.4 GHz flux density less than 1.0 mJy. 
Hence, we conclude that, in general, radio emission in NLS1s is weak, and a much deeper radio survey 
is required to detect the radio emission in most of our NLS1s. 
It is worth to point out that our study yields the largest sample of radio-detected NLS1s among which 
a substantial fraction of NLS1s have flux density of $\sim$ 1.0$-$2.5 mJy. 
Most of the radio-detected NLS1s reported in previous studies are relatively bright 
with S$_{\rm 1.4~GHz}$ $>$ 2.0 mJy \citep[see][]{Foschini15,Gu15}. 
We also examine the distribution of radio flux densities {\it w.r.t.} redshift. 
In Fig.~\ref{fig:FluxHist} (right panel) we plot 1.4 GHz radio flux density versus redshift for our radio-detected NLS1s detected.  
It is evident that both faint as well as bright sources detected in different radio surveys are distributed across all redshifts. 
\subsection{Radio luminosities}
\label{sec:luminosities}
\begin{figure*}
\includegraphics[angle=0,width=9.0cm,trim={0.5cm 0.5cm 0.5cm 0.5cm},clip]{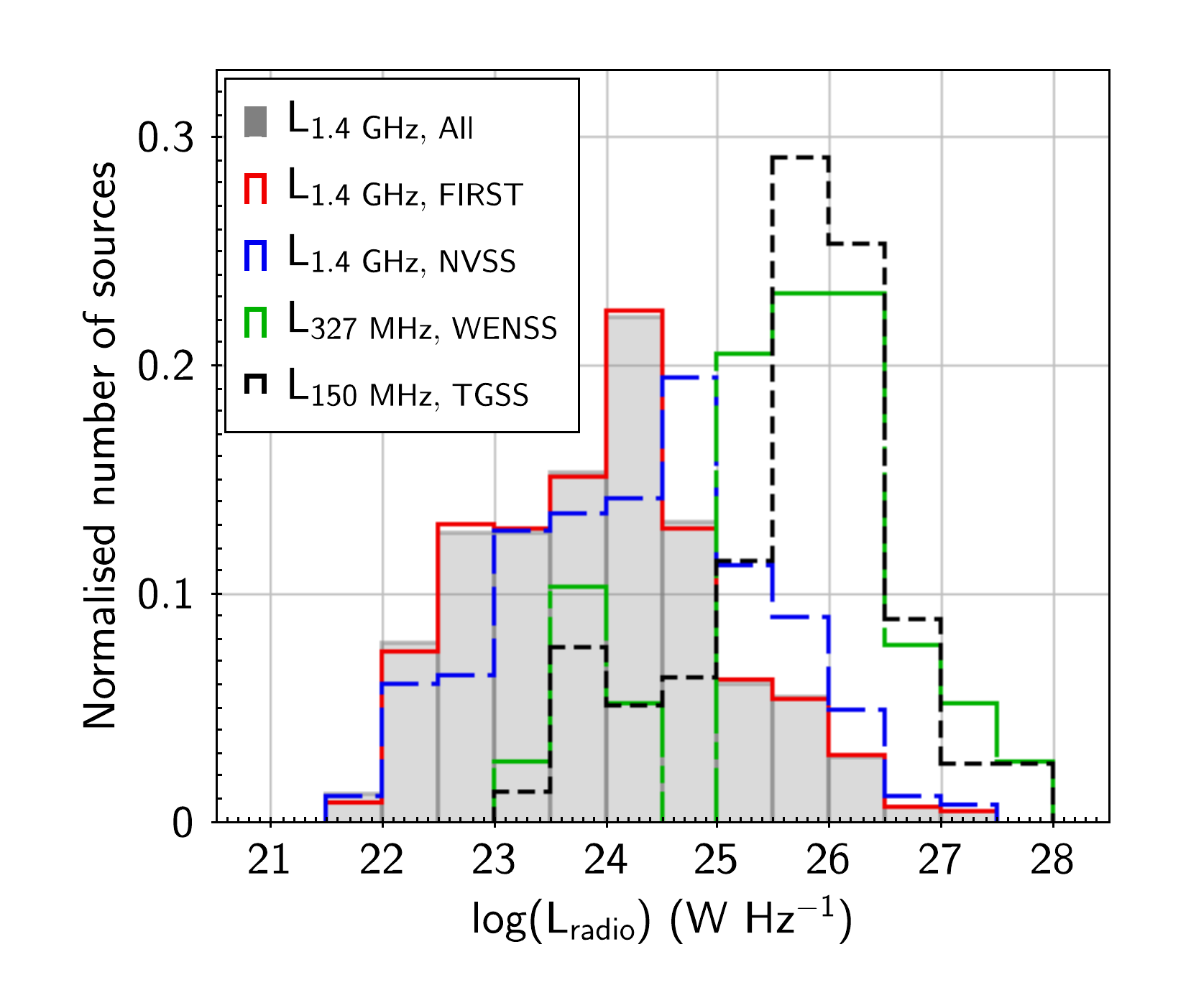}{\includegraphics[angle=0,width=9.0cm,trim={0.5cm 0.5cm 0.5cm 0.5cm},clip]{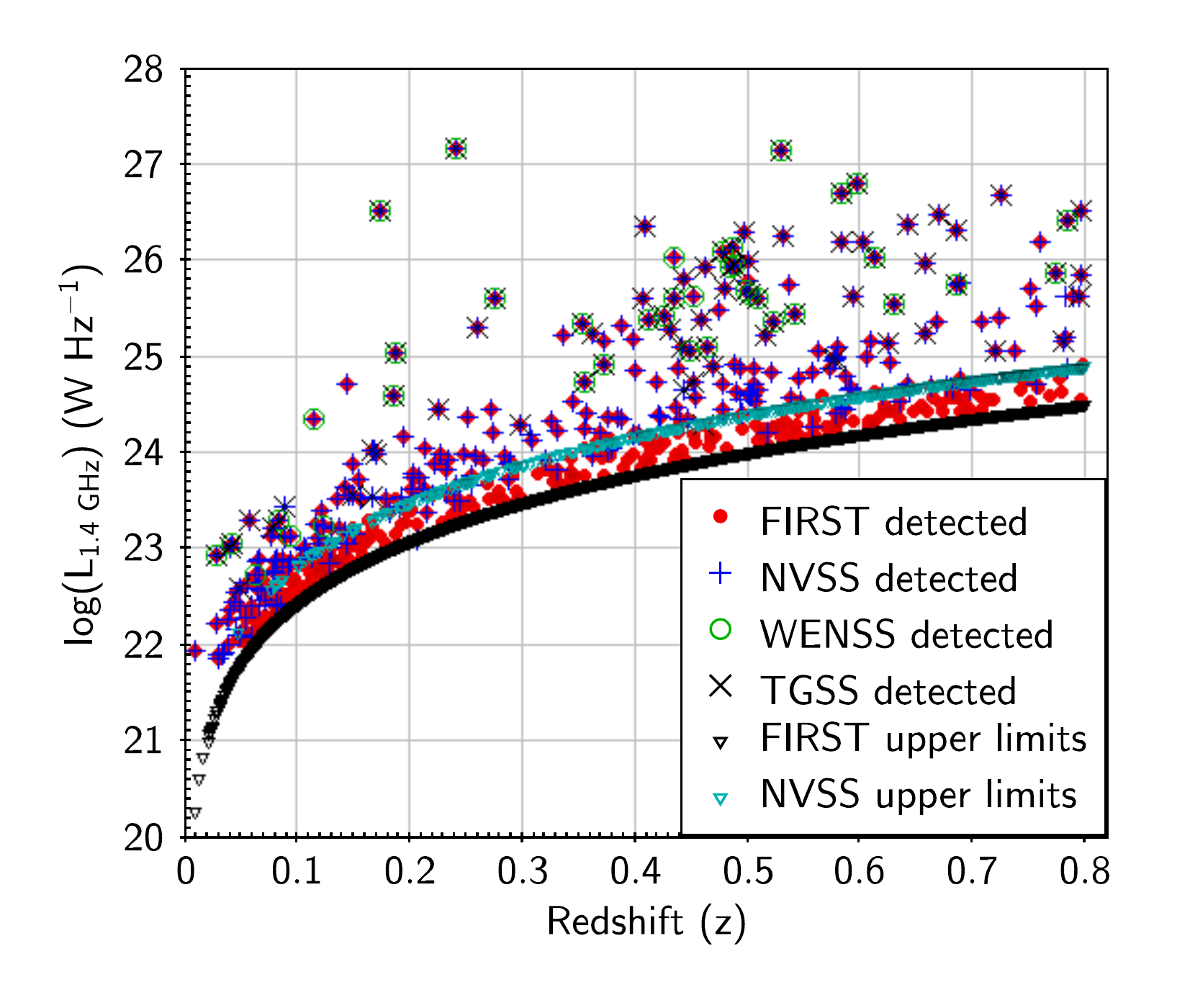}}
\caption{{\it Left panel} : Radio luminosity distributions for our all radio-detected NLS1s (in grey shades) and 
NLS1s detected in the FIRST (in red solid lines), NVSS (in dashed blue lines), WENSS (in broken long-dashed green lines) 
and TGSS (small-dashed black lines). {\it Right panel} : Redshift versus radio luminosity plot. 
The sizes of error bars are smaller than the sizes of symbols. 
The lower and upper curves represent the upper limits from FIRST and NVSS, respectively. 
}
\label{fig:LuminHist} 
\end{figure*}

Radio luminosity is a distance independent parameter and can give us insight into the nature of a source. 
Therefore, we investigate the radio luminosity distributions for our NLS1s. 
In Fig.~\ref{fig:LuminHist} (left panel) we show 1.4 GHz, 327 MHz and 150 MHz radio luminosity distributions 
of our NLS1s. The range, median, first quartile, third quartile, and interquartile range for different radio 
luminosity distributions are listed in Table~\ref{table:HistProp}. 
The radio luminosities are k-corrected using a simple power law radio spectrum with 
the spectral index measured between 1.4 GHz and 150 MHz whenever available, otherwise an average spectral index of 
$-$0.7 is used. The average radio spectral index of $-$0.7 is generally found in Seyfert galaxies \citep{Singh13}.
We note that the 1.4 GHz luminosity distribution of our NLS1s is primarily based on the FIRST measurements 
({\ie}483/498 detections are from FIRST), and NVSS flux density is used only when FIRST detection is not available. 
Our preference for the FIRST survey is due to the fact that it is better suited to detect nuclear radio emission owing 
to its relatively higher resolution ($\sim$ 5$\arcsec$.4). 
We find that the 1.4 GHz radio luminosity distribution of our 498 radio-detected NLS1s span across a wide range of luminosities 
ranging from 7.2 $\times$ 10$^{21}$ W Hz$^{-1}$ to 1.5 $\times$ 10$^{27}$ W Hz$^{-1}$ with 
a median value of 1.6 $\times$ 10$^{24}$ W Hz$^{-1}$. 
The 483 FIRST-detected NLS1s have radio luminosity distribution similar to the 498 radio-detected NLS1s. 
The 274 NVSS-detected NLS1s are distributed over a similar range 
(7.8 $\times$ 10$^{21}$ W Hz$^{-1}$ to 1.5 $\times$ 10$^{27}$ W Hz$^{-1}$) but with a little higher median value 
of 2.9 $\times$ 10$^{24}$ W Hz$^{-1}$. 
Therefore, the NVSS detections are slightly biased towards more radio-luminous sources. 
The NLS1s detected in the WENSS and TGSS consist of mainly radio-powerful sources (L$_{\rm 1.4~GHz}$ $\geq$ 10$^{24}$ W Hz$^{-1}$).    
The 1.4 GHz radio luminosity distribution of 39 WENNS-detected NLS1s spans 
over 5.2 $\times$ 10$^{22}$ W Hz$^{-1}$ to 1.5 $\times$ 10$^{27}$ W Hz$^{-1}$ with a median 
value of 4.0 $\times$ 10$^{25}$ W Hz$^{-1}$. 
While, 327 MHz luminosities of 39 WENNS-detected NLS1s are distributed over a range of 
1.65 $\times$ 10$^{23}$ W Hz$^{-1}$ to 3.94 $\times$ 10$^{27}$ W Hz$^{-1}$ with a median 
value of 5.45 $\times$ 10$^{25}$ W Hz$^{-1}$. 
The 79 TGSS-detected NLS1s have 1.4 GHz luminosities in the range of 
3.9 $\times$ 10$^{22}$ W Hz$^{-1}$ to 1.5 $\times$ 10$^{27}$ W Hz$^{-1}$ with a median of 2.6 $\times$ 10$^{25}$ W Hz$^{-1}$. 
The 150 MHz radio luminosity distribution of 79 TGSS-detected NLS1s ranges from 
1.89 $\times$ 10$^{23}$ W Hz$^{-1}$ to 4.89 $\times$ 10$^{27}$ W Hz$^{-1}$ with a median of 6.2 $\times$ 10$^{25}$ W Hz$^{-1}$. 
\par
We find that 224/498 NLS1s detected in the FIRST survey with no counterparts in the NVSS 
have 1.4 GHz luminosities in the range of 1.1 $\times$ $10^{22}$ W Hz$^{-1}$ to 7.9 $\times$ 10$^{24}$ W Hz$^{-1}$ 
with a median of 6.1 $\times$ 10$^{23}$ W Hz$^{-1}$. 
Therefore, our radio-faint NLS1s that fall below the NVSS detection limit of 2.5 mJy, contain 
a substantial fraction of sources with moderate radio luminosity.  
We also derive upper limits on the 1.4 GHz luminosities for NLS1s with no detected counterparts in the FIRST and NVSS. 
We use FIRST detection limit (1.0 mJy) if a source lie within the FIRST coverage, otherwise NVSS 
detection limit (2.5 mJy) is used. 
We find that the 1.4 GHz radio luminosity distribution for the upper limits of 10603/11101 ($\sim$ 95.5 per cent) NLS1s 
ranges from 1.8 $\times$ $10^{20}$ W Hz$^{-1}$ 
to 3.1 $\times$ 10$^{24}$ W Hz$^{-1}$ with a median of 8.7 $\times$ 10$^{23}$ W Hz$^{-1}$. 
Therefore, a substantial fraction of NLS1s with no detected radio counterparts in the FIRST and NVSS can 
have moderate radio luminosities even if their average true flux densities are nearly one order-of-magnitude 
lower than the FIRST detection limit 
{\ie}if the luminosity distribution for the upper limits is shifted by one order-of-magnitude. 
\par
In Fig.~\ref{fig:LuminHist} (right panel) we plot the 1.4 GHz radio luminosity versus redshift for our NLS1s detected in different surveys. 
It is apparent that the NLS1s with low radio power are only present at lower redshifts ({\ie}sources with L$_{\rm 1.4~GHz}$ $\leq$ 10$^{23}$ W Hz$^{-1}$ are 
lying at $z$ $\leq$ 0.2), while high luminosity sources (L$_{\rm 1.4~GHz}$ $\geq$ 10$^{24}$ W Hz$^{-1}$) are 
preferentially found at higher redshifts. The number of high luminosity sources increases with the increase in redshift. 
The increasing prevalence of radio-powerful NLS1s with the redshift is consistent with the fact the 
co-moving density of radio-powerful AGN increases with redshift over $z$ $\sim$ 0.0$-$2.0 \citep{Dunlop90,Willott01}. 
The lack of low-luminosity sources at higher redshifts is due to an observational bias, where the flux densities of 
low-luminosity sources at higher redshifts fall below the survey detection limit. 
In radio luminosity versus redshift plot we also show the upper limits for NLS1s with no radio counterparts in 
the FIRST and NVSS. 
It is evident that at $z$ $\geq$ 0.5, even powerful radio sources (L$_{\rm 1.4~GHz}$ $\geq$ 10$^{24}$ W Hz$^{-1}$) fall below the FIRST detection limit. 
Therefore, with a deeper radio survey we can expect to unveil a new population of radio-powerful NLS1s at higher redshifts ($z$ $\geq$ 0.5). 
\par
The 1.4~GHz luminosity distribution of our NLS1s spans across nearly seven order of magnitude ranging from radio-weak 
($\sim$ 7.1 $\times$ 10$^{21}$ W Hz$^{-1}$) to very radio-powerful sources ($\sim$ 1.5 $\times$ 10$^{27}$ W Hz$^{-1}$). 
At lower luminosity range ($\sim$ 10$^{21}$$-$10$^{24}$ W Hz$^{-1}$) it has substantial overlap with the radio luminosity distributions 
of Seyfert galaxies and quasars \citep[see][]{Ho08,Rafter11,Singh13}, while at higher luminosity range 
($\sim$ 10$^{24}$$-$10$^{27}$ W Hz$^{-1}$) it overlaps with radio galaxies \citep[see][]{An12}. 
Moreover, from the non-detection of majority of our NLS1s and 
the corresponding luminosity distribution for upper limits it is evident that most of our 
NLS1s are, in general, weak in radio, similar to Seyfert galaxies. And, only a small fraction of NLS1s are radio powerful. 
We note that for our radio-detected NLS1s the 1.4 GHz radio luminosity distribution of our NLS1s straddles across the dividing line 
(L$_{\rm 1.4~GHz}$ $\simeq$ 10$^{24.5}$ W Hz$^{-1}$) of Fanaroff-Riley (FR) type I and type II radio galaxies \citep{Fanaroff74}. 
FR II radio galaxies displaying highly collimated bipolar jets terminating into lobes with hotspots are found to be more 
radio powerful (L$_{\rm 1.4~GHz}$ $\geq$ 10$^{24.5}$ W Hz$^{-1}$) than FR I radio galaxies \citep{Rafter11}.  
We find that there are (152/498) $\sim$ 30.5 per cent radio-detected NLS1s 
that have 1.4 GHz radio luminosity similar to FR II radio galaxies 
(L$_{\rm 1.4~GHz}$ $\geq$ 10$^{24.5}$ W Hz$^{-1}$). 
Indeed, it appears surprising to find a large number of NLS1s with radio luminosities similar to that of FR II radio galaxies.
However, the fraction of NLS1s with L$_{\rm 1.4~GHz}$ $\geq$ 10$^{24.5}$ W Hz$^{-1}$ in our full optically-selected sample is 
merely (152/11101) $\sim$ 1.5 per cent. 
Also, unlike FR II radio galaxies most of the radio-powerful NLS1s exhibit compact radio structures (see Sect.~\ref{sec:Structures}). 
The compact and powerful radio emission in NLS1s infers that these sources may be either 
similar to blazars in which radio jet is directly pointed towards the observer and the projected linear radio size 
is small \citep{Urry95}, or similar to Compact Steep Spectrum (CSS) radio sources that are intrinsically small 
in radio size and are believed to represent the early phase of the evolution of radio galaxies \citep{Giroletti09,Orienti16}. 
A detailed discussion on the radio powerful NLS1s is deferred to Sect.~\ref{sec:Discussion}.
\subsection{Radio spectra}
\label{sec:Spectra}
Radio spectra are useful in unveiling the nature of radio sources. 
We investigate radio spectral properties of our NLS1s using their flux densities at 1.4 GHz from FIRST and NVSS, 
327 MHz from WENSS and 150 MHz from TGSS. 
\subsubsection{150~MHz - 1.4~GHz two-point spectral indices (${\alpha}_{\rm 150~MHz}^{\rm 1.4~GHz}$)}
We derive two-point radio spectral index between 150 MHz and 1.4 GHz (${\alpha}_{\rm 150~MHz}^{\rm 1.4~GHz}$) for our NLS1s, 
where radio spectrum is assumed to be a power law (S$_{\nu}$ $\propto$ ${\nu}^{\alpha}$). 
To estimate ${\alpha}_{\rm 150~MHz}^{\rm 1.4~GHz}$ we use 150 MHz flux density 
from TGSS, and 1.4 GHz flux density from FIRST, whenever available, otherwise NVSS flux density is considered. 
We note that only 79/498 ($\sim$ 16 per cent) of our 1.4 GHz-detected NLS1s have counterparts in 150 MHz TGSS, therefore, 
${\alpha}_{\rm 150~MHz}^{\rm 1.4~GHz}$ estimates are available only for 16 per cent of our radio-detected NLS1s. 
For remaining 419/498 ($\sim$ 84 per cent) 1.4~GHz-detected NLS1s with no counterparts in 150 MHz TGSS we derive the upper limits on ${\alpha}_{\rm 150~MHz}^{\rm 1.4~GHz}$ 
by using the upper limit on 150 MHz flux density which is set equal to the flux density cut-off $\sim$ 24.5 mJy in the TGSS catalogue. 
The lack of 150 MHz counterparts for majority of 1.4~GHz-detected NLS1s can mainly be attributed to much lower sensitivity 
($\sigma$ $\sim$ 3.5 mJy) of TGSS combined with a relatively high cut-off of 7$\sigma$ ($\sim$ 24.5 mJy) imposed in 
the TGSS source catalogue \citep[see][]{Intema17}. 
We also caution that our spectral index estimates can be affected by the resolution bias as the spatial resolutions at 1.4 GHz 
and 150 MHz are different {\ie}5$\arcsec$.4 and 25$\arcsec$, respectively. 
The resolution bias can result a spectrum steeper than the actual, if some radio emission is 
missed in the high-resolution FIRST observations. In case of compact sources the resolution bias has no effect on the spectral 
index estimates.  
Since majority of our sources are compact in the FIRST survey (see Sect.~\ref{sec:Structures}), and therefore, resolution bias 
is unlikely to have any significant impact on our spectral index estimates. 
This is further vindicated by the fact that ${\alpha}_{\rm 150~MHz}^{\rm 1.4~GHz}$ estimates 
are nearly same if we use NVSS flux densities (resolution $\sim$ 45$\arcsec$) instead of FIRST flux densities 
(resolution $\sim$ 5$\arcsec$.4).    
\par
\begin{figure*}
\includegraphics[angle=0,width=9.0cm,trim={0.5cm 0.5cm 0.5cm 0.5cm},clip]{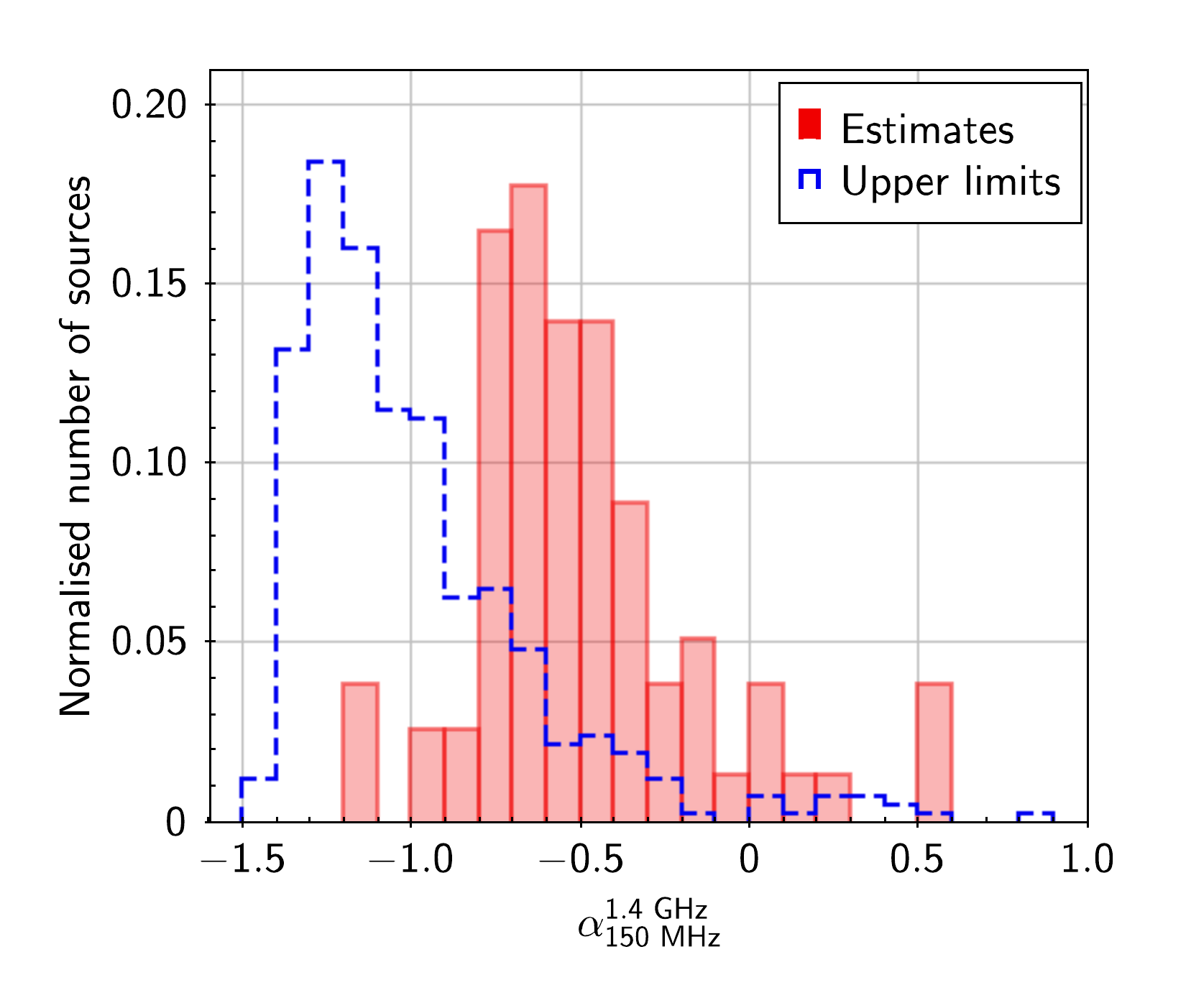}{\includegraphics[angle=0,width=9.0cm,trim={0.5cm 0.5cm 0.5cm 0.5cm},clip]{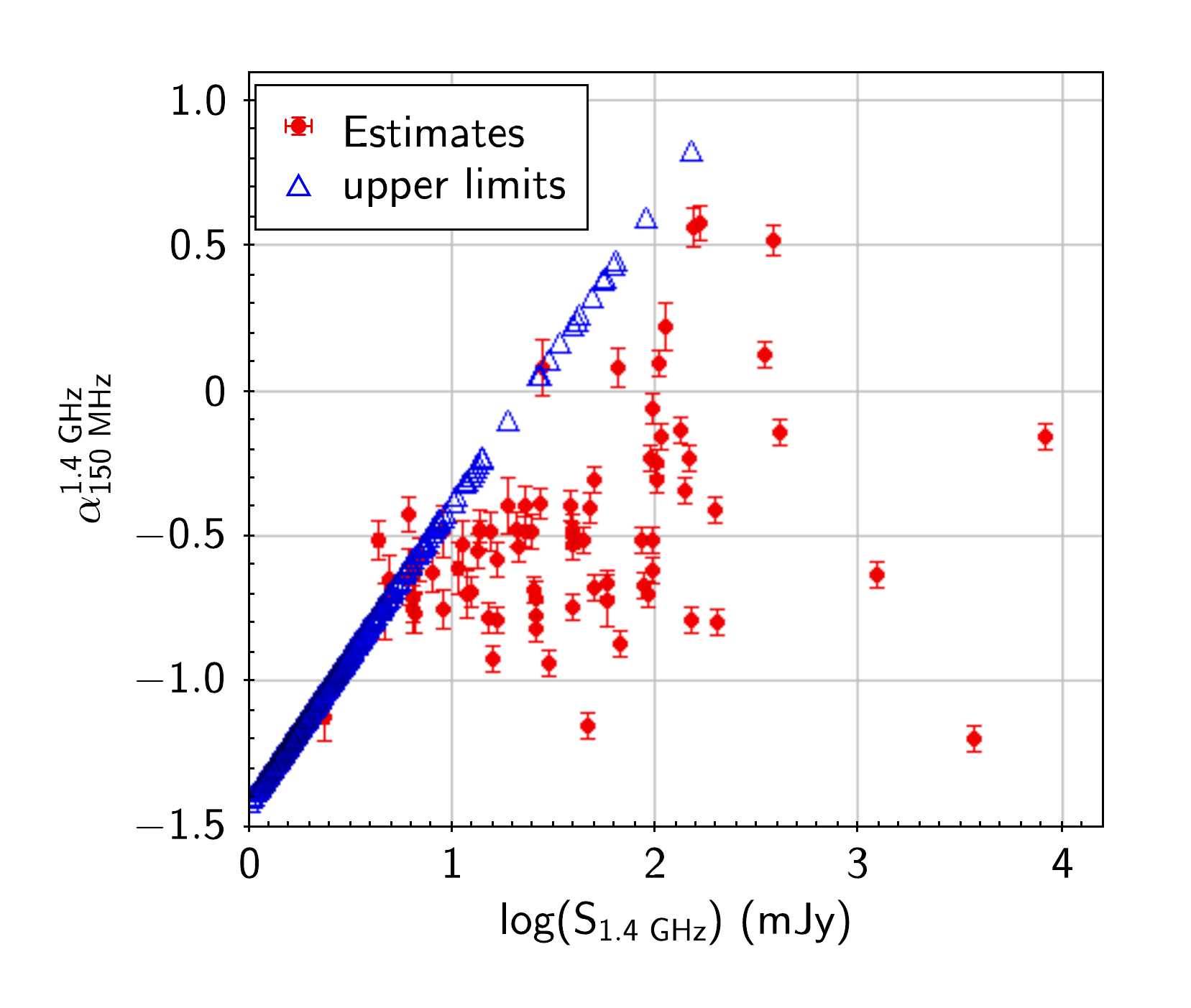}}
\caption{{\it Left panel} : The distribution of the spectral index between 150 MHz and 1.4 GHz (${\alpha}_{\rm 150~MHz}^{\rm 1.4~GHz}$) 
for our all radio-detected NLS1s. 
{\it Right panel} : 1.4 GHz flux density versus ${\alpha}_{\rm 150~MHz}^{\rm 1.4~GHz}$ for our all radio-detected NLS1s.}
\label{fig:SpInHist} 
\end{figure*}

In Fig.~\ref{fig:SpInHist} (left panel) we show the distribution of ${\alpha}_{\rm 150~MHz}^{\rm 1.4~GHz}$ 
including the upper limits for our all radio-detected NLS1s. 
We find that ${\alpha}_{\rm 150~MHz}^{\rm 1.4~GHz}$ estimates range from $-$1.2 to 0.58 with a median value of $-$0.53. 
While the upper limits on ${\alpha}_{\rm 150~MHz}^{\rm 1.4~GHz}$ span from $-$1.43 to 0.8 with a median value of $-$1.08. 
Based on ${\alpha}_{\rm 150~MHz}^{\rm 1.4~GHz}$ we can classify radio sources into flat spectrum 
(${\alpha}_{\rm 150~MHz}^{\rm 1.4~GHz}$ $\geq$ $-$0.5), steep spectrum ($-$1.0 $<$ ${\alpha}_{\rm 150~MHz}^{\rm 1.4~GHz}$ $\leq$ $-$0.5) 
and Ultra-Steep Spectrum (USS, ${\alpha}_{\rm 150~MHz}^{\rm 1.4~GHz}$ $\leq$ $-$1.0) sources.  
We find that our NLS1s consist of both flat spectrum sources (34/79 $\sim$ 43 per cent) as well as 
steep spectrum sources (42/79 $\sim$ 53 per cent). And, the fraction of USS sources is very small (03/79 $\sim$ 4 per cent). 
It is important to note that unlike a general radio population detected in a radio survey our NLS1s 
have substantially higher fraction of flat spectrum sources which is also vindicated with the fact that 
the median of ${\alpha}_{\rm 150~MHz}^{\rm 1.4~GHz}$ distribution ($\sim$ $-$0.53) for our NLS1s is higher ({\ie}flatter) 
than that found for a general radio population at low-frequency. 
For instance, \cite{Mahony16} reported that the radio sources detected in 150 MHz survey of Lockman Hole field 
have a median value of ${\alpha}_{\rm 150~MHz}^{\rm 1.4~GHz}$ $\sim$ $-$0.78, with 
only 5.7 per cent radio sources classified as flat spectrum sources (${\alpha}_{\rm 150~MHz}^{\rm 1.4~GHz}$ $>$ $-$0.5). 
In 325 MHz deep survey of ELAIS-N1 field \cite{Sirothia09} found a median value of $\sim$ $-$0.83 for the distribution of 
spectral index between 325 MHz and 1.4 GHz (${\alpha}_{\rm 325~MHz}^{\rm 1.4~GHz}$). 
The higher value of the median of ${\alpha}_{\rm 150~MHz}^{\rm 1.4~GHz}$ for our NLS1s can be explained with two possible scenarios : 
(i) unlike a general radio population the NLS1s tend to exhibit flat radio spectra, or (ii) average spectral index varies 
with flux density such that the radio-bright NLS1s exhibit relatively flat radio spectra, 
while the radio-faint NLS1s are dominated by steep spectrum sources.  
\par 
In order to probe the variation of average spectral index with flux density we plot 1.4 GHz flux density 
(S$_{\rm 1.4~GHz}$) versus ${\alpha}_{\rm 150~MHz}^{\rm 1.4~GHz}$ for our NLS1s (Fig.~\ref{fig:SpInHist}, right panel). 
From S$_{\rm 1.4~GHz}$ versus ${\alpha}_{\rm 150~MHz}^{\rm 1.4~GHz}$ plot 
it is apparent that the NLS1 with flat/inverted spectra tend to be found at higher flux densities, 
for instance, the median 1.4 GHz flux density for flat spectrum NLS1s (${\alpha}_{\rm 150~MHz}^{\rm 1.4~GHz}$ $\geq$ $-$0.5) is 
$\sim$ 95.5 mJy in compared to $\sim$ 25.2 mJy for steep spectrum NLS1s (${\alpha}_{\rm 150~MHz}^{\rm 1.4~GHz}$ $<$ $-$0.5). 
Also, from ${\alpha}_{\rm 150~MHz}^{\rm 1.4~GHz}$ upper limits it is evident 
that all radio-bright NLS1s at S$_{\rm 1.4~GHz}$ $>$ 10 mJy with no detected 150 MHz counterparts have flat/inverted spectra 
(see Fig.~\ref{fig:SpInHist}, right panel). 
Therefore, unconfirmed steep spectrum NLS1s among ${\alpha}_{\rm 150~MHz}^{\rm 1.4~GHz}$ upper limits, if exist,  
can only be present at the fainter end. 
Thus, we conclude that the average radio spectral index of our NLS1s seems to depend on the flux density such that 
the radio-bright NLS1s tend to show flat radio spectra. 
It is worth to mention that the trend shown by our NLS1s in the flux density versus spectral index plot is 
somewhat opposite to that found for a general radio population. 
In fact, bright radio population detected in a radio survey is dominated by powerful radio galaxies that often show 
steep radio spectra. Thus, in a general radio population the average spectral index 
becomes higher ({\ie}flatter) with decrease in flux density \citep[see][]{Ishwara-Chandra10,Intema11,Williams13}.    
The opposite trend shown by our NLS1s can be understood if most of our radio-bright NLS1s exhibit 
predominantly flat spectra resulted from synchrotron self-absorption, 
while radio-faint NLS1s are powered by optically thin synchrotron emission from AGN or star-formation.
\subsubsection{Radio colour-colour plot (${\alpha}_{\rm 150~MHz}^{\rm 327~MHz}$ versus ${\alpha}_{\rm 327~MHz}^{\rm 1.4~GHz}$)}
We note that the two-point spectral index measured between 150 MHz and 1.4 GHz for our NLS1s 
provides important information about the predominant population, however, it 
does not probe the spectral curvature, if exists. To probe the radio spectral curvature in our NLS1s we compare 
the two-point spectral index measured between 150 MHz and 327 MHz (${\alpha}_{\rm 150~MHz}^{\rm 327~MHz}$), with 
the two-point spectral index measured between 327 MHz and 1.4 GHz (${\alpha}_{\rm 327~MHz}^{\rm 1.4~GHz}$). 
Among our 498 radio-detected NLS1s only 79 sources are detected at 150~MHz as well 1.4~GHz. 
And, only 35/79 NLS1s are detected at 327 MHz in WENSS. 
Therefore, only 35 NLS1s have ${\alpha}_{\rm 150~MHz}^{\rm 327~MHz}$ as well as ${\alpha}_{\rm 327~MHz}^{\rm 1.4~GHz}$ estimates. 
Among the remaining 44 NLS1s that are detected at 150 MHz as well as 1.4 GHz, 
only 11 sources lie in the WENSS coverage area with no detection at 327 MHz. 
Therefore, for 11 NLS1s we have upper limits on ${\alpha}_{\rm 150~MHz}^{\rm 327~MHz}$ 
and lower limits on ${\alpha}_{\rm 327~MHz}^{\rm 1.4~GHz}$. 
There are only four NLS1s that are detected at 327 MHz and 1.4 GHz but not at 150 MHz, and thus, for these four sources we 
have ${\alpha}_{\rm 327~MHz}^{\rm 1.4~GHz}$ estimates and only upper limits on ${\alpha}_{\rm 150~MHz}^{\rm 327~MHz}$.  
\par
\begin{figure}
\includegraphics[angle=0,width=9.0cm,trim={0.0cm 0.0cm 0.5cm 0.5cm},clip]{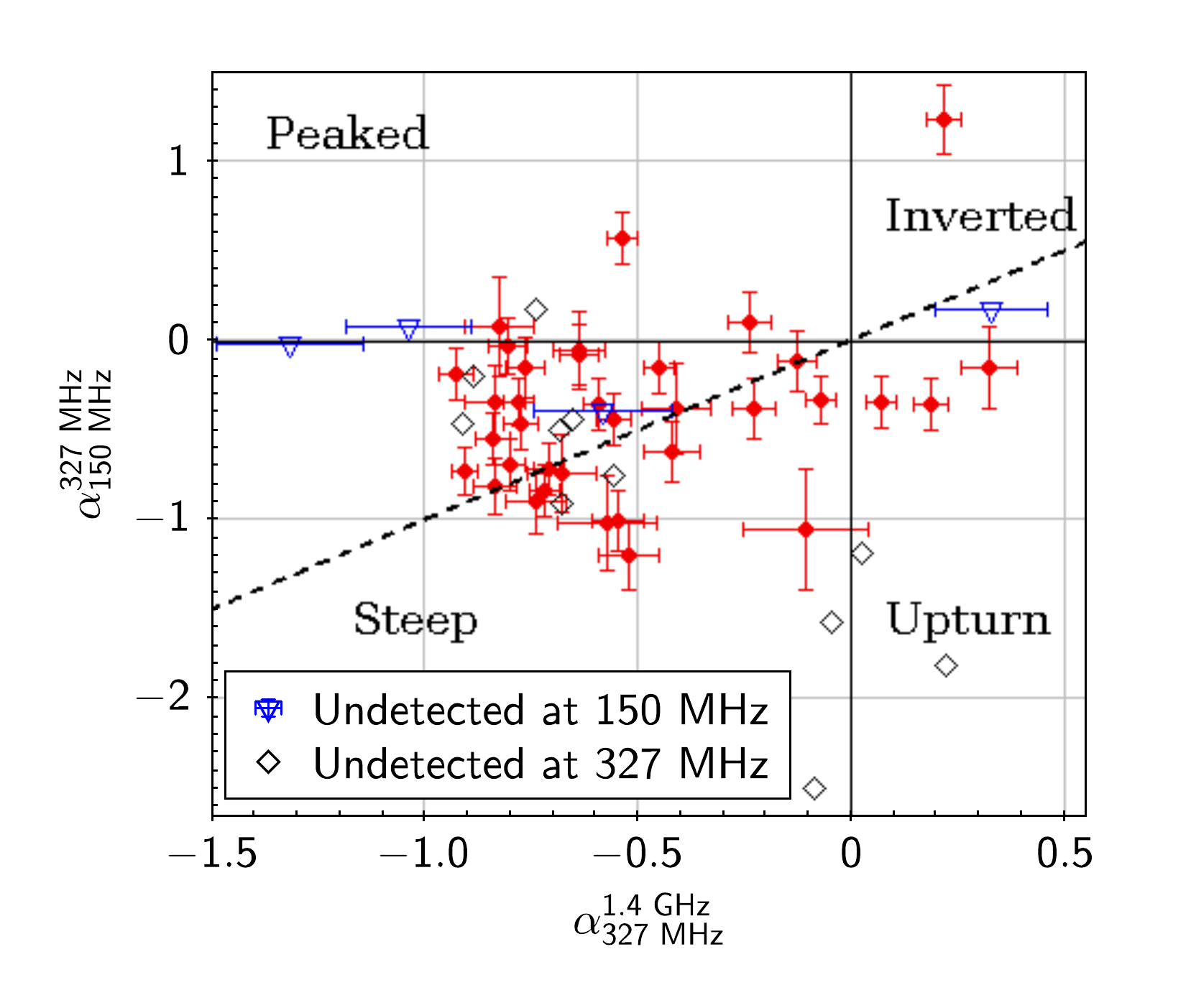}
\caption{Radio colour-colour plot for the subsample of 50 NLS1s.}
\label{fig:RadioColor} 
\end{figure}

In Fig.~\ref{fig:RadioColor} we show ${\alpha}_{\rm 150~MHz}^{\rm 327~MHz}$ versus ${\alpha}_{\rm 327~MHz}^{\rm 1.4~GHz}$ 
{\ie}radio colour-colour plot for our 50 NLS1s.   
We divide radio colour-colour plot into four quadrants {\ie}steep and flat spectrum sources 
(${\alpha}_{\rm 150~MHz}^{\rm 327~MHz}$ $<$0 and ${\alpha}_{\rm 327~MHz}^{\rm 1.4~GHz}$ $<$ 0), 
peaked spectrum sources (${\alpha}_{\rm 150~MHz}^{\rm 327~MHz}$ $>$0 and ${\alpha}_{\rm 327~MHz}^{\rm 1.4~GHz}$ $<$ 0), 
inverted spectrum sources (${\alpha}_{\rm 150~MHz}^{\rm 327~MHz}$ $>$0 and ${\alpha}_{\rm 327~MHz}^{\rm 1.4~GHz}$ $>$ 0), and 
upturn spectrum sources (${\alpha}_{\rm 150~MHz}^{\rm 327~MHz}$ $<$0 and ${\alpha}_{\rm 327~MHz}^{\rm 1.4~GHz}$ $>$ 0). 
It is apparent that majority of our NLS1s lie in the steep and flat spectrum quadrant. 
And, sources with no spectral curvature in 150 MHz - 1.4 GHz frequency range lie on or close to the diagonal line 
of ${\alpha}_{\rm 150~MHz}^{\rm 327~MHz}$ = ${\alpha}_{\rm 327~MHz}^{\rm 1.4~GHz}$. 
It is evident that a fraction of our NLS1s tend to exhibit spectral curvature as they deviate from the diagonal line 
(see Fig.~\ref{fig:RadioColor}). 
Furthermore, the scatter around the diagonal line is asymmetric with relatively more number of sources lying on 
the upper left side of the diagonal line 
({\ie}at ${\alpha}_{\rm 327~MHz}^{\rm 1.4~GHz}$ $<$ 0, there are 16 and 9 sources above and below the diagonal line, respectively). 
All the sources lying on the upper left side of the diagonal line indicate that their spectra are 
steep over 1.4 GHz - 327 MHz but become relatively flatter over 327 MHz - 150 MHz. 
Therefore, we infer that NLS1s tend to exhibit spectral flattening at lower frequencies.
Also, few NLS1s fall into peaked spectrum quadrant suggesting that they show spectral peak {\ie}turnover 
in 150 MHz - 327 MHz range. Sources falling into inverted quadrant possess inverted spectra across 150 MHz - 327 MHz - 1.4 GHz, 
and can exhibit peak at $>$ 1.4 GHz. 
There is a small fraction of sources showing upturn {\ie}spectra become steeper at lower frequencies. 
The upturn in the spectra can be caused of by variability. 
In fact, there are three NLS1s (J093323-001052 alias PMN J0933-0012, J084958+510828 alias SBS0846+513, 
J094857+002226 alias PMN J0948+0022) with upturn spectra that are confirmed 
to show variability based on the comparison between FIRST and NVSS flux densities 
{\ie}S$_{\rm 1.4~GHz,~NVSS}$ $<$ S$_{\rm 1.4~GHz,~FIRST}$ (see Sect.~\ref{sec:Variability}). 
Also, two (PMN 0948+0022, and SBS0846+513) of these three NLS1s are well known $\gamma$-ray detected 
blazar-like NLS1s \citep[see][]{Abdo09a,Paliya16}. 
So, we caution that spectral curvature found in our NLS1s is based on non-contemporaneous 
observations at different frequencies, and therefore, it may be affected in case of strongly variable sources.     
\subsection{Radio structures}
\label{sec:Structures}
Radio structures seen in AGN are important in understanding the evolutionary stage of radio-jets. 
Therefore, we investigate the nature of radio structures in our NLS1s. 
To detect the radio structures we prefer to use FIRST survey owing to its better sensitivity and resolution. 
Whenever FIRST detection is unavailable we use NVSS as it is more sensitive than the WENSS and TGSS. 
\subsubsection{Identification of resolved and unresolved sources} 
First we attempt to identify resolved and unresolved radio sources. 
Unresolved sources have spatial extent less than the size of synthesized beam and appear point-like, 
while resolved sources are larger than the synthesized beam.    
We note that the size of a single component radio source can be estimated from the deconvolved values of major and minor axis of the fitted 
Gaussian. However, values of major and minor axis of the Gaussian fit can be affected by the presence of positive 
and negative noise spikes, in particular, at faint end. 
Therefore, to identify resolved and unresolved sources we use a method based on 
the ratio of integrated to peak flux density (S$_{\rm int}$/S$_{\rm peak}$) that allows us to account for the error introduced by 
the noise \citep[see][]{Prandoni06,Sirothia09}.    
Since the flux density of a radio source is derived from the fitted Gaussian, the ratio of integrated to peak flux density can be expressed as 
S$_{\rm int}$/S$_{\rm peak}$ = ${\theta}_{\rm maj}~{\theta}_{\rm min}$/b$_{\rm maj}$~b$_{\rm min}$, where 
${\theta}_{\rm maj}$ and ${\theta}_{\rm min}$ are the fitted major and minor axis of radio source, respectively, 
while b$_{\rm maj}$ and b$_{\rm min}$ are the major and minor axis of the synthesized beam, respectively. 
For unresolved sources S$_{\rm int}$/S$_{\rm peak}$ = 1 as the fitted source size is equal to the beam-size, 
while for resolved sources S$_{\rm int}$/S$_{\rm peak}$ $>$ 1 as the fitted source size is larger than the beam-size. 
\par
\begin{figure}
\includegraphics[angle=0,width=8.5cm,trim={0.5cm 0.0cm 0.5cm 0.5cm},clip]{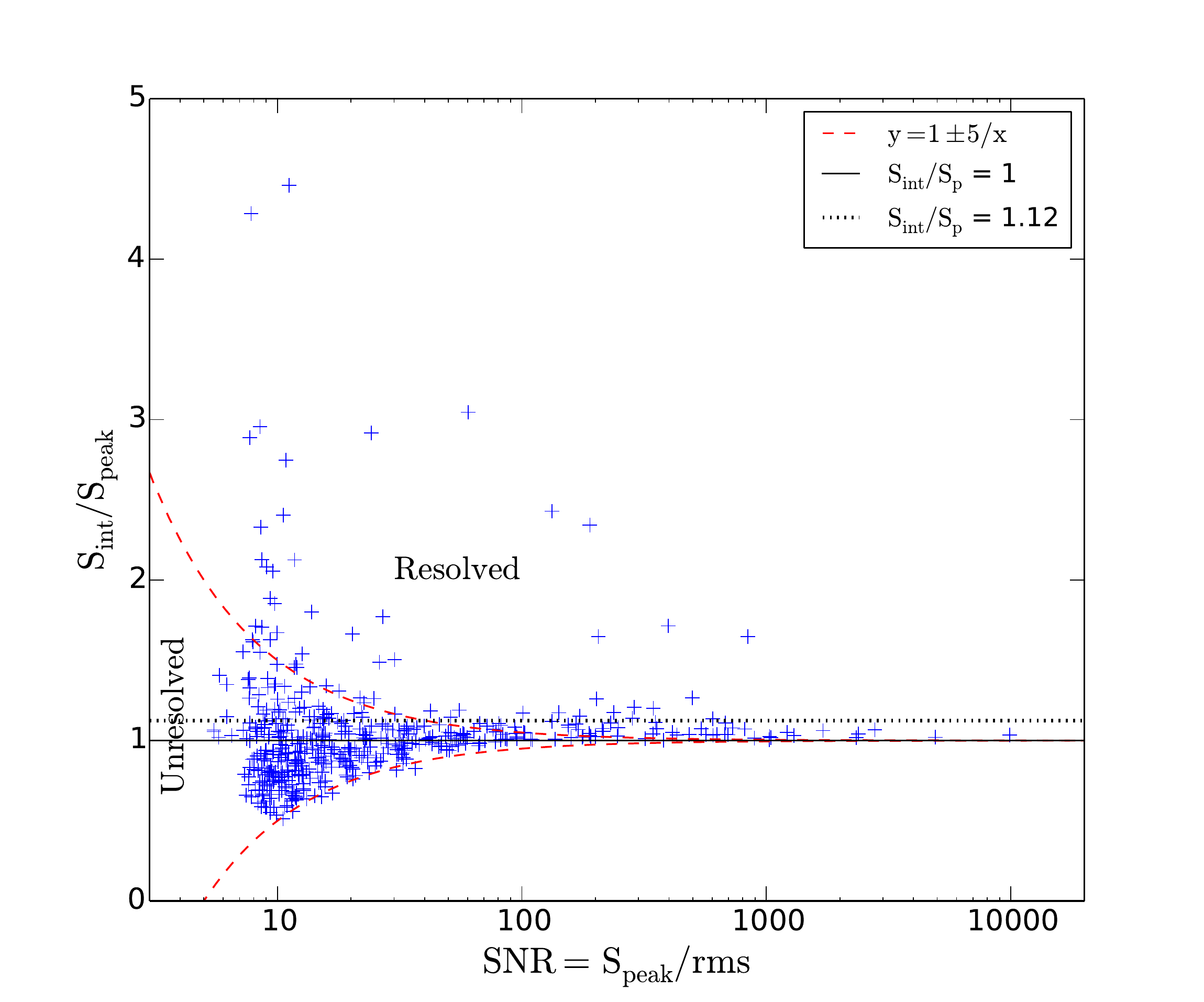}
\caption{S$_{\rm int}$/S$_{\rm peak}$ versus SNR plot for our all radio-detected NLS1s.}
\label{fig:FluxRatio} 
\end{figure}

In Fig.~\ref{fig:FluxRatio} we plot S$_{\rm int}$/S$_{\rm peak}$ versus signal-to-noise ratio (SNR) 
for our all 498 radio-detected NLS1s, where SNR is defined as the ratio of peak flux density to local noise-rms 
({\ie}SNR = S$_{\rm peak}$/${\sigma}_{\rm local}$). 
We note that at faint end (low SNR) there are many sources with S$_{\rm int}$/S$_{\rm peak}$ $<$ 1. 
The unexpected lower value of S$_{\rm int}$ as compared to S$_{\rm peak}$ is due to the fact that a radio source fitted with a Gaussian 
is affected by the presence of negative noise spikes. 
Indeed, the effect of noise is more pronounced at the fainter end. 
In order to account for the errors involved in the estimation of fitted size and flux density we follow an approach described 
in \cite{Sirothia09}. 
In the S$_{\rm int}$/S$_{\rm peak}$ versus SNR plot we first fit a curve with a functional form ${\rm f(x) = 1 - {\frac{5}{x}}}$ 
such that all the points with S$_{\rm int}$/S$_{\rm peak}$ $<$ 1 lie above the curve (see Fig.~\ref{fig:FluxRatio}). 
The Reflection of this curve {\ie}${\rm f(x) = 1 + {\frac{5}{x}}}$ about S$_{\rm int}$/S$_{\rm peak}$ $=$ 1 line renders an upper limit 
for the unresolved sources. 
Therefore, to account for the errors due to noise we consider all the sources lying between the two curves 
(${\rm f(x) = 1 {\pm} {\frac{5}{x}}}$) as unresolved. 
We find that there are a total of 396/498 ($\sim$ 79.5 per cent) unresolved sources that lie in between the two curves. 
We note that, to find unresolved sources, the functional form of the curve in S$_{\rm int}$/S$_{\rm peak}$ versus SNR plot 
is chosen to be ${\rm f(x) = 1 {\pm} {\frac{3.22}{x}}}$ in \cite{Sirothia09}, 
which can be understood as the radio sources in their catalogue are at SNR $\geq$ 6$\sigma$, 
in compared to the sources with SNR $\geq$ 5$\sigma$ in the FIRST and NVSS catalogues. 
It is important to note that the essence lies in accounting for errors due to the noise 
rather than using a particular functional form of the curve fitted to unresolved sources in S$_{\rm int}$/S$_{\rm peak}$ versus SNR plot.
\par
After accounting for the errors due to noise, we apply another criterion based on S$_{\rm int}$/S$_{\rm peak}$ 
to segregate resolved and unresolved sources. 
We note that \cite{Kimball08} proposed that the FIRST radio sources can be segregated into resolved and unresolved sources based on 
a separating line at $\sqrt{\rm S_{int}/S_{peak}}$ = 1.06, which is empirically found by using the distribution 
of FIRST radio sources in the plot of concentration parameter $\theta$ = $\sqrt{\rm S_{int}/S_{peak}}$ versus extended flux density. 
Therefore, a radio source in the FIRST survey can be considered as unresolved if ${\rm S_{int}/S_{peak}}$ $\leq$ 1.12 and resolved if ${\rm S_{int}/S_{peak}}$ $>$ 1.12. 
Using this criterion (shown by a dotted horizontal line ${\rm S_{int}/S_{peak}}$ $=$ 1.12 in Fig.~\ref{fig:FluxRatio}) 
we obtain 47 additional unresolved sources. Therefore, in total, we find 443 ($\sim$ 89 per cent) unresolved sources and 55 ($\sim$ 11 per cent) resolved 
radio sources in our 498 radio-detected NLS1s. 
\subsubsection{Projected linear radio sizes}
\label{sec:Size}
\begin{figure*}
\includegraphics[angle=0,width=9.0cm,trim={0.5cm 0.5cm 0.5cm 0.5cm},clip]{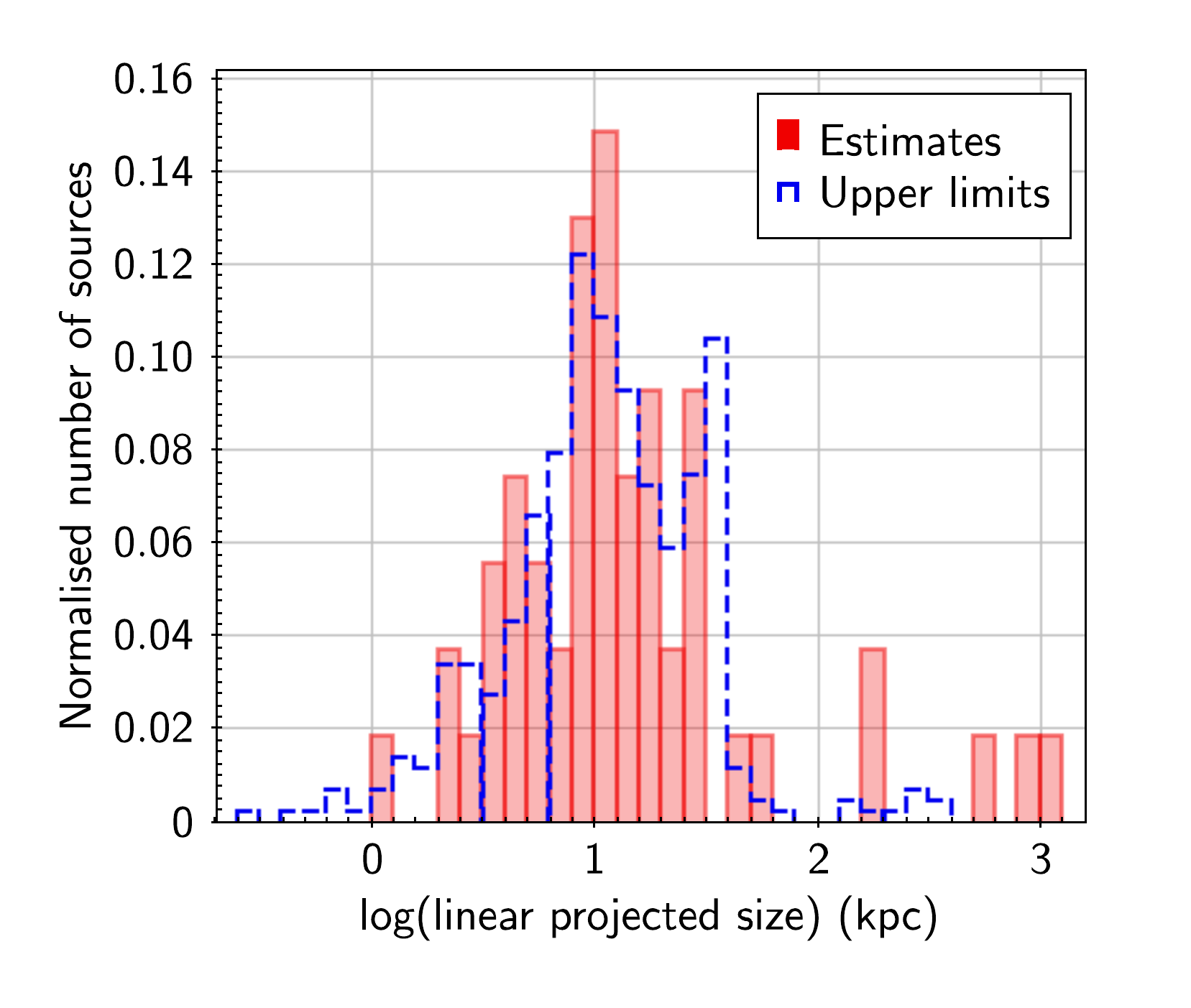}{\includegraphics[angle=0,width=9.0cm,trim={0.5cm 0.0cm 0.5cm 0.5cm},clip]{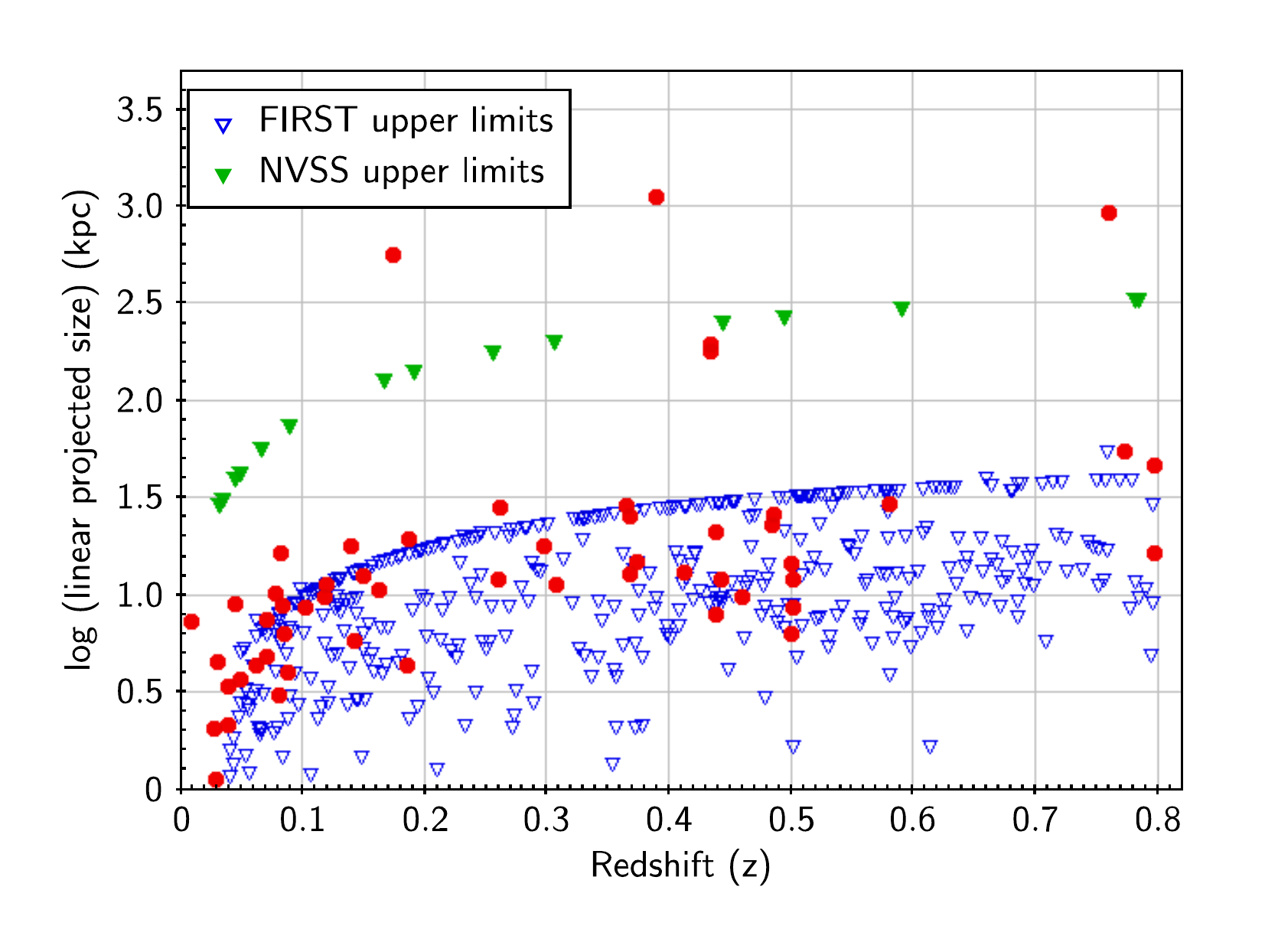}}
\caption{{\it Left panel} : The distribution of the projected linear radio-size for resolved sources 
and the upper limits on the radio-size for unresolved sources. {\it Right panel} : Radio-size versus redshift. 
The sizes of error bars are smaller than the sizes of symbols.}
\label{fig:LinearSizeHist} 
\end{figure*}

We estimate the radio size of our NLS1s. 
Since the apparent source size is projected in the plane of sky, and hence, we obtain only projected linear source size.  
The radio size of a resolved extended source is defined as the geometric mean of the deconvolved major axis 
and minor axis {\ie}$\sqrt{\rm (major~axis) \times (minor~axis)}$. 
There are few sources which display multicomponent radio structures and we derive their linear sizes by measuring 
the distance between the two extreme points along the direction of maximum extension.  
For all 443/498 ($\sim$ 89 per cent) unresolved sources we put an upper limit on their radio sizes which is set equal to 
the deconvolved major axis if minor axis is zero or equal to the beam-size 
(5$\arcsec$.4 in the FIRST and 45$\arcsec$ in the NVSS) if both major and minor axis are zero. 
In Fig.~\ref{fig:LinearSizeHist} (left panel) we plot the distribution of the projected linear sizes for resolved sources, 
and the upper limits for unresolved sources. 
We note that the projected linear sizes for our resolved sources are distributed from 1.13 kpc to 1.11 Mpc 
with a median of $\sim$ 11.9 kpc (see Table~\ref{table:HistProp}). 
Interestingly, radio-size distribution sharply declines beyond 30 kpc {\ie}87.3 per cent sources have projected linear 
size $<$ 30 kpc. In fact, there are only 07/55 (12.7 per cent) NLS1s 
that have projected linear sizes $>$ 30 kpc and display jet-lobe structures similar to radio galaxies. 
\par
The distribution of upper limits on the radio sizes for unresolved sources ranges from 0.28 kpc to 336 kpc with a median of 11.4 kpc. 
We note that the upper limits for unresolved sources are primarily based on the FIRST measurements (for 428/443 sources), 
and include upper limits from NVSS when FIRST measurements are not available. 
Since the NVSS beam-size (45$\arcsec$) is much larger than the FIRST beam-size (5$\arcsec$.4), 
the upper limits on the radio-sizes derived from the NVSS are much larger than that from the FIRST. 
In fact, the apparent tail at higher side ($>$ 50 kpc) seen in the distribution of upper limits is due to 
the large upper limits derived from the NVSS. 
For 428/443 FIRST-detected unresolved sources the distribution of upper limits ranges from 0.28 kpc to 55.5 kpc 
with a median of 10.9 kpc. The wide range of upper limits is partly due to the fact that our NLS1s are distributed 
across a wide range of redshifts from 0.009 to 0.8. 
\par
\begin{figure*}
\includegraphics[angle=0,width=5.2cm,trim={0.0cm 0.0cm 0.0cm 0.0cm},clip]{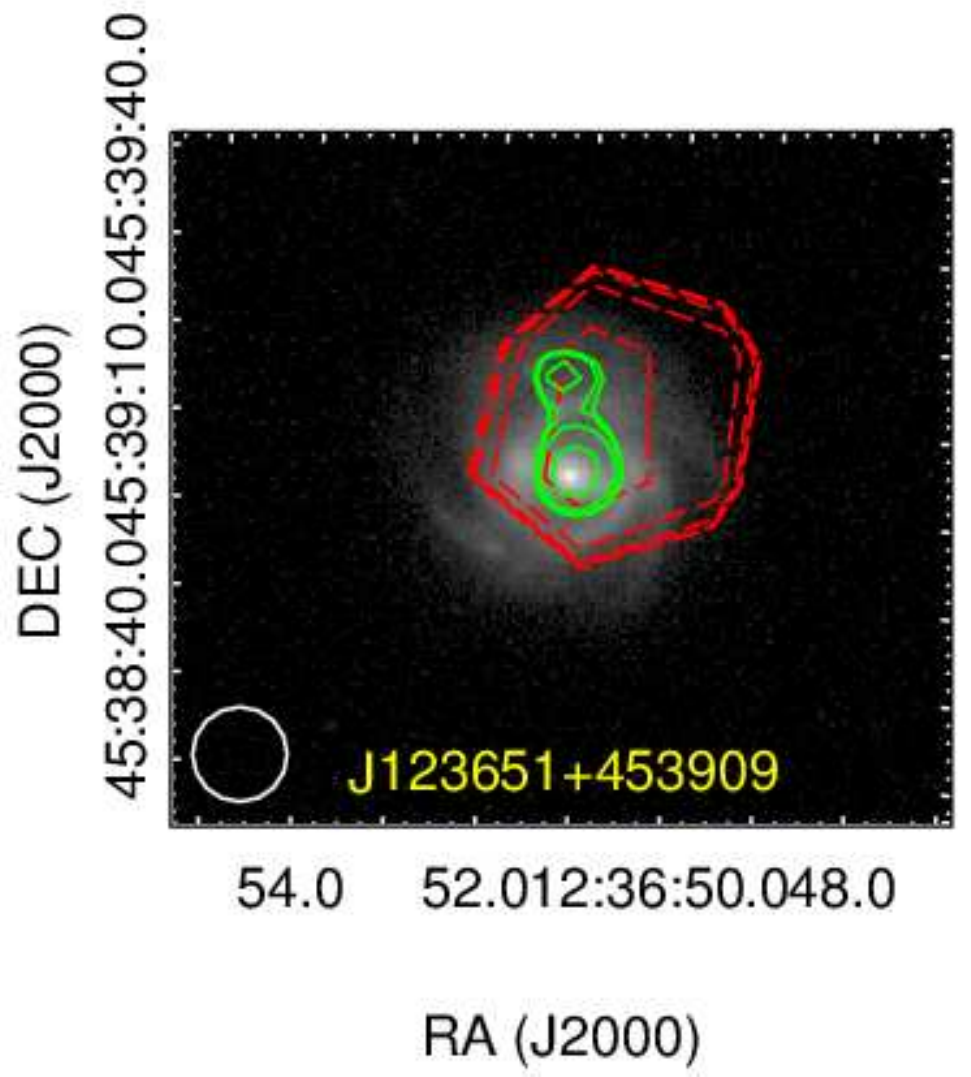}
{\includegraphics[angle=0,width=5.3cm,trim={0.0cm 0.0cm 0.0cm 0.0cm},clip]{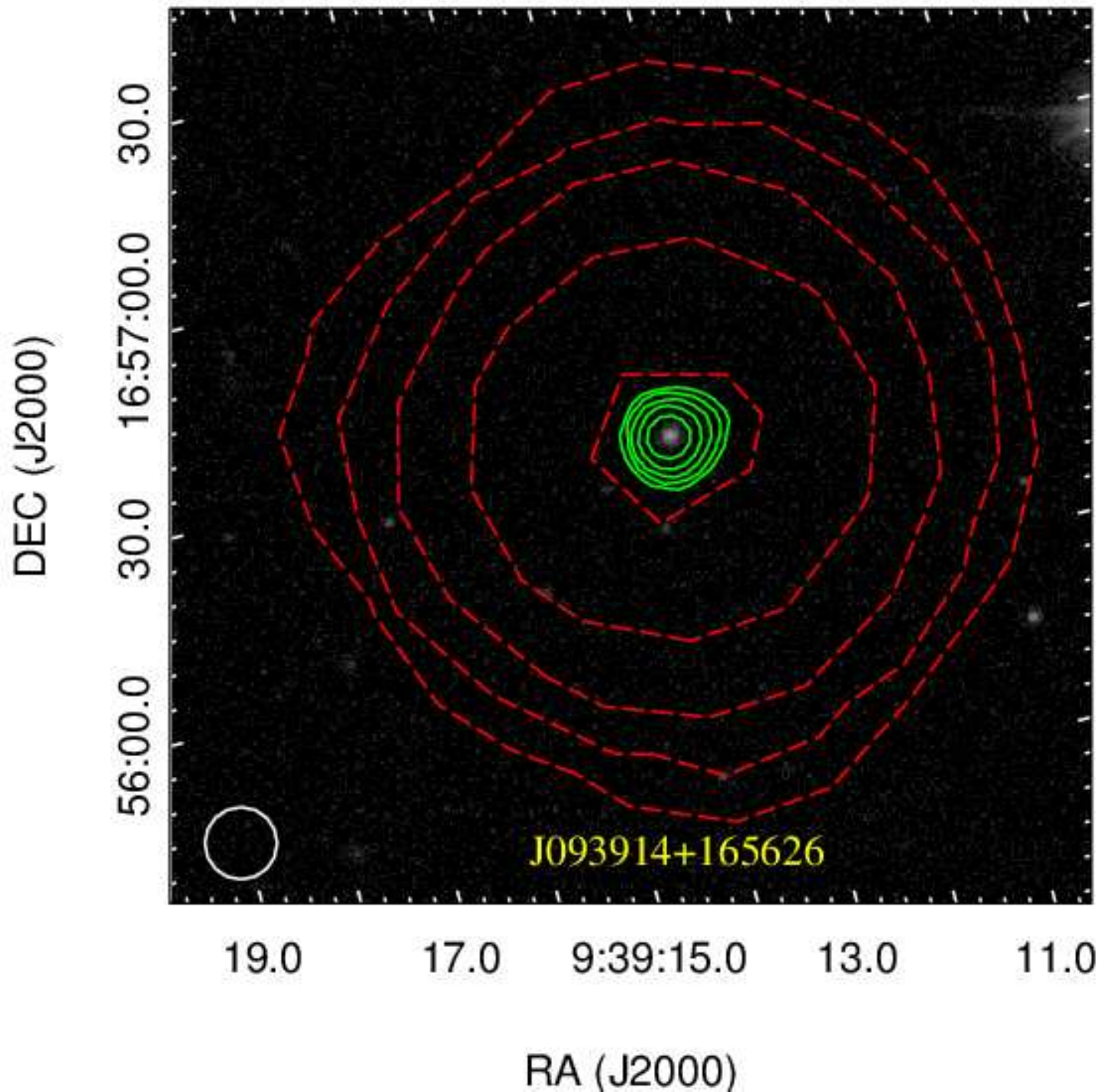}}
{\includegraphics[angle=0,width=5.5cm,trim={0.0cm 0.0cm 0.0cm 0.0cm},clip]{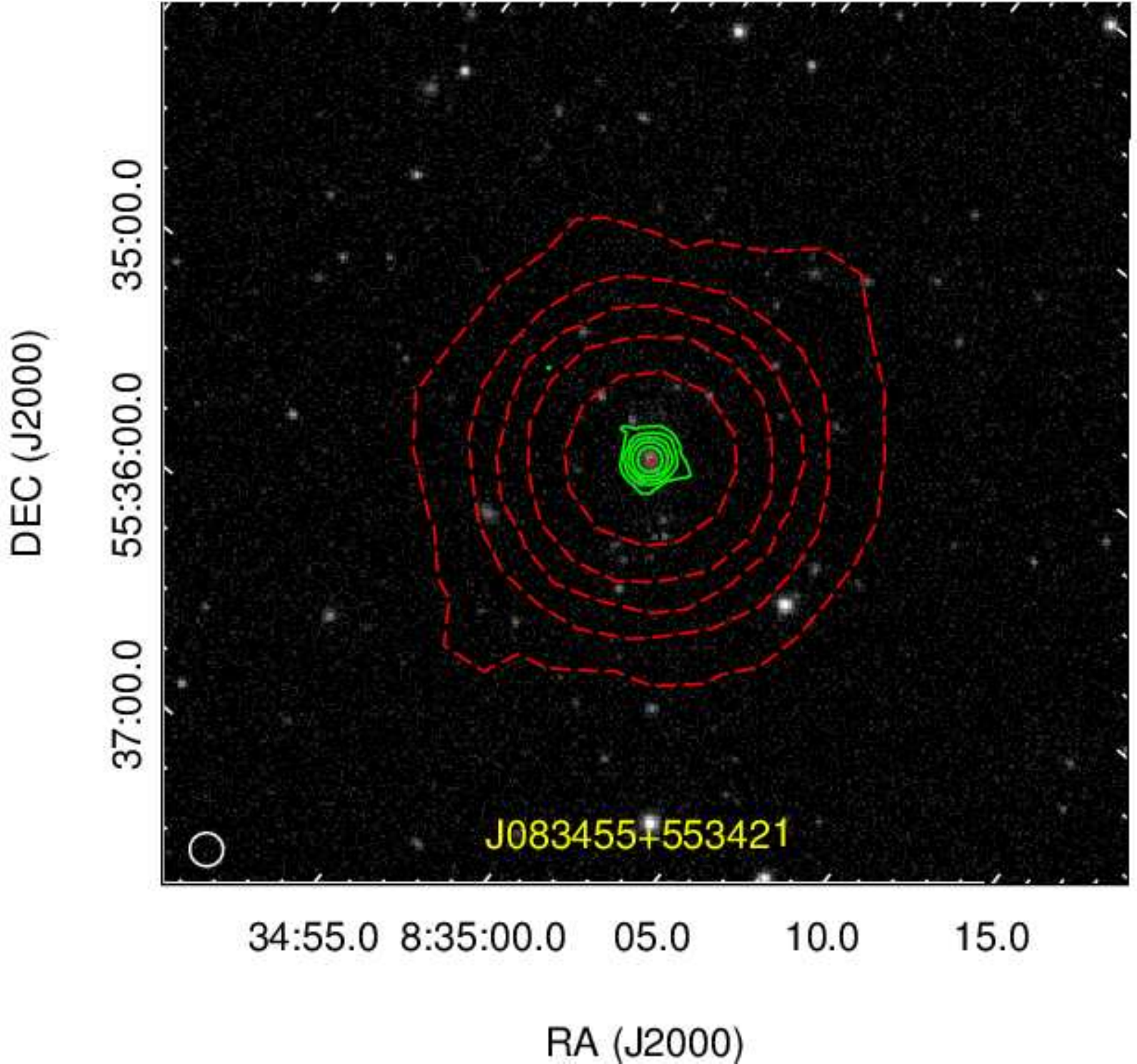}}
\caption{FIRST (solid green curves) and NVSS (dotted red curves) radio contours overlaid on to the SDSS r-band images. 
The outermost lowest radio contour begins at 5$\sigma$ level, except for the FIRST contours in NLS1 J123651+453904. 
The FIRST synthesized beam of 5$\arcsec$.4 $\times$ 5$\arcsec$.4 is shown at the bottom left in each panel.  
{\it Left panel} : A nearby ($z$ $\sim$ 0.0303) NLS1 J123651+453904 with a faint (S$_{\rm 1.4~GHz,~FIRST}$ $\sim$ 3.43$\pm$0.34 mJy) 
and low luminous (log(L$_{\rm 1.4~GHz}$) $\sim$ 21.85$\pm$.04 W~Hz$^{-1}$) radio counterpart. 
The lowest FIRST contour begins at 2.5$\sigma$ level to show the faint extended radio emission seen in the FIRST image. 
NVSS counterpart have offset of 10$\arcsec$.1 due to poor astrometry in faint NVSS sources. 
{\it Middle panel} : A RL-NLS1 J093914+165626 at $z$ $\sim$ 0.5016 with a bright (S$_{\rm 1.4~GHz,~FIRST}$ $\sim$ 97.8$\pm$0.16 mJy) 
and luminous (log(L$_{\rm 1.4~GHz}$) $\sim$ 25.93$\pm$0.01 W~Hz$^{-1}$) radio counterpart. 
{\it Right panel} : A RL-NLS1 J083455+553421 at $z$ $\sim$ 0.2415 with a very bright (S$_{\rm 1.4~GHz,~FIRST}$ $\sim$ 8360$\pm$0.82 mJy) 
and very luminous (log(L$_{\rm 1.4~GHz}$) $\sim$ 27.17$\pm$0.01 W~Hz$^{-1}$) radio counterpart.
}
\label{fig:RadioIm} 
\end{figure*}

In Fig.~\ref{fig:LinearSizeHist} (right panel) we plot redshift versus projected linear radio sizes including 
the upper limits. It is apparent that the upper limits increases with the redshift. 
So, the large upper limits on the radio sizes can partly be attributed to sources at higher redshifts. 
For instance, all FIRST-detected NLS1s having upper limits $\geq$ 20 kpc are at $z$ $\geq$ 0.25.
For nearby NLS1s {\ie}$z$ $\leq$ 0.1 the upper limits on their radio sizes are limited to 12 kpc. 
This infers that the radio size in most of our NLS1s are likely to be compact ({\ie}smaller than few kpc) 
which appear unresolved in the FIRST. 
The compactness of radio size in NLS1s is further vindicated by 
the distribution of radio sizes for resolved sources where nearly 87 per cent sources 
have radio size $<$ 30 kpc. 
Furthermore, a trend of increasing radio size with the increase of redshift is apparent for resolved sources 
(see Fig.~\ref{fig:LinearSizeHist}, right panel), which can be understood due to an observational bias 
where only large radio sources at higher redshifts become resolved and the smaller radio sources remain unresolved.    
In Fig.~\ref{fig:RadioIm} we show the FIRST and NVSS radio contours overlaid on to the corresponding SDSS r-band images for 
three representative sources : (i) a radio-quiet nearby NLS1 with faint extended radio emission, 
(ii) a RL-NLS1 with marginally resolved bright radio counterpart, and 
(iii) a strong RL-NLS1 with extremely bright radio counterpart.
\par
We note that the prevalence of compact radio size in our NLS1s is consistent with 
the high resolution (sub-arcsec to milli-arcsec) radio observations that have shown that NLS1s mostly exhibit 
compact parsec-scale radio emission \citep{Doi13,Gu15}. 
In general, the Kpc-Scale Radio structures (KSRs) in NLS1s are believed to be uncommon \citep[see][]{Doi12,Doi15,Richards15}. 
However, recently, \cite{Berton18} carried out deep 5 GHz JVLA A array observations of a sample of 74 NLS1s and 
found that 21/74 ($\sim$ 28 per cent) NLS1s exhibit KSRs. 
The large number of NLS1s with KSRs found by \cite{Berton18} can be attributed to better sensitivity
(5$\sigma$ $\sim$ 50 $\mu$Jy) and resolution (0$\arcsec$.5). 
Also, most of the NLS1s with KSRs tend to be radio-quiet and/or steep spectrum sources.      
It is worth to note that the number of resolved sources at arcsec-scale (55/498 NLS1s likely having KSRs) 
found in our study is several times more than those found in all previous studies.    
The finding of large number of NLS1s with KSRs can be attributed to the fact 
that : (i) unlike high-resolution sub-arcsec to milli-arcsec (VLA A array and VLBA) observations, the FIRST survey 
with resolution of $\sim$ 5$\arcsec$.4 is better suited to detect extended kpc-scale low-surface-brightness emission, 
(ii) and we carry out a systematic study of the largest sample of 11101 NLS1s. 
The smaller synthesized beam in high frequency and high resolution observations tend to miss the detection of 
kpc-scale extended emission of low-surface-brightness. The detailed study of KSRs in our NLS1s is deferred to a forthcoming paper. 
\subsection{Radio-loudness}
\label{sec:Radioloudness}
Radio-loudness parameter (R), conventionally defined as the ratio of radio luminosity at 5~GHz to optical luminosity 
at 4400{\AA} {\ie}R = L$_{\rm 5~GHz}$/L$_{\rm {\lambda}4400~{\AA}}$ \citep{Kellermann94}, 
can be considered as the proxy of jet production efficiency in AGN. 
In fact, based on the value of R, AGN are divided into Radio-Quiet (RQ) and Radio-Loud (RL) categories, 
where R $<$ 10 for RQ-AGN and R $\geq$ 10 for RL-AGN. 
In general, RL-AGN possess powerful radio-jets often spanning from a few kpc to several hundreds of kpc, 
while radio-jets in RQ-AGN are much weaker and mostly remain confined within the host galaxy 
\citep[see][]{Gallimore06,Kellermann16,Padovani16}. 
Therefore, to understand the nature of radio-jets in our NLS1s we investigate their radio-loudness parameters.   
We estimate radio-loudness parameter (R$_{\rm 1.4~GHz}$) for our NLS1s 
using 1.4 GHz luminosity, instead of 5 GHz luminosity, as the radio counterparts of our NLS1s are primarily detected at 1.4 GHz. 
We note that the dividing line between RQ-AGN and RL-AGN R = 10 defined at 5~GHz corresponds 
to R = 19 at 1.4 GHz \citep{Komossa06}, considering the same assumption ({\ie}both radio and optical spectra of AGN have average 
spectral index of $-$0.5) that was used to defined R = 10 at 5~GHz \citep[see][]{Kellermann89}. 
\begin{figure*}
\includegraphics[angle=0,width=8.0cm,trim={0.5cm 0.5cm 0.5cm 0.5cm},clip]{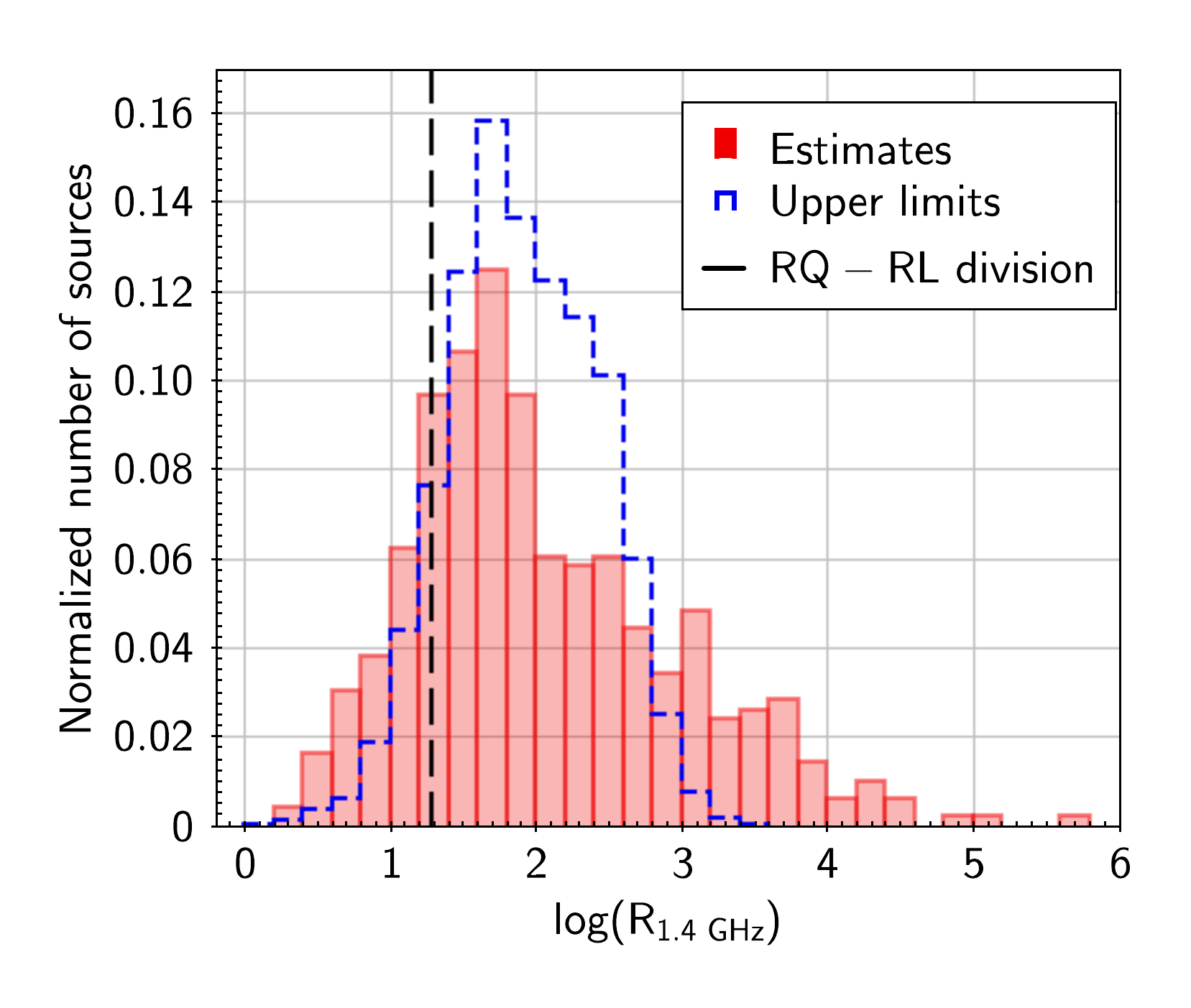}{\includegraphics[angle=0,width=8.0cm,trim={0.5cm 0.5cm 0.5cm 0.5cm},clip]{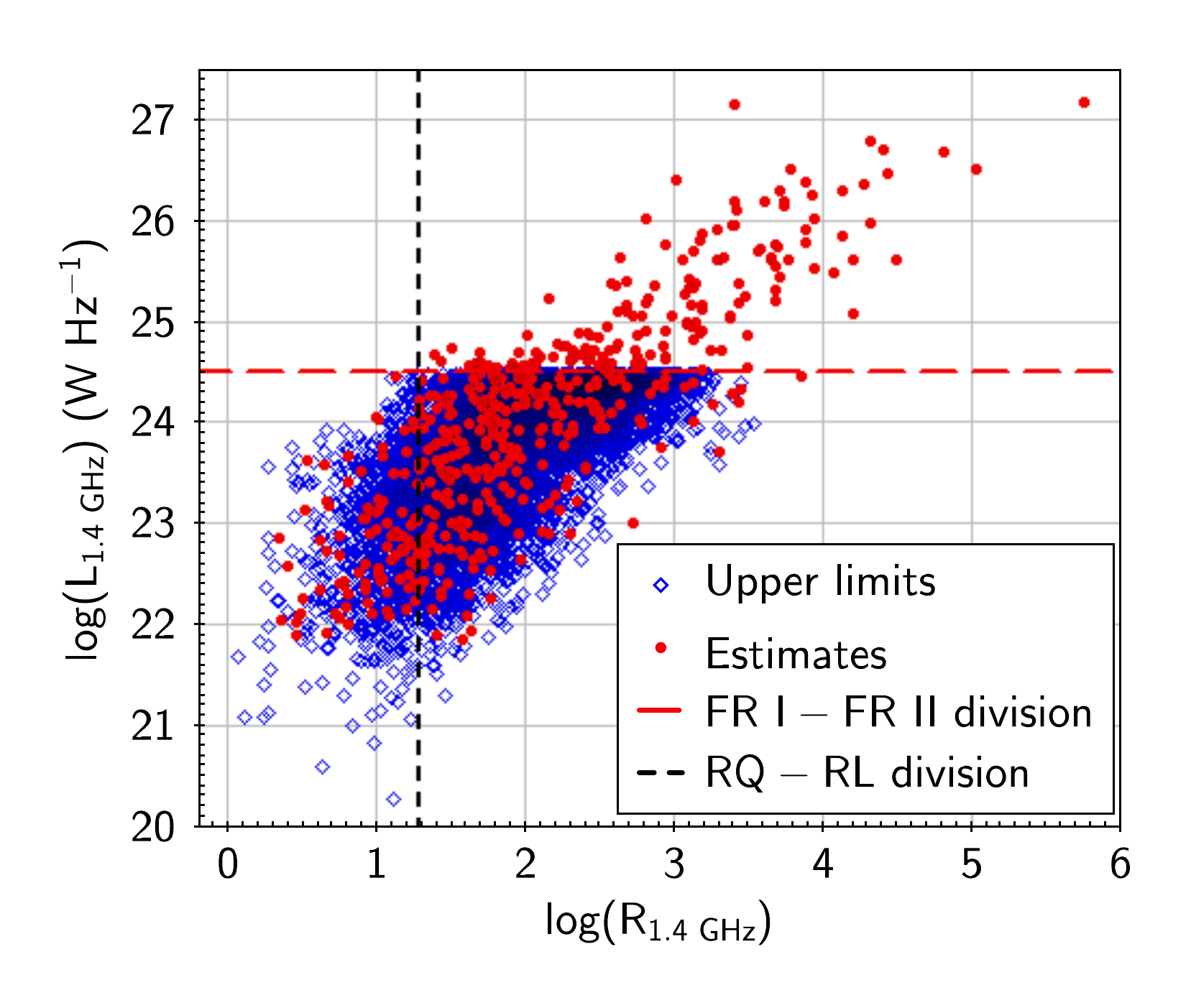}
}
\caption{{\it Left panel} : The distributions of the radio-loudness parameter at 1.4 GHz (R$_{\rm 1.4~GHz}$) for our NLS1s. 
{\it Right panel} : R$_{\rm 1.4~GHz}$ versus redshift for our NLS1s. The sizes of error bars are smaller than the sizes of symbols.}
\label{fig:RLHist} 
\end{figure*}
To estimate R$_{\rm 1.4~GHz}$ we use 1.4 GHz luminosity from the FIRST whenever available, otherwise NVSS luminosity is considered. 
With its higher resolution the FIRST survey is more sensitive to nuclear radio emission and is less susceptible to contamination 
from a neighbouring source. 
The monochromatic optical continuum luminosity at 4400{\AA} (L$_{\rm {\lambda}4400{\AA}}$) for our NLS1s is derived from 5100{\AA} 
luminosity (L$_{\rm {\lambda}5100{\AA}}$) assuming a power law optical spectrum with an average spectral index of $-$0.5 \citep{Sikora07}. 
The 5100{\AA} optical continuum luminosity (L$_{\rm {\lambda}5100{\AA}}$) 
is taken from \cite{Rakshit17} who derived it by fitting the SDSS spectra. 
\par
In Fig.~\ref{fig:RLHist} (left panel) we show the distribution of R$_{\rm 1.4~GHz}$ parameter for our 498/11101 radio-detected NLS1s, 
and the distribution of upper limits on R$_{\rm 1.4~GHz}$ for remaining NLS1s with no detected radio counterparts. 
The upper limit on R$_{\rm 1.4~GHz}$ is derived by using the FIRST flux density limit (1.0 mJy) 
if NLS1 lies within the FIRST coverage otherwise NVSS flux density limit (2.5 mJy) is used. 
We note that R$_{\rm 1.4~GHz}$ for our 498 radio-detected NLS1s ranges from 2.2 to 5.8 $\times$ 10$^{5}$ with a median value of 71.6 
(see Table~\ref{table:HistProp}). It is apparent that R$_{\rm 1.4~GHz}$ distribution is skewed 
towards higher values of R$_{\rm 1.4~GHz}$ due to the presence of few strong radio-loud NLS1s. 
Importantly, R$_{\rm 1.4~GHz}$ distribution does not show bimodality, which is contrary 
to some of the early results \citep[{\eg}][]{Kellermann89}. 
Although, the lack of bimodality seen for our NLS1s is consistent with 
the recent studies based on large AGN samples \citep{Rafter09,Singh15a} that show no demarcation between RQ and RL sources.  
In literature, there have been attempts to explain the bimodality as well as the lack of bimodality \cite[see][]{Kellermann16}. 
The lack of bimodality can be understood if nuclear radio luminosity scales smoothly with the mass of SMBH 
and accretion rate \citep{Lacy01}. And, the bimodality in radio-loudness found in early studies can be explained due to the sample 
selection bias, as the samples contain RL sources hosted in ellipticals and RQ-AGN hosted in spirals \citep{Sikora07}. 
\par
Using R$_{\rm 1.4~GHz}$ = 19 as the dividing line between RQ and RL sources we find that 
407/498 (81.7 per cent) of our radio-detected NLS1 can be classified as RL-NLS1s, 
and only 91/498 (18.3 per cent) are RQ-NLS1s. 
At first glance, the high fraction of RL-NLS1s may appear unexpected, however, 
we note that the overall fraction of confirmed RL-NLS1s is merely 3.7 per cent (407/11101) 
if we consider our full sample of 11101 optically-selected NLS1s.   
Also, the fraction of RL-NLS1s found in our radio-detected NLS1s is similar to that reported 
in \cite{Rakshit17} who considered only FIRST-detected NLS1s and used g-band magnitude to estimate the radio-loudness. 
Notably, our study yields the largest sample of 407 RL-NLS1s till date. 
All the previous samples of RL-NLS1s are limited to less than a few dozens of sources \citep[see][]{Yuan08,Komossa08,Foschini15}.  
The large number of RL-NLS1s found in our study can be attributed to the fact that we use the largest sample of optically-selected 
NLS1s. 
\par 
For NLS1s with no radio counterparts in the FIRST and NVSS the upper limits on 
R$_{\rm 1.4~GHz}$ range from 1.2 to 3423 with a median of 77.9. 
Using R$_{\rm 1.4~GHz}$ = 19 as the RQ-RL dividing line we find that the upper limit on the fraction of RL-NLS1 is 9556/10603 
(90.1 per cent) among the NLS1s with no detected radio-counterparts, and 9556/11101 (86.1 per cent) among the full sample. 
Since the radio flux density of NLS1s with no detected radio counterparts can be much lower than the FIRST and NVSS flux density 
limits, therefore, a very high fraction of RL-NLS1s is unlikely. 
However, a significant fraction of RL-NLS1s can be expected even if the radio flux densities of 
NLS1s with no detected radio counterparts are 1$-$2 order of magnitude lower than the FIRST flux density limit of 1.0 mJy. 
Indeed, the fraction of RL-NLS1s {\ie}3.7 per cent found in our sample is only a lower limit as a fraction NLS1s with no detected radio 
counterparts can turn out to be RL. 
Thus, our study demonstrates that the upcoming deeper radio surveys ({\eg} Square Kilometer Array (SKA) surveys) 
have potential to unveil a new population of RL-NLS1s. 
\par
Interestingly, the fraction of RL-NLS1s ({\ie}3.7 per cent) in our sample is much lower than 
that found in the samples of BL-AGN. 
In general, 10 - 20 per cent sources are found to be RL in optically-selected samples of BL-AGN \citep{Ivezic02,Kellermann16}. 
The lower fraction of RL sources among NLS1s, in compared to BL-AGN, seems to be consistent with the previous results. 
For example, \cite{Komossa06} found only 7.0 per cent RL sources in their sample of 128 NLS1s. 
Moreover, it is important to note that 
the fraction of RL sources in a sample depends on the redshift distribution of sample sources and the sensitivity of radio observations. 
We find that the fraction of RL-NLS1s in our sample is $\sim$ 7 per cent if we consider only nearby 
sources at $z$ $\leq$ 0.1, while it decreases to 3.0$-$4.0 per cent for NLS1s at $z$ $\sim$ 0.1$-$0.8. 
The higher fraction of RL-NLS1s in the lowest redshift bin can be understood as due to an observational bias where 
relatively fainter sources are being detected in a flux limited radio survey. 
The lower fraction of RL sources in NLS1s, in compared to BL-AGN, can be understood 
if NLS1s possess low power radio-jets or are powered by star-formation.  
\par
To get more insights into the nature of RL-NLS1s we plot the radio loudness (R$_{\rm 1.4~GHz}$) versus 1.4 GHz radio luminosity 
(L$_{\rm 1.4~GHz}$) for our NLS1s (see Fig.~\ref{fig:RLHist}, right panel). 
R$_{\rm 1.4~GHz}$ and L$_{\rm 1.4~GHz}$ are expected to be correlated, as 
by definition, R$_{\rm 1.4~GHz}$ is L$_{\rm 1.4~GHz}$ normalized with the 4400{\AA} optical continuum luminosity. 
Indeed, we find that R$_{\rm 1.4~GHz}$ and L$_{\rm 1.4~GHz}$ are correlated 
({\ie}Spearman rank correlation coefficient is $\sim$ 0.8). 
The apparent scatter in R$_{\rm 1.4~GHz}$ versus L$_{\rm 1.4~GHz}$ plot 
{\ie}varying L$_{\rm 1.4~GHz}$ in NLS1s with same L$_{\rm 4400~{\AA}}$ and vice-versa, can be attributed to 
the differences in intrinsic parameters such as the mass and spin of SMBH, accretion process etc.  
From R$_{\rm 1.4~GHz}$ versus L$_{\rm 1.4~GHz}$ plot it is evident that RQ-NLS1s are of low luminosity 
(L$_{\rm 1.4~GHz}$ $\sim$ 10$^{22}$$-$10$^{24}$ W Hz$^{-1}$), while NLS1s with 
high radio luminosity (L$_{\rm 1.4~GHz}$ $\geq$ 10$^{24}$ W Hz$^{-1}$) are very radio-loud. 
Notably, there is a substantial population of radio-intermediate NLS1s with R$_{\rm 1.4~GHz}$ 
$\sim$ 10$-$100 and L$_{\rm 1.4~GHz}$ $\sim$ 10$^{22}$$-$10$^{24}$ W Hz$^{-1}$. 
\subsection{Radio variability}
\label{sec:Variability}
Variability is a one of the common features of AGN, in particular, among blazars \citep{Urry96,Marscher16}. 
We investigate radio flux variability in our NLS1s by comparing 1.4 GHz flux densities from the FIRST and NVSS. 
We note that the FIRST and NVSS observations are taken apart by few months to several years, 
and thus, the comparison of FIRST and NVSS flux densities allows us to probe variability in the same time-scale. 
While comparing the NVSS and FIRST flux densities we must keep in mind that 
the NVSS with larger beam-size (45$\arcsec$), as compared to the FIRST (5$\arcsec$.4), can yield higher integrated 
flux density for sources having extended low-surface-brightness radio emission. 
Therefore, NLS1s having NVSS integrated flux density (S$_{\rm NVSS}^{\rm int}$) higher than their FIRST peak or integrated flux density 
(S$_{\rm FIRST}^{\rm peak}$ or S$_{\rm FIRST}^{\rm int}$) may not always be due to variability.   
However, NLS1s having S$_{\rm NVSS}^{\rm int}$ lower than the S$_{\rm FIRST}^{\rm peak}$ can only be 
due to variability. Therefore, we adopt a conservative approach and identify radio-variable NLS1s as those sources 
for which S$_{\rm NVSS}^{\rm int}$ $<$ S$_{\rm FIRST}^{\rm peak}$.  
The significance of variability is defined as the ratio of the difference between S$_{\rm FIRST}^{\rm peak}$ 
and S$_{\rm NVSS}^{\rm int}$ to the combined uncertainties of FIRST and NVSS flux densities and the systematic uncertainty 
in the FIRST flux density 
{\ie}${\sigma}_{\rm var}$ $=$ (S$_{\rm FIRST}^{\rm peak}$ $-$ S$_{\rm NVSS}^{\rm int}$)
/${\sqrt{{{\sigma}_{\rm FIRST}}^2 + {{\sigma}_{\rm NVSS}}^2 + {\rm (0.05~S_{\rm FIRST}^{\rm peak})}^2}}$ \citep[see][]{Wang06}.   
We note that the FIRST flux densities suffer with an additional systematic uncertainties 
of 5 per cent that are not included in ${\sigma}_{\rm FIRST}$ \citep{Becker95}, 
while systematic uncertainties of 3 per cent present in the NVSS flux densities are already included in ${\sigma}_{\rm NVSS}$ \citep{Condon98}.  
To identify radio-variable NLS1s we use ${\sigma}_{\rm var}$ $>$ 3 cut-off and find only nine ($\sim$ 3.5 per cent) variable sources 
among 259 NLS1s that are detected in both FIRST and NVSS. 
We note that this is only a lower limit on the number of radio-variable sources as a fraction of NLS1s with 
S$_{\rm NVSS}^{\rm int}$ $>$ S$_{\rm FIRST}^{\rm peak}$ can also be variable. 
In fact, there are 57 NLS1s that have S$_{\rm NVSS}^{\rm int}$ $>$ S$_{\rm FIRST}^{\rm peak}$ with ${\sigma}_{\rm var}$ $<$ $-$3, 
which can be identified as candidate variables, since the excess NVSS flux density can be due to extended 
low-surface-brightness emission which is missed in the FIRST survey. 
By adding confirmed and candidate variable sources we obtain an upper limit of 66/259 ($\sim$ 25 per cent) 
on the number of radio variable NLS1s.    
\par 
\begin{figure}
\includegraphics[angle=0,width=9.0cm,trim={0.5cm 0.5cm 0.5cm 0.5cm},clip]{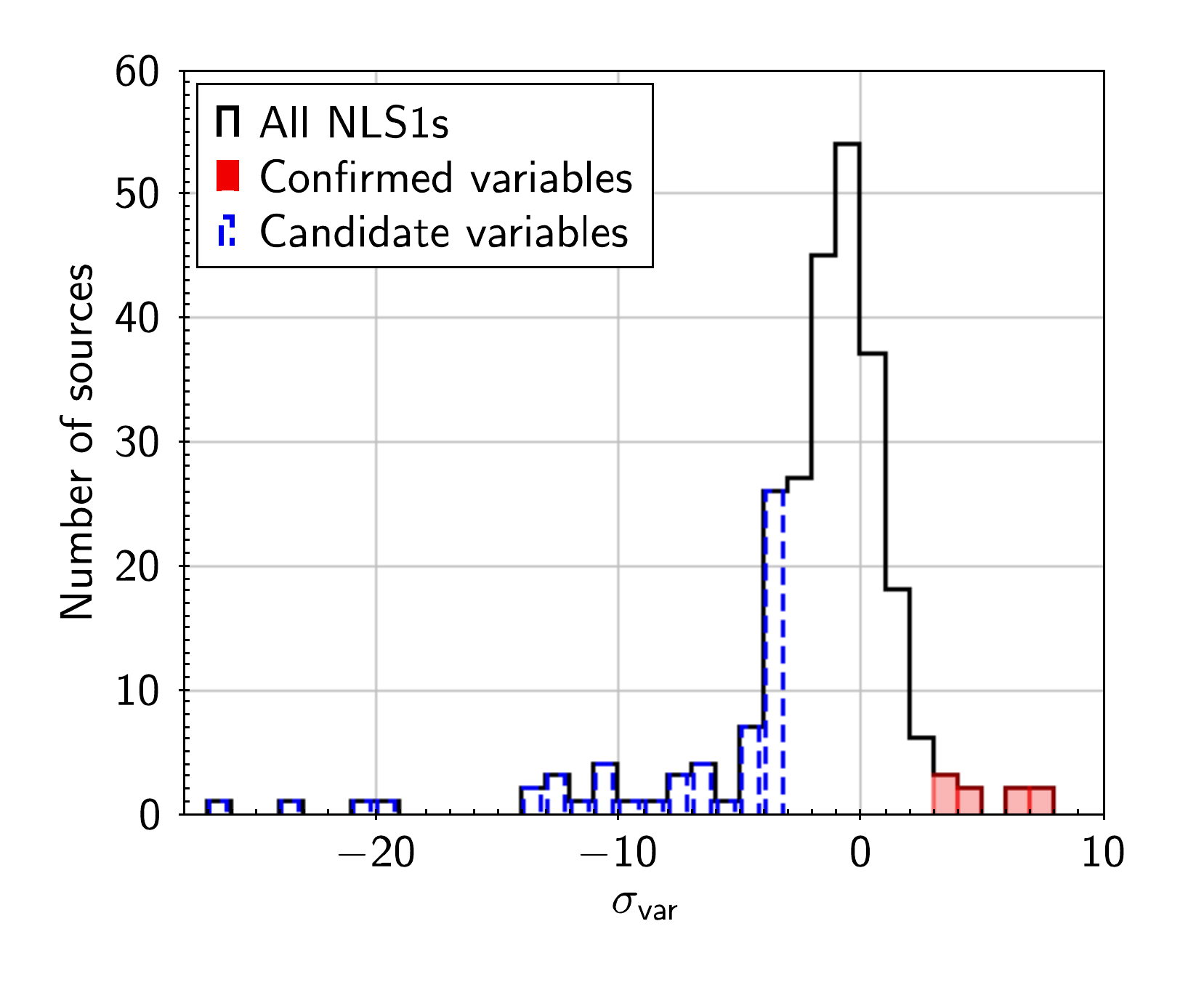}
\caption{Histogram of the variance measured between the FIRST peak flux density and NVSS integrated flux density.}
\label{fig:VarHist} 
\end{figure}
\begin{figure*}
\includegraphics[angle=0,width=9.0cm,trim={0.5cm 0.5cm 0.5cm 0.5cm},clip]{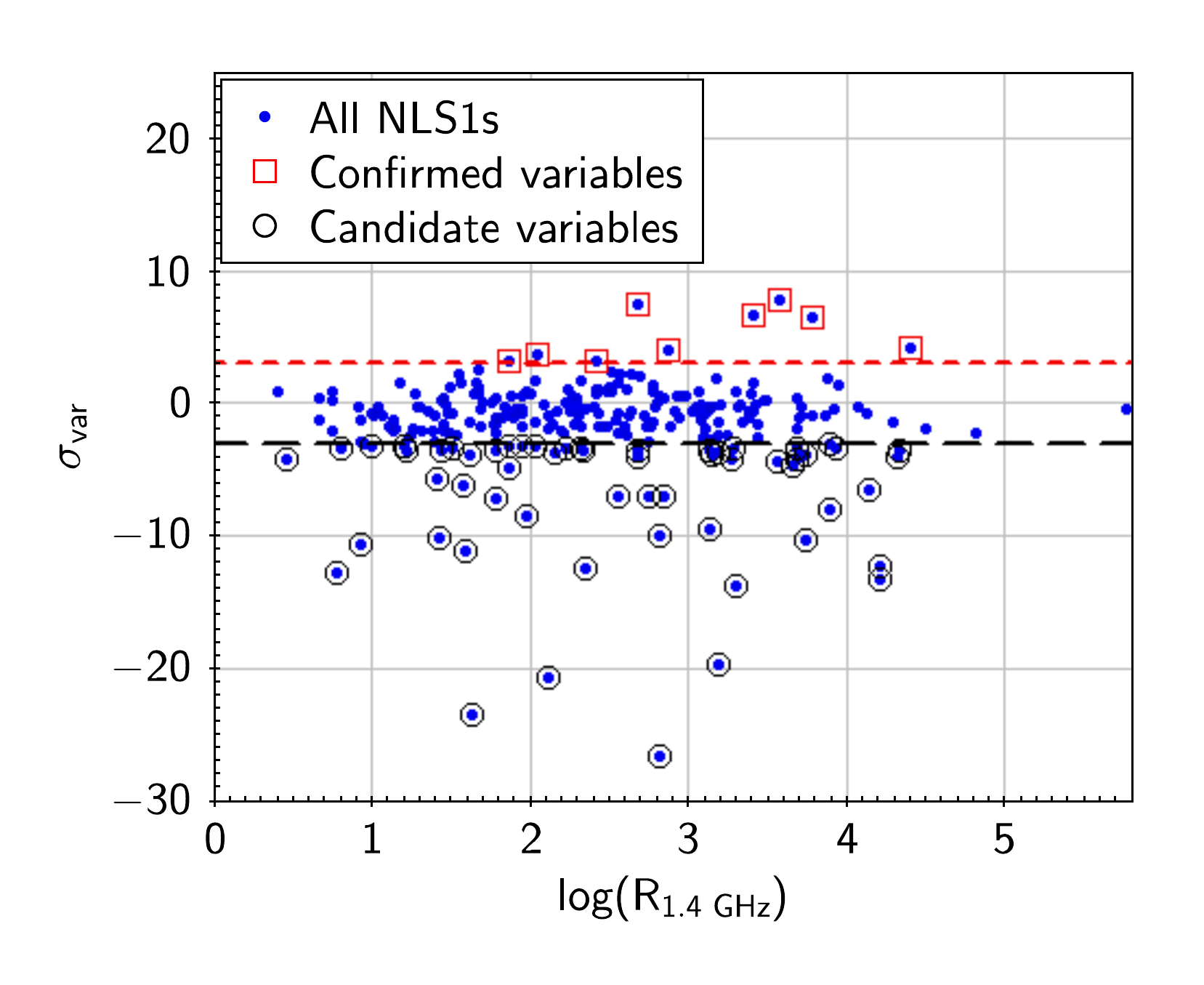}{\includegraphics[angle=0,width=9.0cm,trim={0.5cm 0.5cm 0.5cm 0.5cm},clip]{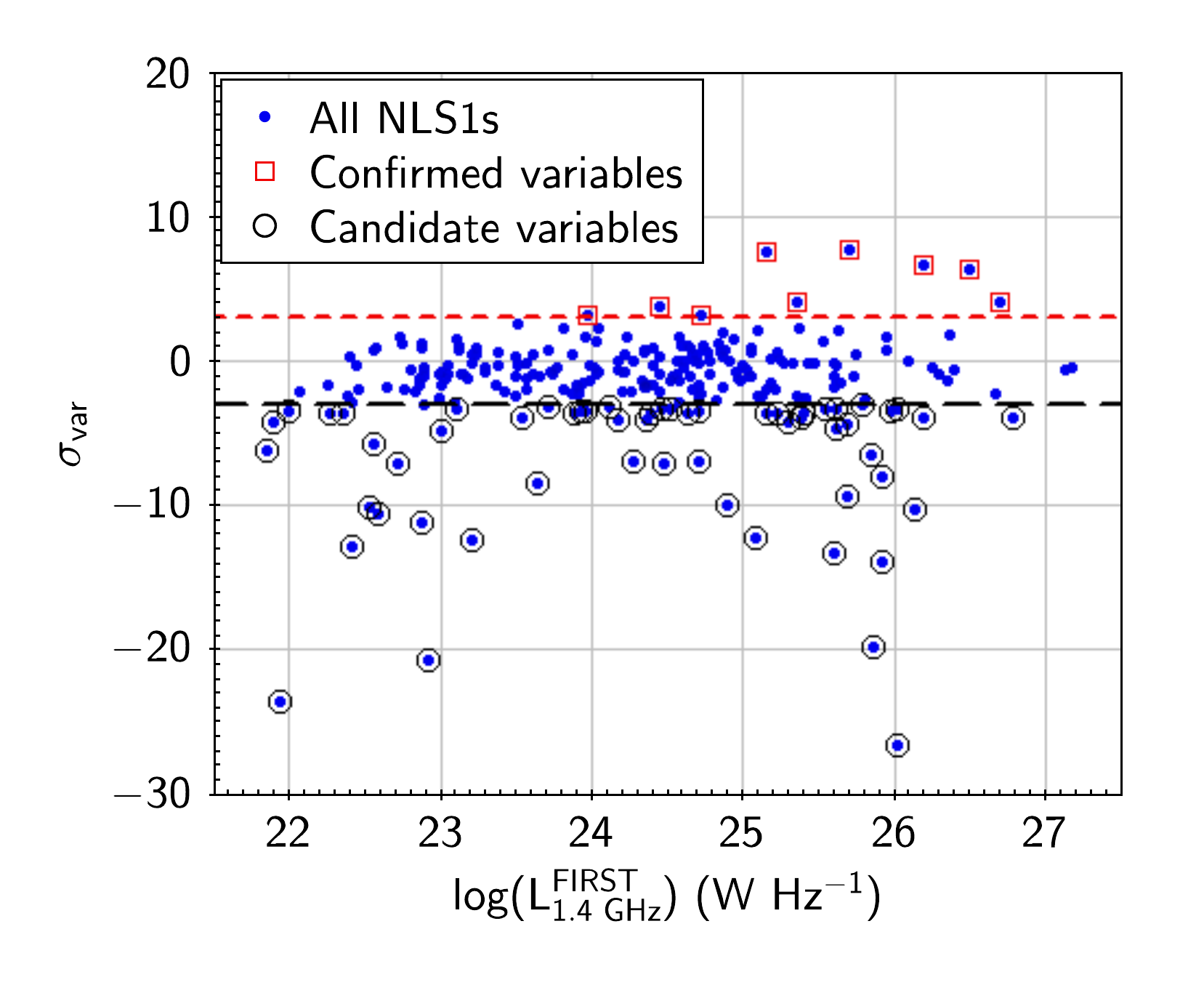}}
\caption{{\it Left panel} : Radio-loudness parameter (R$_{\rm 1.4~GHz}$) versus variance. {\it Right panel} : 1.4 GHz radio luminosity versus variance. 
The sizes of error bars are smaller than the sizes of symbols.}
\label{fig:VarVsR} 
\end{figure*}
In Fig.~\ref{fig:VarHist} (left panel) we plot the distribution of variance (${\sigma}_{\rm var}$) which is ranging 
from $-$26.64 to 7.74 with a median of $\sim$ $-$1.0 (see Table~\ref{table:HistProp}). 
For non-variable and compact sources with no diffuse extended emission, we can expect same flux densities in both NVSS and FIRST, 
and thus, zero variance. The slight leftward shift of the variance distribution, {\it w.r.t.} zero median, can be understood 
as S$_{\rm NVSS}^{\rm int}$ is higher than S$_{\rm FIRST}^{\rm peak}$ for many sources that contain faint 
extended radio emission. As expected, the median value of the variance distribution shifts towards zero if we use integrated instead of 
peak flux densities of the FIRST. Sources at the extreme left in the variance distribution are multicomponent extended sources for which 
S$_{\rm NVSS}^{\rm int}$ is much higher than S$_{\rm FIRST}^{\rm peak}$. 
\par
To investigate the nature of radio-variable NLS1s we plot variance versus radio-loudness parameter (R$_{\rm 1.4~GHz}$) and 
variance versus 1.4~GHz luminosity (see Fig.~\ref{fig:VarVsR}). 
It is evident that all the confirmed variable sources are RL-NLS1s 
with R$_{\rm 1.4~GHz}$ ranging from 31.3 to 11367.8. Also, all the variable NLS1s are radio 
luminous (L$_{\rm 1.4~GHz}$ $\geq$ 10$^{24}$ W Hz$^{-1}$). Both variance versus R$_{\rm 1.4~GHz}$ plot and 
variance versus L$_{\rm 1.4~GHz}$ plot show similar trend, which is not surprising as both R$_{\rm 1.4~GHz}$ and L$_{\rm 1.4~GHz}$ are correlated 
(see Fig.~\ref{fig:RLHist}, right panel). 
We find that the confirmed variable NLS1s are fairly bright {\ie}S$_{\rm FIRST}^{\rm peak}$ is distributed over 4.9 mJy to 344 mJy with 
a median value of 29.4 mJy. Interestingly, all the confirmed radio variable NLS1s appear unresolved in both the FIRST and NVSS. 
Five out of nine confirmed radio variable NLS1s have counterparts at 150 MHz in the TGSS. 
The spectra of all but one of these five confirmed radio variable NLS1s are flat/inverted 
{\ie}${\alpha}_{\rm 150~GHz}^{\rm 1.4~GHz}$ ranges from $-$0.72 to +0.21 with a median of 0.09. 
The remaining four confirmed radio variable NLS1s are also likely to have flat/inverted spectra as the upper limits on ${\alpha}_{\rm 150~GHz}^{\rm 1.4~GHz}$ 
are $-$0.71, $-$0.49, $-$0.32, and 0.09.   
Therefore, we conclude that, in general, confirmed radio variable NLS1s are compact, radio-powerful AGN with flat/inverted radio spectra.    
Furthermore, the redshifts of our nine confirmed radio variable NLS1s are relatively higher {\ie}$z$ $\sim$ 0.25$-$0.8 with a median of 0.48. 
We note that the properties of our confirmed radio variable NLS1s infer them to be blazar-like {\ie}core-dominated RL-AGN 
that are believed to possess relativistically beamed jets on pc-scales \citep[see][]{Doi11,Doi12,Wajima14,Gabanyi18}. 
Unlike confirmed radio variable NLS1s, the candidate radio variable NLS1s include both radio-loud and radio-quiet 
sources distributed across low radio luminosity to high radio luminosity. Also, the candidate radio variable NLS1s often tend 
show extended radio emission, for instance, 28 out of 55 extended radio sources identified 
by comparing the integrated and peak flux densities of the FIRST (see Sect.~\ref{sec:Structures}) are classified as 
the candidate radio variable sources. This infers that a substantial fraction among the candidate variable sources may not be variable.
\subsection{Radio Polarization}
\label{sec:Polarization}
\begin{figure*}
\includegraphics[angle=0,width=9.0cm,trim={0.5cm 0.5cm 0.5cm 0.5cm},clip]{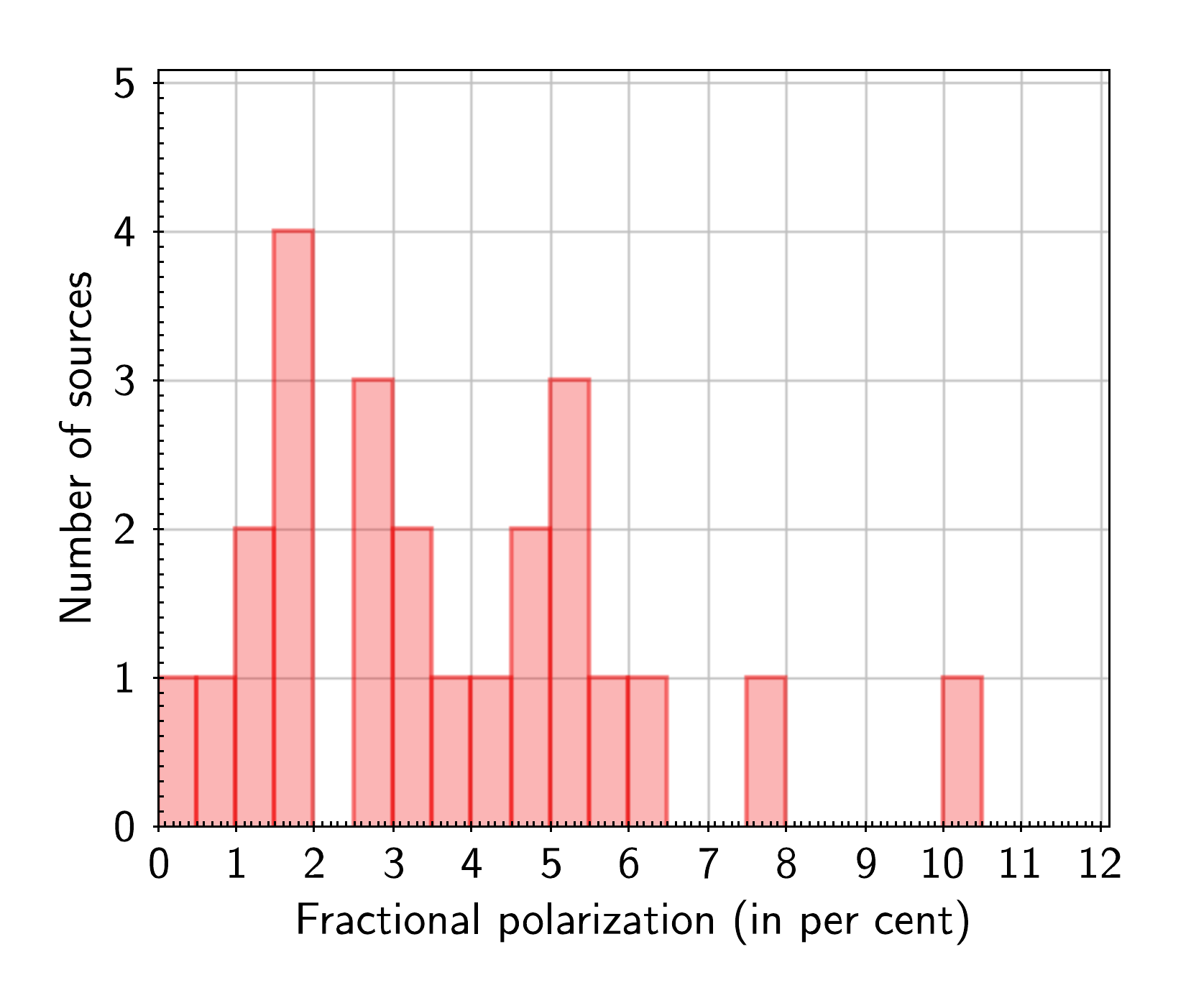}{\includegraphics[angle=0,width=9.0cm,trim={0.5cm 0.5cm 0.5cm 0.5cm},clip]{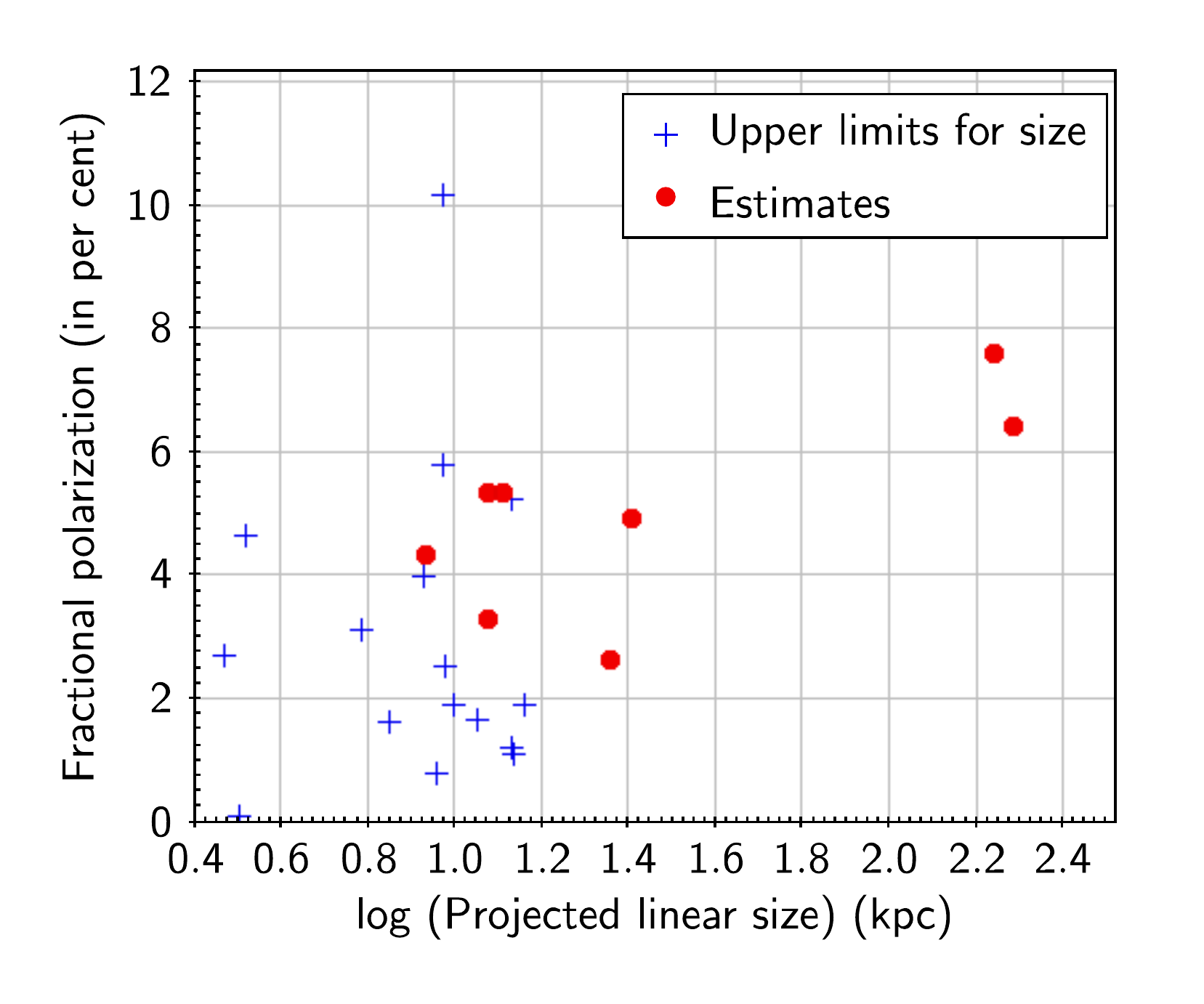}}
\caption{{\it Left panel} : Histogram of the fractional radio polarization in our NLS1s. 
{\it Right panel} : Projected linear radio-size versus fractional radio polarization. 
The sizes of error bars are smaller than the sizes of symbols.} 
\label{fig:PolHist} 
\end{figure*}

Radio polarization is one of the tools to understand the nature of radio-jet in AGN. 
We examine the degree of polarization in our NLS1s using 1.4~GHz NVSS measurements. 
NVSS gives linear polarized flux densities along with uncertainties. 
We consider the presence of linear polarization in only those sources that show polarized flux density with 
SNR (S$_{\rm pol}$/${\sigma}_{\rm pol}$) $\geq$ 3. 
We find that there are only 24/274 ($\sim$ 9 per cent) NVSS-detected NLS1s showing radio polarization with SNR $\geq$ 3. 
For NLS1s showing polarized emission we define degree of polarization (or fractional polarization) 
as the percentage of polarized flux density {\it w.r.t.} 
the integrated flux density {\ie}100 $\times$ (S$_{\rm pol}$/S$_{\rm int}$). 
In Fig.~\ref{fig:PolHist} (left panel) we show the distribution of the degree of polarization that ranges from 0.1 per cent to 10.2 per cent 
with a median of 3.3 per cent. 
Therefore, among a small fraction ($\sim$ 9 per cent) of NVSS-detected NLS1s, we detect polarized radio emission, 
while remaining large fraction ($\sim$ 91 per cent) of NLS1s do not show polarized emission. 
All the 24 NLS1s showing polarized emission are fairly bright {\ie}1.4 GHz flux densities range from 25.9 mJy to 8283 mJy 
with a median of 115 mJy. 
However, we caution that radio polarization found only in radio-bright sources can be due to that fact that 
we tend to miss radio-faint sources as they generally show lower SNR for the polarized flux.
We also find that all 24 NLS1s exhibiting radio polarization are of high luminosity (L$_{\rm 1.4~GHz}$ $\geq$ 10$^{24.5}$ W Hz$^{-1}$) similar 
to powerful FR II radio galaxies. Therefore, all the NLS1s exhibiting radio polarization are radio-loud with powerful radio-jets. 
\par
We also examine the radio sizes of all 24 NLS1s showing radio polarization. 
We find that 16/24 NLS1s are unresolved and remaining 08/24 have resolved structures in the FIRST. 
The upper limits of the projected linear sizes of 16/24 unresolved NLS1s range from 2.9 kpc to 14.5 kpc with a median of 9.4 kpc. 
While, the projected linear radio-sizes of 08/24 resolved sources span from 8.6 kpc to 193.2 kpc with a median of 17.9 kpc. 
From the fractional polarization versus projected linear size plot (see Fig.~\ref{fig:PolHist}, right panel), it appears that the extended sources tend 
to show higher degree of polarization, however, a much larger statistics is required to confirm it. 
The trend seen in the fractional polarization versus projected linear size plot is consistent with previous studies which report 
that compact radio sources having sizes smaller than a few kpc show either little or no polarization \citep{Cotton03,Fanti04}. 
It is believed that the dense inter-stellar medium (ISM) enshrouding compact radio sources act as a Faraday screen and cause depolarization. 
And, the change in the ISM properties at larger distances ($>$ 10 kpc) causes the change in the degree of polarization for large radio sources. 
We note that blazars that have long radio jets but appear small due to projection effect, are known to show significant polarization \citep{Saikia99}. 
Therefore, a few of our NLS1s having compact radio size but relatively higher fractional polarization 
can be understood as blazar-like radio sources. 
\section{Can radio emission in NLS1s be powered by star-formation?}
\label{sec:SF}
Recently, \cite{Caccianiga15} reported that a substantial fraction of radio emission in several RL-NLS1s, in particular those showing 
red MIR colours, can arise from star-formation. 
In view of this we investigate if radio emission in our radio-detected NLS1s can have significant contribution from star-formation. 
\subsection{Radio-IR correlation : q$_{\rm 22~{\mu}m}$ parameter}
\label{sec:q22} 
To distinguish the emission from AGN and star-formation we use the fact that star-forming galaxies (SFGs) 
are known to exhibit radio-IR correlation across a wide range of luminosities and 
redshifts \citep[see][]{Condon92,Appleton04,Basu15}.
The correlation between 1.4 GHz luminosity and IR luminosity 
(monochromatic IR luminosity as well as bolometric IR luminosity between 8.0 $\mu$m to 1000 $\mu$m) in SFGs is 
attributed to the fact that both radio and IR emission are closely related to star-formation \citep{Ivison10}. 
AGN with predominant radio emission from jet deviate from radio-IR correlation by showing radio-excess \citep{Moric10,DelMoro13}.  
Therefore, we examine if our NLS1s show radio-excess in the radio-IR correlation. 
We note that radio-IR correlation can be represented as the ratio of IR flux to 1.4 GHz radio flux density 
{\ie}q = log(S$_{\rm IR}$/S$_{\rm 1.4~GHz}$) \citep[see][]{Appleton04}. 
For our radio-detected NLS1s we estimate q$_{\rm 22~{\mu}m}$ = log(S$_{\rm 22~{\mu}m}$/S$_{\rm 1.4~GHz}$), where IR flux at 22 $\mu$m is 
taken from Wide-field Infrared Survey Explorer (WISE)\footnote{http://irsa.ipac.caltech.edu/Missions/wise.html}, 
and 1.4 GHz flux density is taken from FIRST whenever available, otherwise from NVSS.    
WISE is an all sky survey carried out at four MIR photometric bands namely W1 [3.6 $\mu$m], W2 [4.6 $\mu$m], W3 [12 $\mu$m] and W4 [22 $\mu$m], 
with 5$\sigma$ sensitivity of 0.08 mJy, 0.11 mJy, 1.0 mJy and 6.0 mJy, and angular resolution of 6$\arcsec$.1, 6$\arcsec$.4, 6$\arcsec$.5 
and 12$\arcsec$.0, respectively \citep{Wright10}. 
Using the most recent version of WISE source catalogue (AllWISE\footnote{http://wise2.ipac.caltech.edu/docs/release/allwise/} data release) 
we obtain MIR counterparts for 481, 480, 440 and 354 of our 498 radio-detected NLS1s with SNR $\geq$ 5.0 in W1, W2, W3 and W4 bands, respectively. 
The WISE counterparts of our radio-detected NLS1s are searched within a circle of 2$\arcsec$.0 radius centred at SDSS optical positions.    
q$_{\rm 22~{\mu}m}$ is estimated using k-corrected fluxes, where IR spectral index is derived using W3 and W4 band fluxes, 
and radio spectral index is derived using 1.4 GHz and 150 MHz flux densities, whenever available, 
otherwise an average radio spectral index of $-$0.7 is considered.   
\par
\begin{figure*}
\includegraphics[angle=0,width=9.0cm,trim={0.5cm 0.5cm 0.5cm 0.5cm},clip]{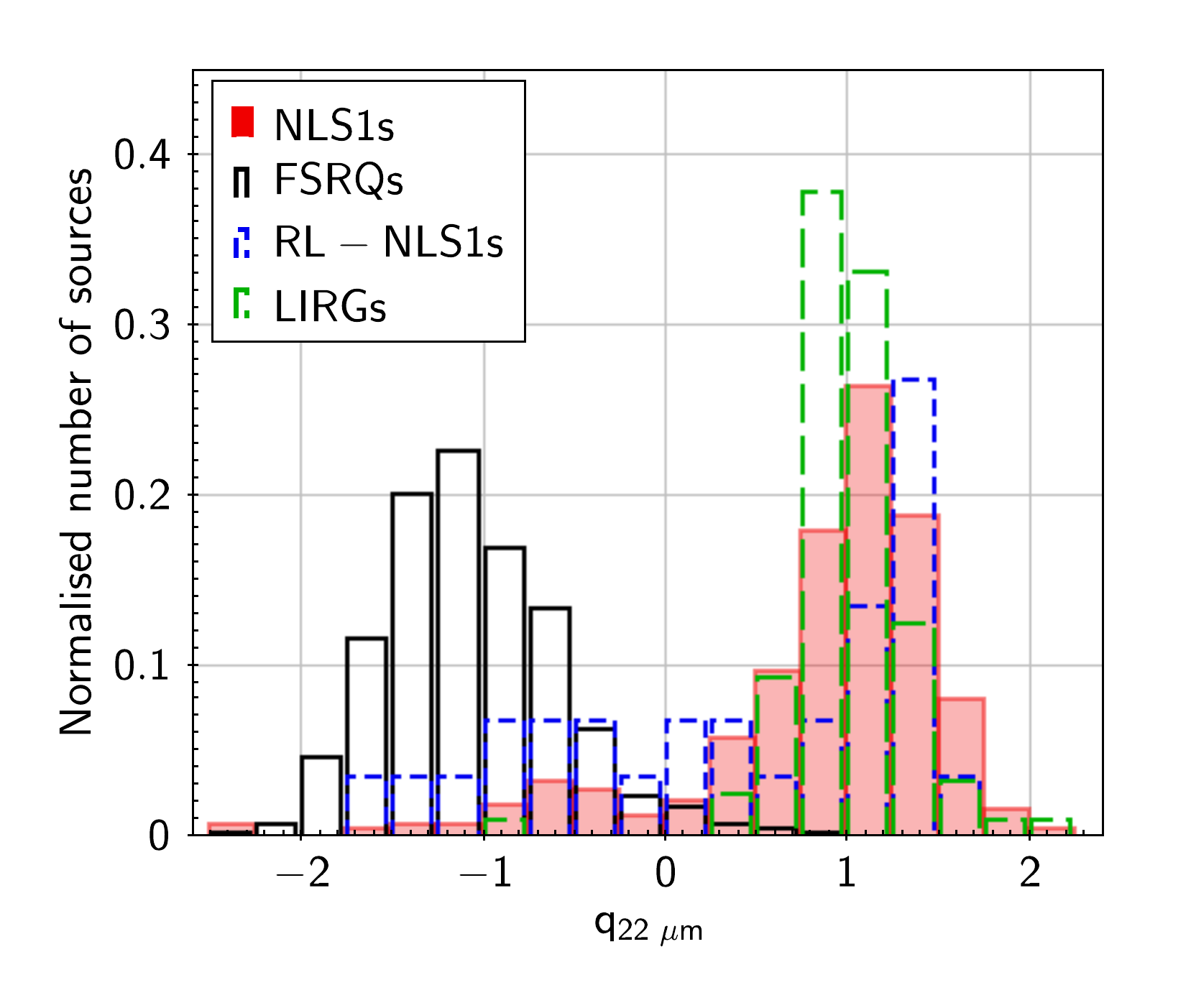}{\includegraphics[angle=0,width=9.0cm,trim={0.5cm 0.5cm 0.5cm 0.5cm},clip]{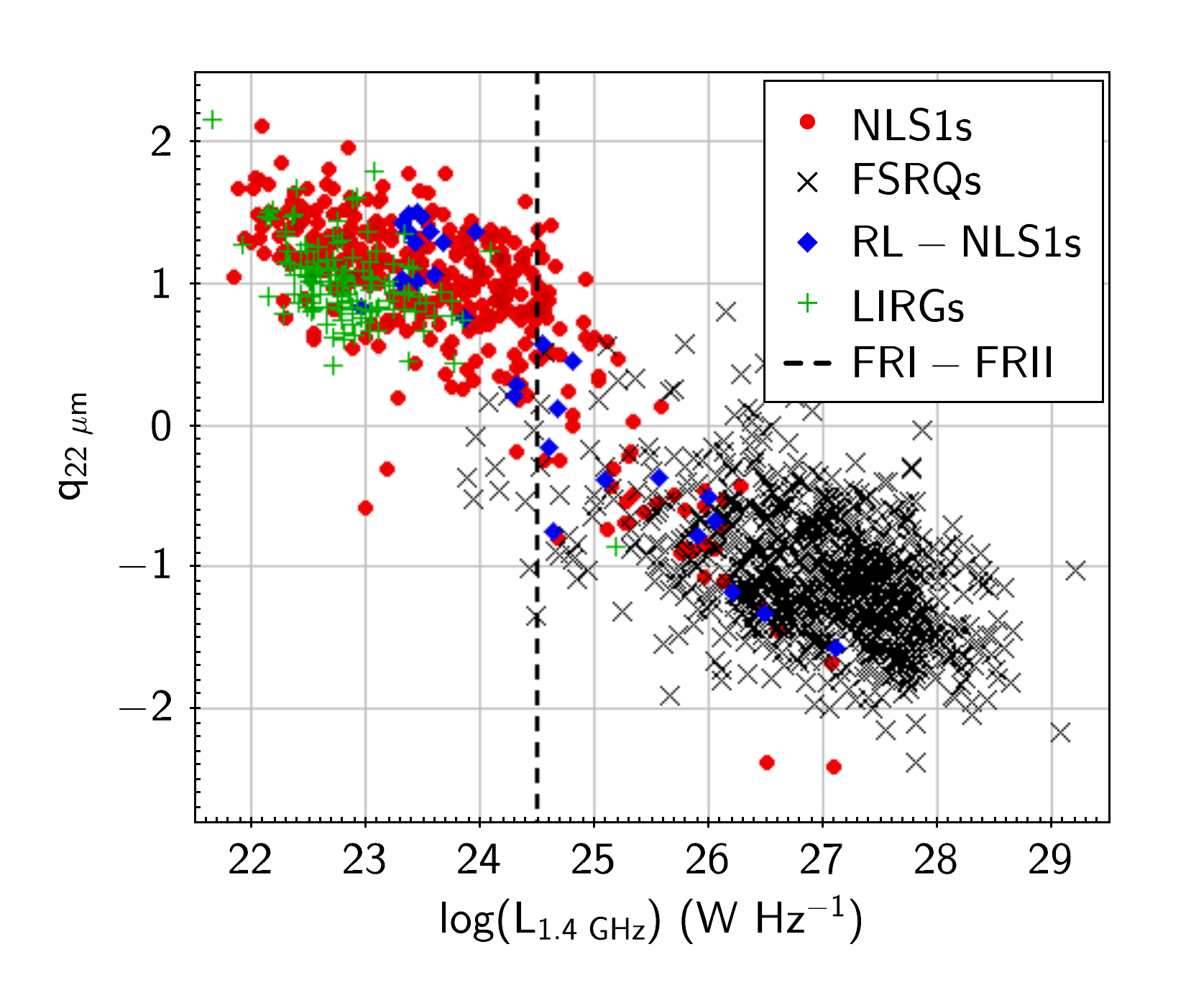}}
\caption{{\it Left panel} : Histograms of q$_{\rm 22~{\mu}m}$ for our NLS1s (in light red shades), FSRQs (in solid lines), LIRGs 
(in dashed green lines) and flat-spectrum RL-NLS1s (in small-dashed blue lines). 
{\it Right panel} : L$_{1.4~GHz}$ versus q$_{\rm 22~{\mu}m}$ for our NLS1s, FSRQs, LIRGs and flat-spectrum NLS1s. 
The sizes of error bars are smaller than the sizes of symbols.} 
\label{fig:q22Hist} 
\end{figure*}

In Fig.~\ref{fig:q22Hist} (left panel) we show the distribution of q$_{\rm 22}$ for our 354 radio-detected NLS1s and 
compare it with that for 34 flat-spectrum RL-NLS1s (taken from \citealt{Foschini15}), 
854 flat-spectrum radio quasars (FSRQs; taken from \citealt{Massaro09}), and 143 nearby luminous IR galaxies 
(LIRGs; taken from \citealt{Armus09}). 
To estimate q$_{\rm 22~{\mu}m}$ we consider only those sources that are detected at 22 $\mu$m and 1.4 GHz with SNR $\geq$ 5. 
We find that q$_{\rm 22~{\mu}m}$ for our radio-detected NLS1s, flat-spectrum RL-NLS1s, FSRQs and LIRGs 
are distributed over $-$2.4 to 2.11 with a median of 1.04, $-$1.58 to 1.50 with a median of 0.76, 
$-$2.38 to 0.80 with a median of $-$1.11, and $-$0.85 to 2.16 with a median of 1.01, respectively (see Table~\ref{table:Comp}). 
As expected FSRQs and LIRGs represent two extreme populations in q$_{\rm 22~{\mu}m}$ distribution. 
LIRGs with high star-formation rates are known to follow radio-IR correlation while FSRQ being RL-AGN show strong radio-excess. 
From Fig.~\ref{fig:q22Hist} (left panel) it is apparent that q$_{\rm 22~{\mu}m}$ distribution for NLS1s is bimodal 
showing one prominent peak in the LIRGs regime and a second small peak in the FSRQs regime. 
The fraction of our radio-detected NLS1s having q$_{\rm 22~{\mu}m}$ ($<$ 0) in the FSRQs regime 
is small (37/354 $\sim$ 10.5 per cent). 
From q$_{\rm 22~{\mu}m}$ versus radio-luminosity plot (Fig.~\ref{fig:q22Hist}, right panel) it is clear that most of 
our NLS1s with q$_{\rm 22~{\mu}m}$ similar to FSRQs have high radio luminosity (L$_{\rm 1.4~GHz}$ $\geq$ 10$^{24.5}$ W Hz$^{-1}$). 
Therefore, our NLS1s with high radio luminosity show clear radio-excess in the radio-IR correlation. 
Hence, similar to FSRQs, the radio emission in our NLS1s with high radio luminosity (L$_{\rm 1.4~GHz}$ $\geq$ 10$^{24.5}$ W Hz$^{-1}$) 
is powered by strong AGN-jet. 
We note that the majority (317/354 $\sim$ 89.5 per cent with q$_{\rm 22~{\mu}m}$ $>$ 0) of our NLS1s are 
distributed around q$_{\rm 22~{\mu}m}$ = 1.0 in the LIRGs regime (see Fig.~\ref{fig:q22Hist}, left panel). 
Therefore, it appears that our NLS1s with lower radio luminosities may have a significant contribution from star-formation. 
However, we caution that unlike RL-AGN, RQ-AGN or even radio-intermediate AGN having relatively low radio luminosity may 
not show clear radio-excess in radio-IR correlation \citep{Roy98,Moric10}, despite the fact that radio emission in RQ-AGN 
is primarily due to AGN \citep{Gallimore06,Singh15a}.  
\par
To get better insights on to the nature of radio emission in NLS1s 
we compare our NLS1s with LIRGs in q$_{\rm 22~{\mu}m}$ versus radio luminosity plot 
(see Fig.~\ref{fig:q22Hist}, right panel). 
It is interesting to note that NLS1s with q$_{\rm 22~{\mu}m}$ (> 0.0) in the LIRGs regime 
have nearly one order-of-magnitude higher average radio luminosity than that for LIRGs. 
The higher radio luminosity of NLS1s with q$_{\rm 22~{\mu}m}$ similar to LIRGs can be explained 
if NLS1s have excess in both radio as well as IR emission such that the ratio between IR and radio emission 
(q$_{\rm 22~{\mu}m}$) does not change significantly. 
The excess radio and IR emission in NLS1s in compared to LIRGs can be understood if both radio and IR emission are 
caused by either due to AGN or due to star-formation. 
We note that the high radio luminosity (L$_{\rm 1.4~GHz}$ $\geq$ 10$^{24}$ W Hz$^{-1}$) is unlikely to be 
powered by star-formation or star-burst alone \citep{Afonso05}. 
Therefore, radio emission in our NLS1s that have higher radio luminosity in compared to LIRGs, is likely to be 
powered by AGN-jet. The excess IR emission can be from dusty tori around AGN. 
Indeed, NLS1s with low radio luminosity (L$_{\rm 1.4~GHz}$ $<$ 10$^{23.5}$ W Hz$^{-1}$) similar to LIRGs tend to 
show higher IR excess in compared to LIRGs (see Fig.~\ref{fig:q22Hist}, right panel). 
This suggests that, in general, NLS1s are likely to differ with LIRGs that have fairly high SFRs 
($\sim$ up to a few 100 M$\odot$ yr$^{-1}$; \citealt{Howell10,Herrero17}) . 
Moreover, we caution that our analysis is based on only 22~${\mu}$m band and a 
detailed SED modelling covering full range of IR (8~$\mu$m$-$1000~$\mu$m) 
is required to characterise the nature of IR emission.    
\begin{table*}
\begin{minipage}{140mm}
\caption{Comparison of our NLS1s with RL-NLS1s, FSRQs and LIRGs.}
\resizebox{16cm}{!}{
\begin{tabular}{@{}cccccc@{}}
\hline
Source type           &  q$_{\rm 22~{\mu}m}$ range & ([3.4] $-$ [4.6]) range & ([4.6] $-$ [12]) range  & ([12] $-$ [22]) range    & log(L$_{\rm 1.4~GHz}$) range  \\    
                      &   (median, Q1, Q3, IQR)    & (median, Q1, Q3, IQR)   & (median, Q1, Q3, IQR)   & (median, Q1, Q3, IQR)    & (median, , Q1, Q3, IQR)    \\
                      &                            &    (Vega)               &     (Vega)              &        (Vega)            &   (W~Hz$^{-1}$) \\ \hline
Radio-undetected NLS1 &                            &  $-$0.11$-$1.69         & 1.85$-$4.48             & 1.40$-$3.71              &                    \\   
                      &                            & (0.93, 0.78, 1.06, 0.28)& (2.91, 2.72, 3.11, 0.39)& (2.46, 2.29, 2.63, 0.34) &                    \\
Radio-detected NLS1s  & $-$2.40$-$2.11             &  $-$0.09$-$1.57         & 1.91$-$4.93             & 1.61$-$3.73              & 21.85$-$27.09    \\
                      &  (1.04, 0.71, 1.30, 0.59)  & (0.97, 0.82, 1.11, 0.29)& (3.02, 2.81, 3.23, 0.42)& (2.52, 2.36, 2.68, 0.32) & (24.02, 23.14, 24.56, 1.42) \\
RL-NLS1s$^{\rm a}$    & $-$1.58$-$1.50             &  0.80$-$1.34            & 1.91$-$4.06             & 2.21$-$3.03              & 22.97$-$27.09    \\
                      &  (0.76, -0.38, 1.29, 1.67) & (1.02, 0.94, 1.11, 0.17)& (2.90, 2.75, 3.16, 0.41)& (2.62, 2.44, 2.69, 0.25) & (24.60, 23.55, 25.39, 1.84) \\
FSRQs$^{\rm b}$       & $-$2.38$-$0.80             &  0.17$-$1.59            & 1.05$-$5.68             & 1.67$-$3.26              & 23.88$-$29.21    \\
                      &($-$1.11, $-$1.38, $-$0.77, 0.66)& (1.10, 0.99, 1.19, 0.20)& (2.95, 2.76, 3.18, 0.42)& (2.45, 2.33, 2.57, 0.23) & (27.09, 26.58, 27.55, 0.97) \\
LIRGs$^{\rm c}$       & $-$0.85$-$2.16             &  0.07$-$2.25            & 1.44$-$5.32             & 1.69$-$4.34              & 21.66$-$25.19     \\
                      &  (1.01, 0.83, 1.15, 0.32)  &(0.40, 0.29, 0.60, 0.31) &(3.98, 3.74, 4.13, 0.39) & (2.80, 2.49, 3.12, 0.69) & (22.78, 22.53, 23.04, 0.51) \\
\hline  
\end{tabular}}
\label{table:Comp} 
\\
{\it Notes}. For our optically selected NLS1s that remain undetected in FIRST survey we have only MIR colours, 
and 1.4 GHz radio luminosity as well as q$_{\rm 22~{\mu}m}$ are not available.  
1.4 GHz luminosity is considered only for MIR detected sources with SNR $\geq$ 5.0 in [3.4] and [4.6] bands.
The two band MIR colours are considered only for sources that are detected with SNR $\geq$ 5.0 in the respective bands. 
For sources flagged as extended in the WISE catalogue we consider magnitude derived using elliptical aperture fitted to source.  
a : \cite{Foschini15}, b : \cite{Massaro09}, c : \cite{Armus09}.
\end{minipage}
\end{table*}
\subsection{MIR colours}
\label{sec:Colours}
\begin{figure}
\includegraphics[angle=0,width=8.0cm,trim={0.0cm 0.0cm 0.5cm 0.5cm},clip]{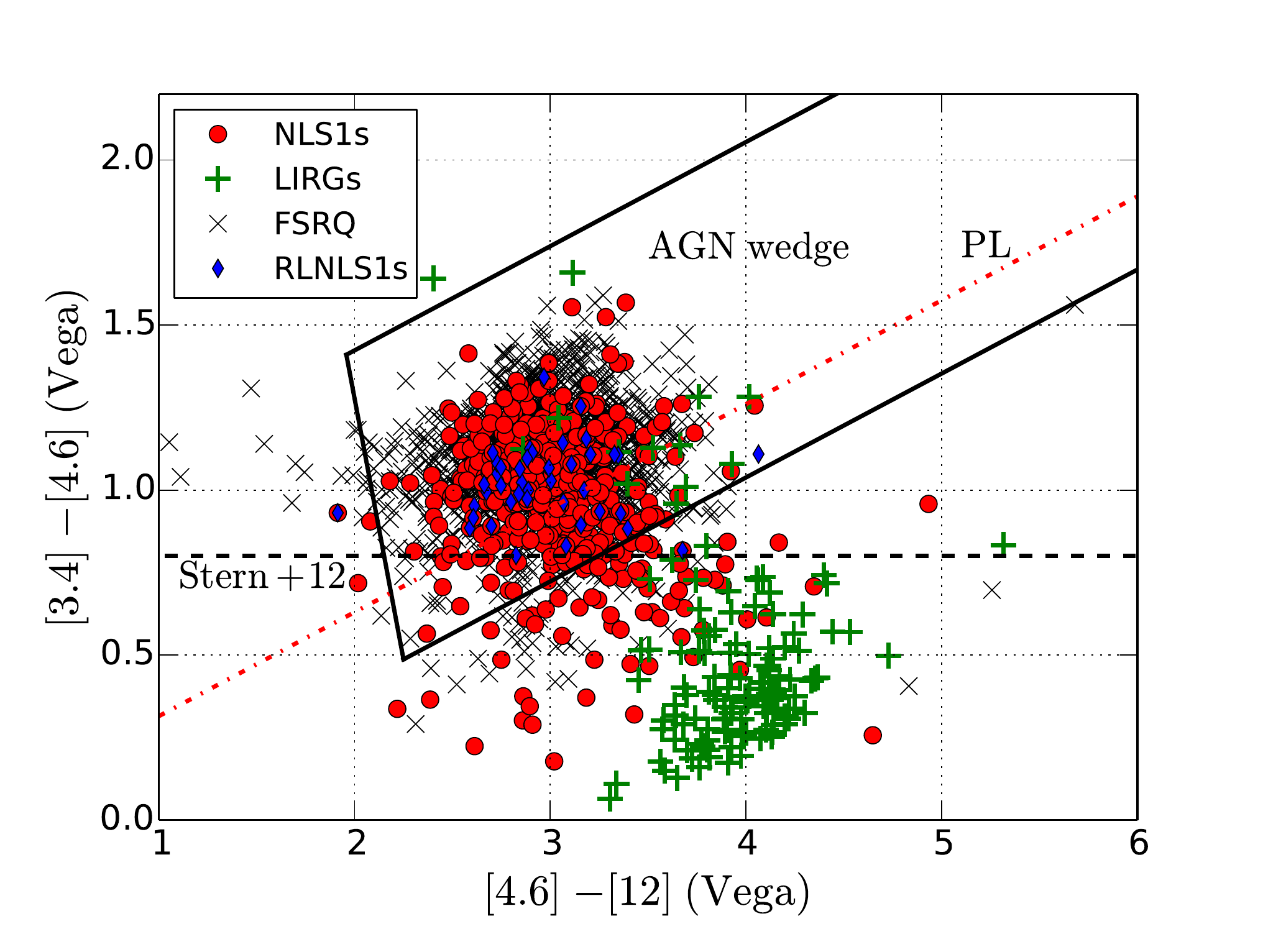}
\caption{WISE colour-colour ([3.4] $-$ [4.6] versus [4.6] $-$ [12]) plot for our radio-detected NLS1s, flat-spectrum RL-NLS1s, 
FSRQs and LIRGs. AGN selection wedge as defined by Mateos et al. (2012) is shown by solid lines. The horizontal dashed 
line represents AGN selection criterion ([3.4] $-$ [4.6] $\geq$ 0.8) given by Stern et al. (2012). 
Red dot-dashed line represents a simple power law of spectral index ranging over 0 to 2.5. 
Sources falling outside the AGN wedge are expected to have MIR emission dominated by star-formation. 
The sizes of error bars are smaller than the sizes of symbols.} 
\label{fig:WISEColor} 
\end{figure}

To get further insights on to the nature of radio and IR emission in our radio-detected NLS1s 
we examine MIR colours for our NLS1s, and compare these with that of RL-NLS1s, FSRQs and LIRGs. 
The MIR emission in AGN is believed to arise from the AGN-heated dust present in putative tori and/or from relativistic jets. 
In general, AGN show MIR spectra different than SFGs. 
Therefore, MIR colours can be used to distinguish AGN and SFGs \citep[see][]{Stern12,Mateos12}. 
In Fig.~\ref{fig:WISEColor} we plot MIR colour-colour diagram in which AGN selection wedge is defined as : 
W1 $-$ W2 $\geq$ 0.315 $\times$ (W2 $-$ W3) $-$ 0.222 and W1 $-$ W2 $<$ 0.315 $\times$ (W2 $-$ W3) + 0.796, 
and W1 $-$ W2 = $-$3.172 $\times$ (W2 $-$ W3) + 7.624, where W1, W2 and W3 are WISE bands 
at 3.4 ${\mu}$m, 4.6 ${\mu}$m and 12 ${\mu}$m, respectively \citep{Mateos12}. 
From MIR colour-colour plot it is evident that most of our radio-detected NLS1s fall within 
the AGN selection wedge. And, only a small fraction (60/440 $\sim$ 13.5 per cent) of our sources, mostly with low radio luminosity 
(L$_{\rm 1.4~GHz}$ $\leq$ 10$^{23}$ W Hz$^{-1}$), fall outside the AGN region (see Fig.~\ref{fig:WISEColor}). 
The fraction of NLS1s lying outside the AGN selection region is even lower if we consider the AGN selection criterion 
({\ie}W1 $-$ W2 $\geq$ 0.8) given by \cite{Stern12} along with that of \cite{Mateos12}. 
In fact, most of our NLS1s including the flat-spectrum RL-NLS1s and FSRQs are clustered around the power law line in 
the MIR colour-colour plot. 
Within the AGN selection wedge the deviation of sources from the power law line can be understood due to varying 
redshifts and extinction for different sources. 
\par
\cite{Mateos12} showed that AGN selection wedge is fairly reliable and only low-luminosity and heavily obscured AGN tend to fall 
outside the AGN selection wedge. Since AGN in our optically-selected NLS1s is unlikely to be obscured as broad-line regions 
are viewed directly in type~1 AGN, and hence NLS1s lying outside the AGN selection wedge 
possibly have a substantial contribution from star-formation.    
Therefore, using MIR colour-colour plot we find that MIR emission in most of our radio-detected NLS1s is primarily dominated by AGN. 
While, a small fraction of NLS1s with low radio luminosity can have a significant contribution from star-formation. 
It is important to note that, in general, the MIR colours of LIRGs are different than that for AGN. 
Although, a small fraction of LIRGs showing MIR colours similar to AGN are likely to host AGN. 
Indeed, \cite{Vardoulaki15} found that a fraction of sources in the LIRGs sample of \cite{Armus09} host AGN.  
In general, in compared to AGN, LIRGs have bluer colour in [3.4] $-$ [4.6] and redder colour in [4.6] $-$ [12], and hence, LIRGs 
occupy different location in MIR colour-colour plot. 
The different colours of LIRGs in MIR colour-colour plot can be explained as the MIR emission in LIRGs 
is mainly powered by star-formation. 
\par
Thus, the comparison of our NLS1s with LIRGs suggests that, unlike in LIRGs, 
the MIR emission in NLS1s is mostly powered by AGN.
Since MIR emission in most of our radio-detected NLS1s is primarily from AGN, and hence it would be inappropriate to estimate 
their radio-excess using radio-IR correlation, unless we separate out 
the contribution from AGN and star-formation.    
Therefore, based on the MIR colours we infer that the radio emission in most of our NLS1s is primarily from AGN. 
The contribution of star-formation to the radio emission may be substantial only in NLS1s with low radio luminosity. 
Although, we note that far-IR (FIR) observations would be more crucial to estimate the contribution from star-formation, 
as the relatively cooler dust present in star-forming regions contribute more at FIR. 
We find that majority of our radio-detected NLS1s are not detected in the Infrared Astronomical Satellite 
(IRAS) faint source catalogue version 2.0 \citep{Moshir90}.  
\subsection{1.4 GHz radio luminosity versus MIR colours}
\label{sec:ColourVsL}
\begin{figure*}
\includegraphics[angle=0,width=8.0cm,trim={0.0cm 0.0cm 0.5cm 0.5cm},clip]{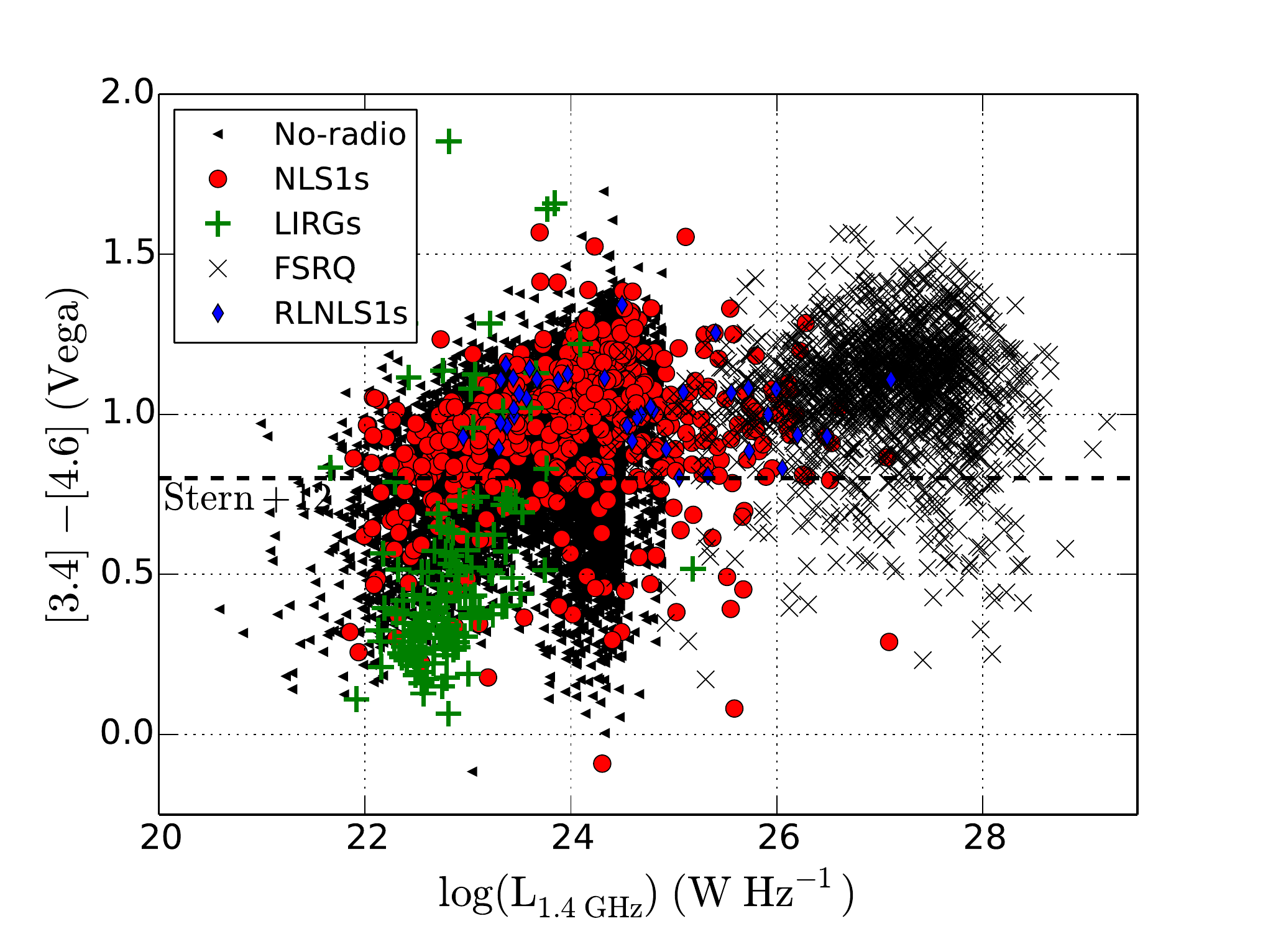}
{\includegraphics[angle=0,width=8.0cm,trim={0.0cm 0.0cm 0.5cm 0.5cm},clip]{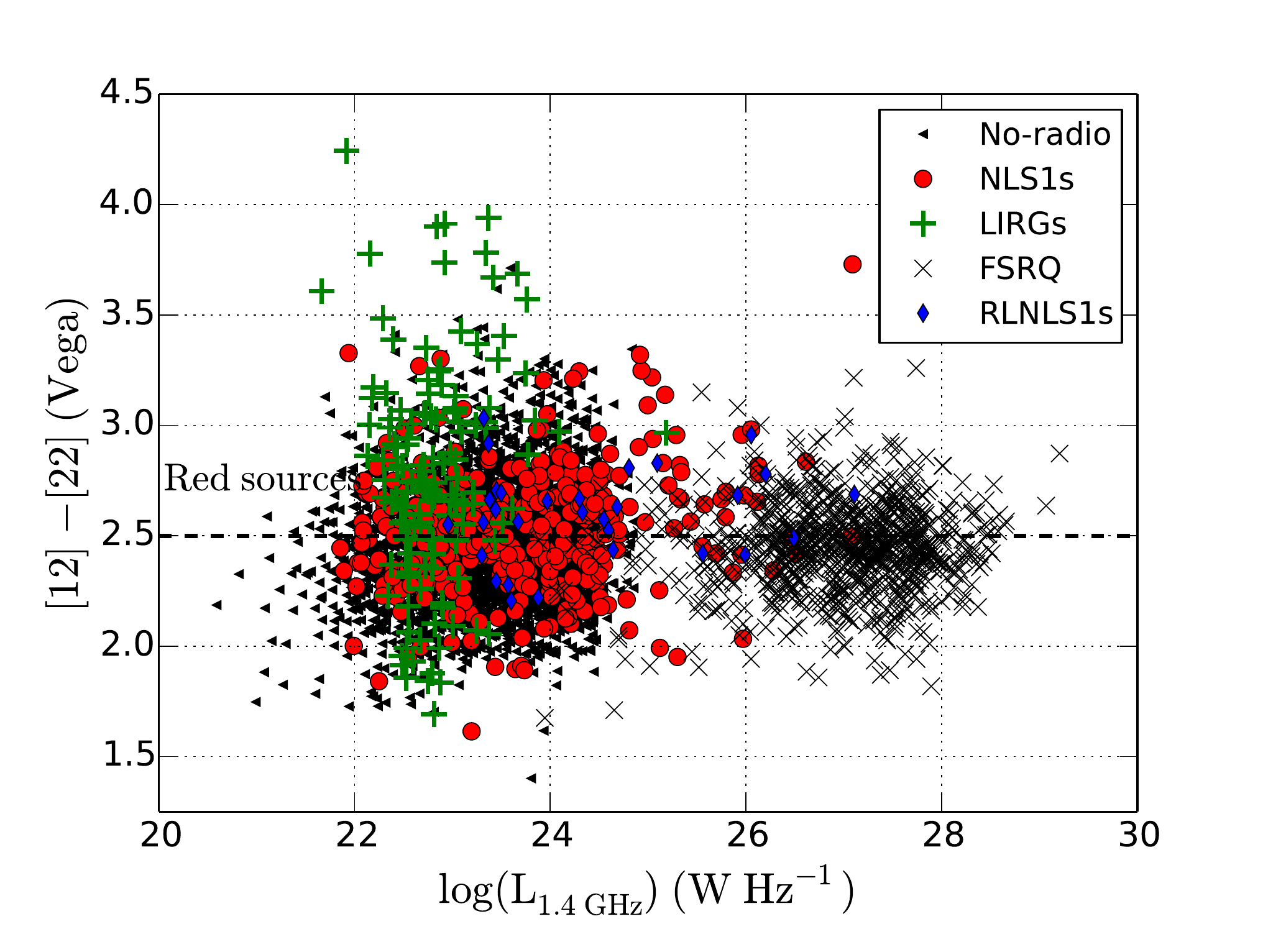}}
\caption{{\it Left panel} : L$_{\rm 1.4~GHz}$ versus [3.4] $-$ [4.6] colour. 
{\it Right panel} : L$_{\rm 1.4~GHz}$ versus [12] $-$ [22] colour. The sizes of error bars are smaller than the sizes of symbols.} 
\label{fig:WISEColorVsRL} 
\end{figure*}

In Fig.~\ref{fig:WISEColorVsRL} we show 1.4 GHz radio luminosity versus MIR colours plot for our radio-detected NLS1s, 
flat-spectrum RL-NLS1s, FSRQs and LIRGs.  
The range, median, first quartile, third quartile and interquartile range of 
1.4 GHz radio luminosity distributions and MIR colours of our NLS1s, flat-spectrum RL-NLS1s, FSRQ and LIRGs 
are listed in Table~\ref{table:Comp}.
We find that, on average, LIRGs have lowest radio luminosity and FSRQs have highest radio luminosity while 
our radio-detected NLS1s have radio luminosities ranging from LIRGs regime to FSRQs regime. 
In 1.4 GHz radio luminosity versus ([3.4] $-$ [4.6]) colour plot our NLS1s, FSRQs and LIRGs 
fall into different locations. 
The FSRQs with higher radio luminosities (L$_{\rm 1.4~GHz}$ $\sim$ 10$^{24}$$-$10$^{29}$ W Hz$^{-1}$) tend to exhibit 
red colour in [3.4] $-$ [4.6], while LIRGs mostly with much lower radio luminosities 
(L$_{\rm 1.4~GHz}$ $\sim$ 10$^{22}$$-$10$^{23.5}$ W Hz$^{-1}$) have bluer colour. 
Our radio-detected NLS1s mostly show red colour in [3.4]$-$[4.6] similar to FSRQs but also show a tendency of [3.4]$-$[4.6] colour
becoming bluer at lower radio luminosities. 
\par
In 1.4 GHz radio luminosity versus [12] $-$ [22] colour plot the segregation of NLS1s, FSRQs and LIRGs becomes less distinct.  
It seems that many LIRGs have [12] $-$ [22] colour similar to NLS1s and FSRQs.
The less distinct difference in [12] $-$ [22] colour 
can be understood if AGN component tend to decrease and star-formation component tend to increase at redder bands. 
However, we note that even if our radio-detected NLS1s have high star-formation similar to LIRGs the relatively higher radio luminosity of 
NLS1s indicates the AGN dominance. 
Only NLS1s with low radio luminosity can have a substantial contribution from star-formation. 
\cite{Caccianiga15} suggest that all NLS1s with [12] $-$ [22] $>$ 2.5 can primarily be powered by star-formation.
However, we note that several LIRGs that are known to show high SFRs have [12] $-$ [22] $<$ 2.5. 
Therefore, MIR colour [12] $-$ [22] $>$ 2.5 may not be considered as a criterion for high SFRs.   
Also, on average, purely star-forming LIRGs have much lower radio luminosity than that for NLS1s. 
Therefore, the radio emission in NLS1s with [12] $-$ [22] colours similar to LIRGs may have contribution from star-formation 
but the star-formation component is likely dominate only in NLS1s with low radio luminosity (L$_{\rm 1.4~GHz}$ $<$ 10$^{23.5}$ W Hz$^{-1}$). 
We note that the fraction of our radio-detected NLS1s having 1.4 GHz radio luminosity lower than 10$^{23.5}$ W Hz$^{-1}$ 
is merely 30 per cent. Therefore, in general, the radio emission in our radio-detected NLS1s is likely to be powered by AGN. 
\par 
Interestingly, our radio-undetected NLS1s also show MIR colours similar to the radio-detected NLS1s 
(see Table~\ref{table:Comp} and Fig.~\ref{fig:WISEColorVsRL}).
Therefore, if the radio emission in NLS1s is powered by star-formation then it is difficult to explain 
the non-detection of the majority (95 per cent) of our NLS1s having MIR colours similar to the radio-detected NLS1s. 
Both radio-detected NLS1s and radio-undetected NLS1s have similar redshift distributions. 
Therefore, similar MIR colours of both radio-detected NLS1s and radio-undetected NLS1s suggests that 
the radio emission in NLS1s is unlikely to be powered only by star-formation.
Furthermore, the linear projected radio-sizes of radio-detected NLS1s having low radio luminosity is smaller 
than the typical optical sizes (30$-$40 kpc) of galaxies (see Fig.~\ref{fig:SizeVsLumin}). 
This further indicates that the radio emission in NLS1s is associated either with AGN activity or with a nuclear star-burst. 
However, the MIR colours for the most of our radio-detected NLS1s do not support the presence of dominant star-burst. 
Therefore, our analysis based on q$_{\rm 22~{\mu}m}$ parameter, MIR colours, radio luminosity and radio sizes suggests 
that, in general, the radio emission in our radio-detected NLS1s is likely to be powered by AGN. 
\section{Discussion : Evolutionary stage of radio-jets}
\label{sec:Discussion}
The evolutionary stage of a radio AGN can be inferred from its radio size. 
In literature, radio luminosity versus linear radio-size plot has been 
used to infer the evolutionary stage of different types of radio AGN \citep[see][]{Kunert10,An12}. 
According to dynamical evolutionary models the compact parsec-scale radio-jets evolve into kpc-scale 
radio-jets which eventually turn into large jet-lobe structures of the size of hundreds of kpc \citep{Readhead96,Snellen2000}.   
However, in a fraction of AGN the radio-jets may be too weak to dispel the ISM, and thus, can remain confined within the nuclear 
region of host galaxy \citep{Marecki03,Kawakatu08}. 
\par
\begin{figure*}
\includegraphics[angle=0,width=8.5cm,trim={0.0cm 0.0cm 0.0cm 0.0cm},clip]{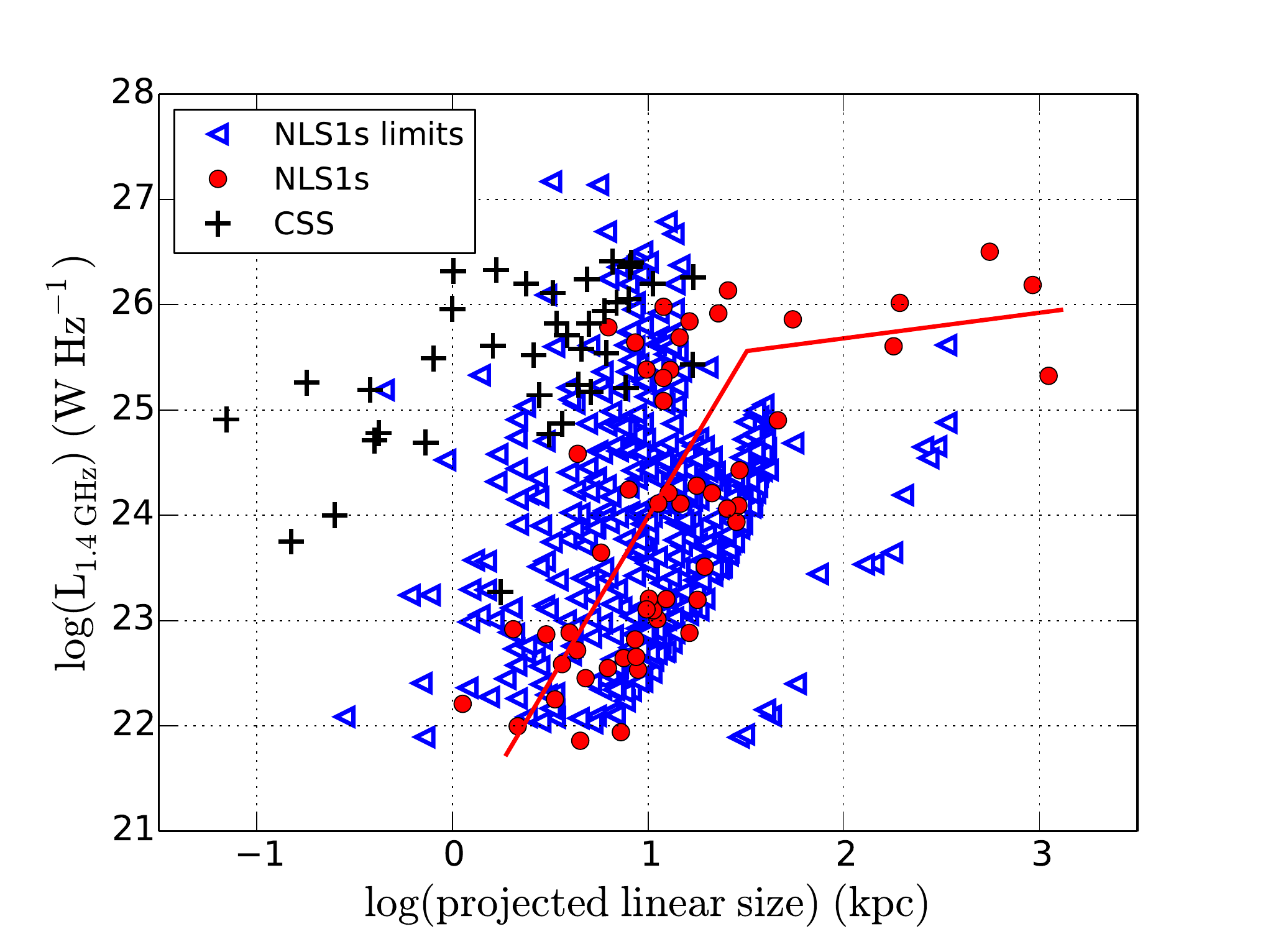}
{\includegraphics[angle=0,width=9.0cm,trim={0.5cm 0.5cm 0.5cm 0.5cm},clip]{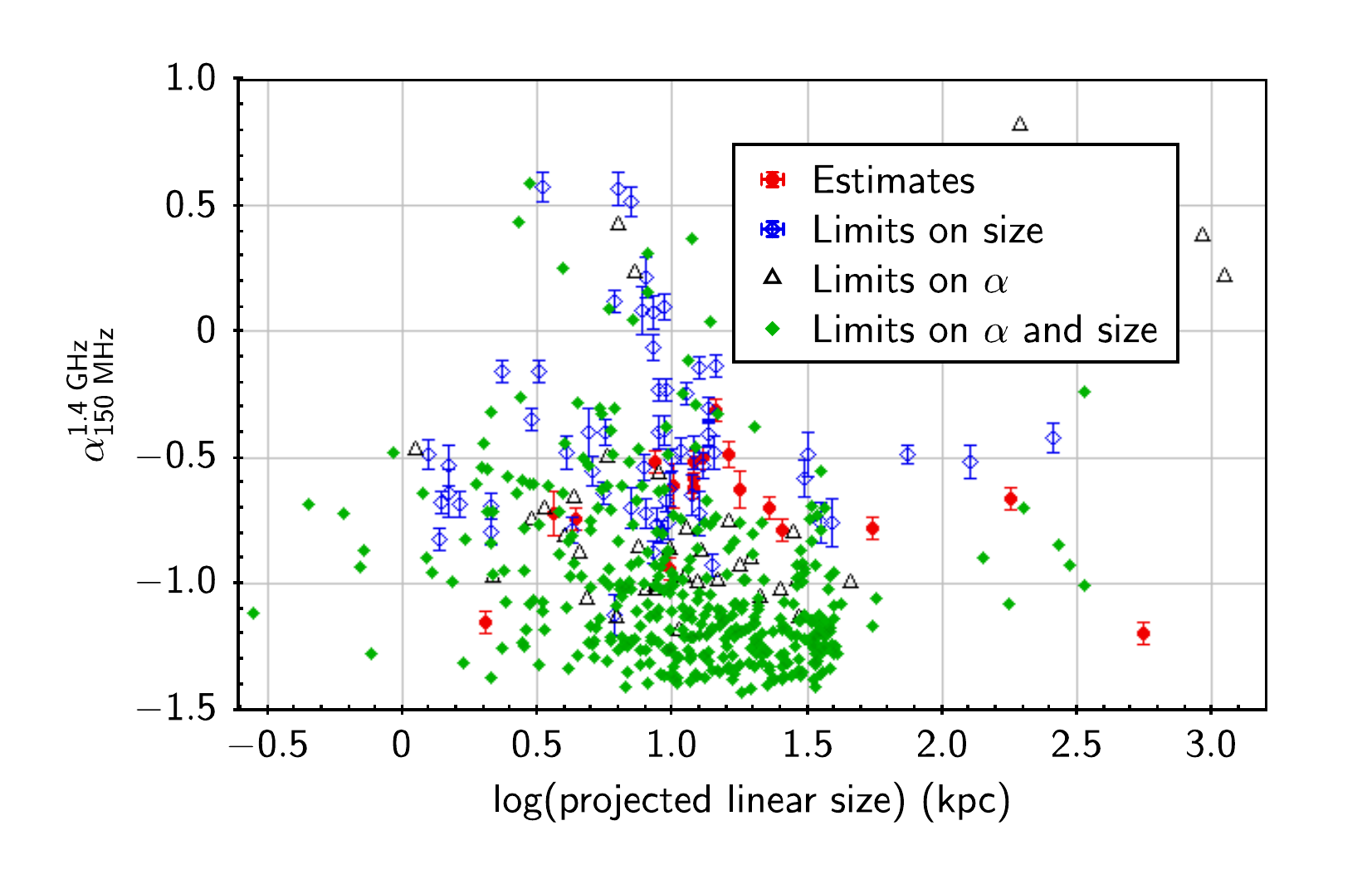}}
\caption{{\it Left panel} : 1.4 GHz radio luminosity versus the projected linear radio-size plot. `+' symbols represent 
low luminosity CSS sources taken from Kunert-Bajraszewska et al. (2010). We have excluded CSS sources with uncertain redshift estimates.   
Red solid line represents the trend shown by NLS1s with radio-size estimates. 
The upper limits on the radio-sizes are used for unresolved sources. The sizes of error bars are smaller than the sizes of symbols. 
{\it Right panel} : Projected linear radio-size versus spectral index plot for our all radio-detected NLS1s. 
Limits on the radio spectral indices and radio sizes are the upper limits.}
\label{fig:SizeVsLumin} 
\end{figure*}

In Fig.~\ref{fig:SizeVsLumin} (left panel) we plot 1.4 GHz radio luminosity versus projected linear radio-size for 
our NLS1s and CSS sources. 
It is apparent that most of our NLS1s have small radio-sizes ($<$ 30 kpc), and only a few NLS1s have 
large radio structures. 
The NLS1s with small radio-sizes are distributed across a wide range of luminosities ($\sim$ 10$^{22}$$-$10$^{26}$ W Hz$^{-1}$), 
and all NLS1s with large radio-sizes have radio luminosity $\geq$ 10$^{25}$ W Hz$^{-1}$ (see Fig.~\ref{fig:SizeVsLumin}, left panel). 
In  1.4 GHz radio luminosity versus linear size plot the CSS sources of high radio luminosity overlap with our NLS1s. 
While, the CSS sources with low radio luminosity tend to have much smaller size ($<$ 1 kpc) than that can 
be measured from the FIRST and NVSS observations. 
Our NLS1s with upper limits on their sizes may overlap with the CSS sources of low radio luminosities. 
Thus, the comparison of our NLS1s and CSS sources in the radio luminosity versus radio-size plot suggests that 
many of our radio-detected NLS1s are similar to the CSS sources. 
\par
Furthermore, NLS1s with small radio-sizes show a trend of increasing radio luminosity with the increase in radio-size. 
While, for NLS1s with large radio-sizes, the radio luminosity remains nearly constant with the increase in radio-size, however, 
a better statistics is required to get a reliable trend. 
It is worth to point out that the trend shown by our NLS1s is consistent with the evolutionary models suggesting 
a steep increase in radio luminosity when compact parsec-scale radio-jets grow into kpc-scale radio-jets, 
and a slow decrease in radio luminosity with the increase in radio-size in case of evolved large radio sources \citep[see][]{An12}. 
In the dynamical evolutionary models both GPS and CSS radio sources are believed to represent early evolutionary phase of radio galaxies, where GPS sources evolve into CSS sources 
which can eventually grow into large radio galaxies \citep{Fanti09}.  
Therefore, radio-jets in our NLS1s are likely to be in the early phase of evolution. 
The wide radio luminosity distribution for our NLS1s with small radio-sizes can be understood if our NLS1s belong to 
different stages of the early evolutionary phase. 
However, we note that NLS1s may also include sources with low jet power that can fail to grow into large radio-sizes. 
We also caution that the radio-sizes of our NLS1s are subject to the projection effect, and hence, 
a fraction of NLS1s having large radio-sizes may appear small due to foreshortening of the radio-jet 
{\ie}blazar-like sources. Also, blazar-like NLS1s are expected to show higher radio luminosity due to beaming effect. 
\par
To obtain more insights into the evolutionary stage of the radio-jets in our NLS1s we also plot 
projected linear radio-size versus radio spectral index (${\alpha}_{\rm 150~MHz}^{\rm 1.4~GHz}$) (see Fig.~\ref{fig:SizeVsLumin}, right panel). 
Previous studies on radio-AGN have shown that the radio spectral peak is anti-correlated with the radio-size so that young radio 
sources with small radio-sizes have spectral turn-over at higher frequencies, and the spectral peak shifts to lower frequency as 
radio-size grows larger \citep{ODea97,Jeyakumar16}. 
We note that our NLS1s with small radio-sizes (< 30 kpc) show a tentative trend of the flattening of spectra with the decrease in 
the linear radio-size (Fig.~\ref{fig:SizeVsLumin}, right panel). 
The spectral index becomes lower ({\ie}steeper) with the increase in radio-size which is consistent with the anti-correlation shown by 
GPS and CSS sources. 
Therefore, the radio-size versus two point spectral index (${\alpha}_{\rm 150~MHz}^{\rm 1.4~GHz}$) plot 
further indicates the similarity of our NLS1s with GPS and CSS radio sources. 
In fact, parsec-scale high-resolution radio observations of small samples have revealed that NLS1s contain both GPS and CSS like radio sources 
\citep{Foschini11,Gu15,Caccianiga17}. 
In general, GPS sources have radio-sizes of $<$ 1 kpc and show spectral peak around 1 GHz, while CSS radio sources 
have radio-sizes of few kpc to $\sim$ 10 kpc and show steep spectra around 1.0 GHz.  
Therefore, we note that our NLS1s with the flat/inverted and steep radio spectra are similar to GPS and CSS radio sources, respectively. 
\par
\begin{figure}
\includegraphics[angle=0,width=9.0cm,trim={0.0cm 0.0cm 0.0cm 0.0cm},clip]{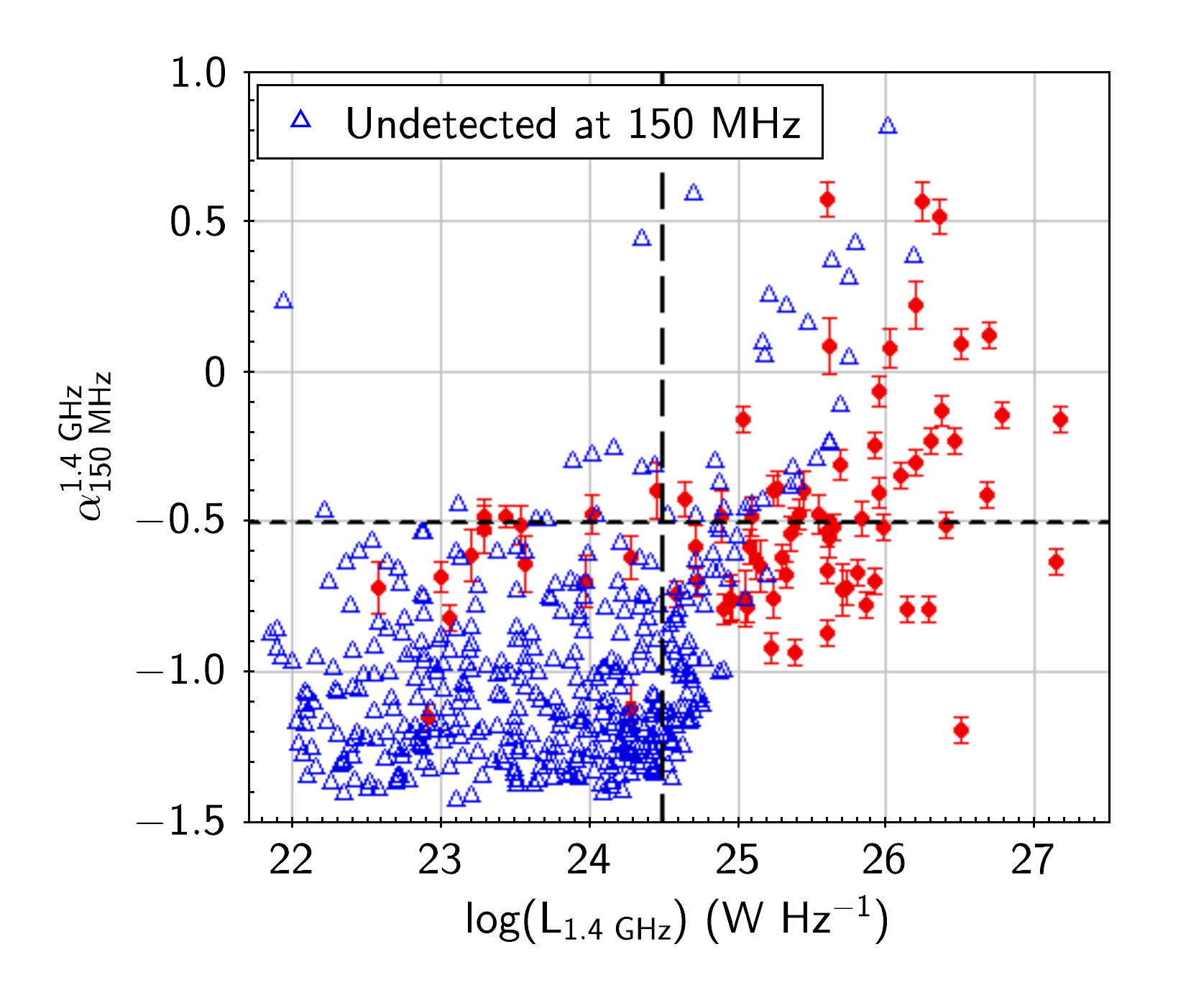}
\caption{1.4 GHz radio luminosity versus spectral index plot for our all radio-detected NLS1s. Dashed horizontal line represents the division between steep and flat spectrum sources, 
and long-dashed vertical line represents the division between low-luminosity and high-luminosity sources.}
\label{fig:SpInVsL} 
\end{figure}
Furthermore, it has been proposed that steep-spectrum NLS1s can be the parent population of flat-spectrum NLS1s 
when the radio-jet is viewed at large angles \citep{Berton15,Berton16}. 
The extended radio emission with very steep spectrum may also be indicative of the relic radio emission from previous episode of 
AGN activity {\eg}MRK 783 \citep{Congiu17}.   
In our sample we find that the steep-spectrum NLS1s are distributed across low to high radio luminosities, 
while flat-spectrum NLS1s are preferentially found at higher radio luminosities (see Fig.~\ref{fig:SpInVsL}). 
In fact, most of the flat/inverted spectrum NLS1s are as powerful as FR II radio galaxies (L$_{\rm 1.4~GHz}$ $\geq$ 10$^{24.5}$ W Hz$^{-1}$). 
The high radio luminosity of the flat/inverted spectrum NLS1s can be understood if they are beamed sources, similar to blazars, 
in which apparent luminosity is Doppler boosted due to relativistic beaming effect \citep{Foschini15}. 
Indeed, high resolution parsec-scale radio imaging of flat-spectrum radio-powerful NLS1s shows 
single core or core-jet radio morphologies and thus indicating them to be blazar-like sources \citep{Gu15}. 
Therefore, the radio properties of our NLS1s at the higher end of radio luminosity suggest them to blazar-like sources in which 
luminosity is boosted due to beaming effect.
\section{Conclusions}
\label{sec:Conclusions}
In this paper we present radio properties of, hitherto, the largest sample of 11101 optically-selected NLS1s, 
using 1.4 GHz FIRST, 1.4 GHz NVSS, 327 MHz WENSS and 150 MHz TGSS. 
The conclusions of our study are outlined as below. 
\begin{itemize}
\item With the combined use of FIRST and NVSS we detect radio counterparts of only 
498/11101 $\sim$ 4.5 per cent of our NLS1s. The radio detection rate of our NLS1s is lower than 
that for BL-Seyfert galaxies ($\sim$ 10 per cent) across all redshift bins. 
\item Our study yields the largest sample of radio-detected NLS1s. 
The 1.4 GHz flux density distribution of our NLS1s peaks in the faintest bin of 1.0$-$2.5 mJy and sharply declines beyond 5.0 mJy.
The flux density distribution and the non-detection of 95.5 per cent of our NLS1s suggest that, in general, NLS1s are weak 
in radio. 
\item The 1.4 GHz radio luminosity distribution of our NLS1s spans across nearly seven order-of-magnitude 
{\ie}from 7.2 $\times$ 10$^{21}$ W Hz$^{-1}$ to 1.5 $\times$ 10$^{27}$ W Hz$^{-1}$ with a median value 
of 1.6 $\times$ 10$^{24}$ W Hz$^{-1}$. Therefore, we primarily detect radio-powerful NLS1s among which more 
that 50 per cent have radio luminosities similar to radio galaxies. 
The 1.4 GHz luminosity distribution of upper limits for NLS1s with no detected radio counterparts indicates 
the possible presence of a substantial fraction of NLS1s with moderate radio luminosities, 
even if their average true flux densities are 
ten times lower than the FIRST detection limit of 1.0 mJy.  
\item Our NLS1s show both flat (${\alpha}_{\rm 150~MHz}^{\rm 1.4~GHz}$ $\geq$ $-$0.5) as well as steep (${\alpha}_{\rm 150~MHz}^{\rm 1.4~GHz}$ $<$ $-$0.5) 
spectral indices measured between 150 MHz and 1.4 GHz. 79 NLS1s with their radio-detections 
at 150 MHz and 1.4 GHz have ${\alpha}_{\rm 150~MHz}^{\rm 1.4~GHz}$ in the range 
of $-$1.2 to 0.58 with a median value of $-$0.53. Using S$_{\rm 1.4~GHz}$ 
versus ${\alpha}_{\rm 150~MHz}^{\rm 1.4~GHz}$ plot we find that, in general, radio-bright NLS1s tend to 
show flat radio spectra. This trend is contrary to the general radio population detected in a survey 
in which radio-bright sources are mostly powerful radio galaxies with steep spectra. 
The flat spectra seen in radio-bright NLS1s can be due to synchrotron self-absorption. 
\item The projected linear radio-sizes in most of our NLS1s tend to be compact (few kpc or less) with 443/498 (89 per cent) radio-detected NLS1s 
appear unresolved in the FIRST and NVSS. The distribution of radio-size of 55/498 
resolved sources ranges from 1.6 kpc to 1.1 Mpc with a median of 11 kpc, and shows a sharp decline beyond 30 kpc. 
Therefore, our study has increased the number of NLS1s with KSRs by several times.
\item Most (407/498 $\sim$ 81.7 per cent) of our radio-detected NLS1s can be classified as RL, however, 
the overall fraction of RL-NLS1s is only 3.7 per cent among the full sample of 11101 optically-selected NLS1s. 
The distribution of upper limits on R$_{\rm 1.4~GHz}$, for the NLS1s with no detected radio counterparts, suggests 
the possible existence of a significant fraction of RL-NLS1s, even if their average true radio flux densities 
are one order of magnitude lower than the FIRST detection limit of 1.0 mJy. 
\item The fraction of RL-NLS1s is highest ($\sim$ 7 per cent) in the lowest redshift bin ($z$ $\sim$ 0.0$-$0.1) 
and decreases to 3$-$4 per cent in the higher redshift bins. 
The overall fraction of RL sources ($\sim$ 3.7 per cent) in our NLS1s is much lower than that found 
in BL-AGN (10$-$20 per cent). This can be understood if NLS1s have lower efficiency of the production of powerful radio-jets. 
\item 
Based on q$_{\rm 22~{\mu}m}$, MIR colours, and radio luminosity versus MIR colours diagnostic plot we infer 
that, in general, the radio emission in our NLS1s of high radio luminosity (L$_{\rm 1.4~GHz}$ $\geq$ 10$^{23.5}$ W Hz$^{-1}$) 
is primarily powered by AGN, while the radio emission in our NLS1s with low radio luminosity 
(L$_{\rm 1.4~GHz}$ $<$ 10$^{23.5}$ W Hz$^{-1}$) can have a substantial contribution from star-formation.
\item Using criteria of S$_{\rm FIRST}^{\rm peak}$ $>$ S$_{\rm NVSS}^{\rm int}$ and ${\sigma}_{\rm var}$ $\geq$ 3.0, we find 
09/259 confirmed radio-variable NLS1s that show similarity with blazars {\ie}compact, RL sources with flat/inverted radio spectra. 
We note that this is only a lower limit as some sources with S$_{\rm FIRST}^{\rm peak}$ $<$ S$_{\rm NVSS}^{\rm int}$ 
may also be variable, but higher flux density in the NVSS can also be due to the detection of 
additional faint radio emission.   
\item A small fraction (24/274 $\sim$ 9 per cent) of our NVSS-detected NLS1s show radio polarization with the degree of polarization 
ranging from 0.1$-$10 per cent with a median of 3.3 per cent. Notably, all NLS1s showing radio polarization 
are powerful radio sources (L$_{\rm 1.4~GHz}$ $\geq$ 10$^{24}$ W Hz$^{-1}$) and consist of both compact as well as extended 
radio sources.   
\item The location of our radio-detected NLS1s in the diagnostic plot of the radio luminosity versus radio-size suggests 
that, similar to CSS sources, many of our NLS1s with small radio-sizes, can be in the early phase of their evolution. 
However, NLS1s with low jet power or intermittent AGN activity in short time scale, may fail 
to grow into large-scale jet-lobe structures. 
\item The L$_{\rm 1.4~GHz}$ versus ${\alpha}_{\rm 150~MHz}^{\rm 1.4~GHz}$ plot for our NLS1s shows that 
high luminosity ($\geq$ 10$^{25}$ W Hz$^{-1}$) NLS1s tend to exhibit flat/inverted spectra and compact radio sizes. 
Therefore, majority of NLS1 at high luminosity end may be blazar-like sources in which luminosity increases due to Doppler boosting.     
\end{itemize}
Our study demonstrates that NLS1s exhibit diverse radio properties resulting the wide range of radio luminosities, 
spectral indices, radio-sizes, radio-loudness, variabilities and polarizations. 
The lower radio detection rate in NLS1s, as compared to BL-AGN, can be related to the low efficiency of the production of powerful radio-jets. 
Radio-jets in NLS1s are possibly in the early evolutionary phase, however, radio-jets 
may remain confined within the nuclear region of the host galaxy due to low jet power or intermittent AGN activity.  
\section*{Acknowledgements}
We thank the anonymous referee for useful comments and suggestions that helped in improving 
the quality of this publication.
This paper uses data from the Giant Metrewave Radio Telescope (GMRT). We thank the staff  of the GMRT that made these
observations possible. GMRT is run by the National Centre for Radio Astrophysics of the Tata Institute of Fundamental
Research. This research has made use of the NASA/IPAC Extragalactic Database (NED) which is operated by the Jet Propulsion Laboratory, California 
Institute of Technology, under contract with the National Aeronautics and Space Administration. 
This publication makes use of data products from WISE, which is a joint project of the University of California, 
Los Angeles and the Jet Propulsion Laboratory/California Institute of Technology, funded by the National 
Aeronautics and Space Administration. 
Funding for SDSS has been provided by the Alfred P. Sloan Foundation, the Participating Institutions, the National Science Foundation, 
and the U.S. Department of Energy Office of Science. The SDSS website is http://www.sdss.org/.
SDSS is managed by the Astrophysical Research Consortium for the Participating Institutions of the SDSS Collaboration 
including the University of Arizona, the Brazilian Participation Group, Brookhaven National Laboratory, Carnegie Mellon University, 
University of Florida, the French Participation Group, the German Participation Group, Harvard University, the Instituto de Astrofisica de Canarias, 
the Michigan State/Notre Dame/JINA Participation Group, Johns Hopkins University, Lawrence Berkeley National Laboratory, 
Max Planck Institute for Astrophysics, Max Planck Institute for Extraterrestrial Physics, New Mexico State University, 
New York University, Ohio State University, Pennsylvania State University, University of Portsmouth, Princeton University, 
the Spanish Participation Group, University of Tokyo, University of Utah, Vanderbilt University, University of Virginia, 
University of Washington, and Yale University. This research has made use of NASA's Astrophysics Data System Bibliographic Services. 
\bibliographystyle{mnras}
\bibliography{NLSY1}

\begin{thebibliography}{}
\makeatletter
\relax
\def\mn@urlcharsother{\let\do\@makeother \do\$\do\&\do\#\do\^\do\_\do\%\do\~}
\def\mn@doi{\begingroup\mn@urlcharsother \@ifnextchar [ {\mn@doi@}
  {\mn@doi@[]}}
\def\mn@doi@[#1]#2{\def\@tempa{#1}\ifx\@tempa\@empty \href
  {http://dx.doi.org/#2} {doi:#2}\else \href {http://dx.doi.org/#2} {#1}\fi
  \endgroup}
\def\mn@eprint#1#2{\mn@eprint@#1:#2::\@nil}
\def\mn@eprint@arXiv#1{\href {http://arxiv.org/abs/#1} {{\tt arXiv:#1}}}
\def\mn@eprint@dblp#1{\href {http://dblp.uni-trier.de/rec/bibtex/#1.xml}
  {dblp:#1}}
\def\mn@eprint@#1:#2:#3:#4\@nil{\def\@tempa {#1}\def\@tempb {#2}\def\@tempc
  {#3}\ifx \@tempc \@empty \let \@tempc \@tempb \let \@tempb \@tempa \fi \ifx
  \@tempb \@empty \def\@tempb {arXiv}\fi \@ifundefined
  {mn@eprint@\@tempb}{\@tempb:\@tempc}{\expandafter \expandafter \csname
  mn@eprint@\@tempb\endcsname \expandafter{\@tempc}}}

\bibitem[\protect\citeauthoryear{{Abdo} et~al.,}{{Abdo}
  et~al.}{2009a}]{Abdo09a}
{Abdo} A.~A.,  et~al., 2009a, \mn@doi [\apj] {10.1088/0004-637X/699/2/976},
  \href {http://adsabs.harvard.edu/abs/2009ApJ...699..976A} {699, 976}

\bibitem[\protect\citeauthoryear{{Abdo} et~al.,}{{Abdo}
  et~al.}{2009b}]{Abdo09b}
{Abdo} A.~A.,  et~al., 2009b, \mn@doi [\apjl] {10.1088/0004-637X/707/2/L142},
  \href {http://adsabs.harvard.edu/abs/2009ApJ...707L.142A} {707, L142}

\bibitem[\protect\citeauthoryear{{Afonso}, {Georgakakis}, {Almeida}, {Hopkins},
  {Cram}, {Mobasher}  \& {Sullivan}}{{Afonso} et~al.}{2005}]{Afonso05}
{Afonso} J.,  {Georgakakis} A.,  {Almeida} C.,  {Hopkins} A.~M.,  {Cram} L.~E.,
   {Mobasher} B.,   {Sullivan} M.,  2005, \mn@doi [\apj] {10.1086/428923},
  \href {http://adsabs.harvard.edu/abs/2005ApJ...624..135A} {624, 135}

\bibitem[\protect\citeauthoryear{{Ai}, {Yuan}, {Zhou}, {Wang}  \& {Zhang}}{{Ai}
  et~al.}{2011}]{Zhang11}
{Ai} Y.~L.,  {Yuan} W.,  {Zhou} H.~Y.,  {Wang} T.~G.,   {Zhang} S.~H.,  2011,
  \mn@doi [\apj] {10.1088/0004-637X/727/1/31}, \href
  {http://adsabs.harvard.edu/abs/2011ApJ...727...31A} {727, 31}

\bibitem[\protect\citeauthoryear{{An} \& {Baan}}{{An} \& {Baan}}{2012}]{An12}
{An} T.,  {Baan} W.~A.,  2012, \mn@doi [\apj] {10.1088/0004-637X/760/1/77},
  \href {http://adsabs.harvard.edu/abs/2012ApJ...760...77A} {760, 77}

\bibitem[\protect\citeauthoryear{{Appleton} et~al.,}{{Appleton}
  et~al.}{2004}]{Appleton04}
{Appleton} P.~N.,  et~al., 2004, \mn@doi [\apjs] {10.1086/422425}, \href
  {http://adsabs.harvard.edu/abs/2004ApJS..154..147A} {154, 147}

\bibitem[\protect\citeauthoryear{{Armus} et~al.,}{{Armus}
  et~al.}{2009}]{Armus09}
{Armus} L.,  et~al., 2009, \mn@doi [\pasp] {10.1086/600092}, \href
  {http://adsabs.harvard.edu/abs/2009PASP..121..559A} {121, 559}

\bibitem[\protect\citeauthoryear{{Basu}, {Wadadekar}, {Beelen}, {Singh},
  {Archana}, {Sirothia}  \& {Ishwara-Chandra}}{{Basu} et~al.}{2015}]{Basu15}
{Basu} A.,  {Wadadekar} Y.,  {Beelen} A.,  {Singh} V.,  {Archana} K.~N.,
  {Sirothia} S.,   {Ishwara-Chandra} C.~H.,  2015, \mn@doi [\apj]
  {10.1088/0004-637X/803/2/51}, \href
  {http://adsabs.harvard.edu/abs/2015ApJ...803...51B} {803, 51}

\bibitem[\protect\citeauthoryear{{Becker}, {White}  \& {Helfand}}{{Becker}
  et~al.}{1995}]{Becker95}
{Becker} R.~H.,  {White} R.~L.,   {Helfand} D.~J.,  1995, \mn@doi [\apj]
  {10.1086/176166}, \href {http://adsabs.harvard.edu/abs/1995ApJ...450..559B}
  {450, 559}

\bibitem[\protect\citeauthoryear{{Berton} et~al.,}{{Berton}
  et~al.}{2015}]{Berton15}
{Berton} M.,  et~al., 2015, \mn@doi [\aap] {10.1051/0004-6361/201525691}, \href
  {http://adsabs.harvard.edu/abs/2015A%26A...578A..28B} {578, A28}

\bibitem[\protect\citeauthoryear{{Berton} et~al.,}{{Berton}
  et~al.}{2016}]{Berton16}
{Berton} M.,  et~al., 2016, \mn@doi [\aap] {10.1051/0004-6361/201628171}, \href
  {http://adsabs.harvard.edu/abs/2016A%26A...591A..98B} {591, A98}

\bibitem[\protect\citeauthoryear{{Berton} et~al.,}{{Berton}
  et~al.}{2018}]{Berton18}
{Berton} M.,  et~al., 2018, \mn@doi [\aap] {10.1051/0004-6361/201832612}, \href
  {http://adsabs.harvard.edu/abs/2018A%26A...614A..87B} {614, A87}

\bibitem[\protect\citeauthoryear{{Caccianiga} et~al.,}{{Caccianiga}
  et~al.}{2015}]{Caccianiga15}
{Caccianiga} A.,  et~al., 2015, \mn@doi [\mnras] {10.1093/mnras/stv939}, \href
  {http://adsabs.harvard.edu/abs/2015MNRAS.451.1795C} {451, 1795}

\bibitem[\protect\citeauthoryear{{Caccianiga}, {Dallacasa}, {Ant{\'o}n},
  {Ballo}, {Berton}, {Mack}  \& {Paulino-Afonso}}{{Caccianiga}
  et~al.}{2017}]{Caccianiga17}
{Caccianiga} A.,  {Dallacasa} D.,  {Ant{\'o}n} S.,  {Ballo} L.,  {Berton} M.,
  {Mack} K.-H.,   {Paulino-Afonso} A.,  2017, \mn@doi [\mnras]
  {10.1093/mnras/stw2471}, \href
  {http://adsabs.harvard.edu/abs/2017MNRAS.464.1474C} {464, 1474}

\bibitem[\protect\citeauthoryear{{Condon}}{{Condon}}{1992}]{Condon92}
{Condon} J.~J.,  1992, \mn@doi [\araa] {10.1146/annurev.aa.30.090192.003043},
  \href {http://adsabs.harvard.edu/abs/1992ARA%26A..30..575C} {30, 575}

\bibitem[\protect\citeauthoryear{{Condon}, {Cotton}, {Greisen}, {Yin},
  {Perley}, {Taylor}  \& {Broderick}}{{Condon} et~al.}{1998}]{Condon98}
{Condon} J.~J.,  {Cotton} W.~D.,  {Greisen} E.~W.,  {Yin} Q.~F.,  {Perley}
  R.~A.,  {Taylor} G.~B.,   {Broderick} J.~J.,  1998, \mn@doi [\aj]
  {10.1086/300337}, \href {http://adsabs.harvard.edu/abs/1998AJ....115.1693C}
  {115, 1693}

\bibitem[\protect\citeauthoryear{{Congiu} et~al.,}{{Congiu}
  et~al.}{2017}]{Congiu17}
{Congiu} E.,  et~al., 2017, \mn@doi [\aap] {10.1051/0004-6361/201730616}, \href
  {http://adsabs.harvard.edu/abs/2017A%26A...603A..32C} {603, A32}

\bibitem[\protect\citeauthoryear{{Cotton} et~al.,}{{Cotton}
  et~al.}{2003}]{Cotton03}
{Cotton} W.~D.,  et~al., 2003, \mn@doi [\pasa] {10.1071/AS02031}, \href
  {http://adsabs.harvard.edu/abs/2003PASA...20...12C} {20, 12}

\bibitem[\protect\citeauthoryear{{D'Ammando} et~al.,}{{D'Ammando}
  et~al.}{2012}]{DAmmando12}
{D'Ammando} F.,  et~al., 2012, \mn@doi [\mnras]
  {10.1111/j.1365-2966.2012.21707.x}, \href
  {http://adsabs.harvard.edu/abs/2012MNRAS.426..317D} {426, 317}

\bibitem[\protect\citeauthoryear{{D'Ammando} et~al.,}{{D'Ammando}
  et~al.}{2015}]{DAmmando15}
{D'Ammando} F.,  et~al., 2015, \mn@doi [\mnras] {10.1093/mnras/stu2251}, \href
  {http://adsabs.harvard.edu/abs/2015MNRAS.446.2456D} {446, 2456}

\bibitem[\protect\citeauthoryear{{Dabhade}, {Gaikwad}, {Bagchi},
  {Pandey-Pommier}, {Sankhyayan}  \& {Raychaudhury}}{{Dabhade}
  et~al.}{2017}]{Dabhade17}
{Dabhade} P.,  {Gaikwad} M.,  {Bagchi} J.,  {Pandey-Pommier} M.,  {Sankhyayan}
  S.,   {Raychaudhury} S.,  2017, \mn@doi [\mnras] {10.1093/mnras/stx860},
  \href {http://adsabs.harvard.edu/abs/2017MNRAS.469.2886D} {469, 2886}

\bibitem[\protect\citeauthoryear{{Del Moro} et~al.,}{{Del Moro}
  et~al.}{2013}]{DelMoro13}
{Del Moro} A.,  et~al., 2013, \mn@doi [\aap] {10.1051/0004-6361/201219880},
  \href {http://adsabs.harvard.edu/abs/2013A%26A...549A..59D} {549, A59}

\bibitem[\protect\citeauthoryear{{Doi} et~al.,}{{Doi} et~al.}{2007}]{Doi07}
{Doi} A.,  et~al., 2007, \mn@doi [\pasj] {10.1093/pasj/59.4.703}, \href
  {http://adsabs.harvard.edu/abs/2007PASJ...59..703D} {59, 703}

\bibitem[\protect\citeauthoryear{{Doi}, {Asada}  \& {Nagai}}{{Doi}
  et~al.}{2011}]{Doi11}
{Doi} A.,  {Asada} K.,   {Nagai} H.,  2011, \mn@doi [\apj]
  {10.1088/0004-637X/738/2/126}, \href
  {http://adsabs.harvard.edu/abs/2011ApJ...738..126D} {738, 126}

\bibitem[\protect\citeauthoryear{{Doi}, {Nagira}, {Kawakatu}, {Kino}, {Nagai}
  \& {Asada}}{{Doi} et~al.}{2012}]{Doi12}
{Doi} A.,  {Nagira} H.,  {Kawakatu} N.,  {Kino} M.,  {Nagai} H.,   {Asada} K.,
  2012, \mn@doi [\apj] {10.1088/0004-637X/760/1/41}, \href
  {http://adsabs.harvard.edu/abs/2012ApJ...760...41D} {760, 41}

\bibitem[\protect\citeauthoryear{{Doi}, {Asada}, {Fujisawa}, {Nagai},
  {Hagiwara}, {Wajima}  \& {Inoue}}{{Doi} et~al.}{2013}]{Doi13}
{Doi} A.,  {Asada} K.,  {Fujisawa} K.,  {Nagai} H.,  {Hagiwara} Y.,  {Wajima}
  K.,   {Inoue} M.,  2013, \mn@doi [\apj] {10.1088/0004-637X/765/1/69}, \href
  {http://adsabs.harvard.edu/abs/2013ApJ...765...69D} {765, 69}

\bibitem[\protect\citeauthoryear{{Doi}, {Wajima}, {Hagiwara}  \& {Inoue}}{{Doi}
  et~al.}{2015}]{Doi15}
{Doi} A.,  {Wajima} K.,  {Hagiwara} Y.,   {Inoue} M.,  2015, \mn@doi [\apjl]
  {10.1088/2041-8205/798/2/L30}, \href
  {http://adsabs.harvard.edu/abs/2015ApJ...798L..30D} {798, L30}

\bibitem[\protect\citeauthoryear{{Dunlop} \& {Peacock}}{{Dunlop} \&
  {Peacock}}{1990}]{Dunlop90}
{Dunlop} J.~S.,  {Peacock} J.~A.,  1990, \mnras, \href
  {http://adsabs.harvard.edu/abs/1990MNRAS.247...19D} {247, 19}

\bibitem[\protect\citeauthoryear{{Falcke}, {Sherwood}  \& {Patnaik}}{{Falcke}
  et~al.}{1996}]{Falcke96}
{Falcke} H.,  {Sherwood} W.,   {Patnaik} A.~R.,  1996, \mn@doi [\apj]
  {10.1086/177956}, \href {http://adsabs.harvard.edu/abs/1996ApJ...471..106F}
  {471, 106}

\bibitem[\protect\citeauthoryear{{Fanaroff} \& {Riley}}{{Fanaroff} \&
  {Riley}}{1974}]{Fanaroff74}
{Fanaroff} B.~L.,  {Riley} J.~M.,  1974, \mn@doi [\mnras]
  {10.1093/mnras/167.1.31P}, \href
  {http://adsabs.harvard.edu/abs/1974MNRAS.167P..31F} {167, 31P}

\bibitem[\protect\citeauthoryear{{Fanti}}{{Fanti}}{2009}]{Fanti09}
{Fanti} C.,  2009, \mn@doi [Astronomische Nachrichten]
  {10.1002/asna.200811137}, \href
  {http://adsabs.harvard.edu/abs/2009AN....330..120F} {330, 120}

\bibitem[\protect\citeauthoryear{{Fanti} et~al.,}{{Fanti}
  et~al.}{2004}]{Fanti04}
{Fanti} C.,  et~al., 2004, \mn@doi [\aap] {10.1051/0004-6361:20040460}, \href
  {http://adsabs.harvard.edu/abs/2004A%26A...427..465F} {427, 465}

\bibitem[\protect\citeauthoryear{{Foschini}}{{Foschini}}{2011}]{Foschini11}
{Foschini} L.,  2011, in Narrow-Line Seyfert 1 Galaxies and their Place in the
  Universe. p.~24 (\mn@eprint {arXiv} {1105.0772})

\bibitem[\protect\citeauthoryear{{Foschini} et~al.,}{{Foschini}
  et~al.}{2012}]{Foschini12}
{Foschini} L.,  et~al., 2012, \mn@doi [\aap] {10.1051/0004-6361/201220225},
  \href {http://adsabs.harvard.edu/abs/2012A%26A...548A.106F} {548, A106}

\bibitem[\protect\citeauthoryear{{Foschini}, {Berton}, {Caccianiga}, {Ciroi},
  {Cracco}, {Peterson}  \& {Rafanelli}}{{Foschini} et~al.}{2015}]{Foschini15}
{Foschini} L.,  {Berton} M.,  {Caccianiga} A.,  {Ciroi} S.,  {Cracco} V.,
  {Peterson} B.~M.,   {Rafanelli} P.,  2015, \mn@doi [\aap]
  {10.1051/0004-6361/201424972}, \href
  {http://adsabs.harvard.edu/abs/2015A%26A...575A..13F} {575, A13}

\bibitem[\protect\citeauthoryear{{Gab{\'a}nyi}, {Frey}, {Paragi},
  {J{\"a}rvel{\"a}}, {Morokuma}, {An}, {Tanaka}  \& {Tar}}{{Gab{\'a}nyi}
  et~al.}{2018}]{Gabanyi18}
{Gab{\'a}nyi} K.~{\'E}.,  {Frey} S.,  {Paragi} Z.,  {J{\"a}rvel{\"a}} E.,
  {Morokuma} T.,  {An} T.,  {Tanaka} M.,   {Tar} I.,  2018, \mn@doi [\mnras]
  {10.1093/mnras/stx2449}, \href
  {http://adsabs.harvard.edu/abs/2018MNRAS.473.1554G} {473, 1554}

\bibitem[\protect\citeauthoryear{{Gallimore}, {Axon}, {O'Dea}, {Baum}  \&
  {Pedlar}}{{Gallimore} et~al.}{2006}]{Gallimore06}
{Gallimore} J.~F.,  {Axon} D.~J.,  {O'Dea} C.~P.,  {Baum} S.~A.,   {Pedlar} A.,
   2006, \mn@doi [\aj] {10.1086/504593}, \href
  {http://adsabs.harvard.edu/abs/2006AJ....132..546G} {132, 546}

\bibitem[\protect\citeauthoryear{{Giroletti} \& {Polatidis}}{{Giroletti} \&
  {Polatidis}}{2009}]{Giroletti09}
{Giroletti} M.,  {Polatidis} A.,  2009, \mn@doi [Astronomische Nachrichten]
  {10.1002/asna.200811154}, \href
  {http://adsabs.harvard.edu/abs/2009AN....330..193G} {330, 193}

\bibitem[\protect\citeauthoryear{{Goodrich}}{{Goodrich}}{1989}]{Goodrich89}
{Goodrich} R.~W.,  1989, \mn@doi [\apj] {10.1086/167586}, \href
  {http://adsabs.harvard.edu/abs/1989ApJ...342..224G} {342, 224}

\bibitem[\protect\citeauthoryear{{Gu} \& {Chen}}{{Gu} \& {Chen}}{2010}]{Gu10}
{Gu} M.,  {Chen} Y.,  2010, \mn@doi [\aj] {10.1088/0004-6256/139/6/2612}, \href
  {http://adsabs.harvard.edu/abs/2010AJ....139.2612G} {139, 2612}

\bibitem[\protect\citeauthoryear{{Gu}, {Chen}, {Komossa}, {Yuan}, {Shen},
  {Wajima}, {Zhou}  \& {Zensus}}{{Gu} et~al.}{2015}]{Gu15}
{Gu} M.,  {Chen} Y.,  {Komossa} S.,  {Yuan} W.,  {Shen} Z.,  {Wajima} K.,
  {Zhou} H.,   {Zensus} J.~A.,  2015, \mn@doi [\apjs]
  {10.1088/0067-0049/221/1/3}, \href
  {http://adsabs.harvard.edu/abs/2015ApJS..221....3G} {221, 3}

\bibitem[\protect\citeauthoryear{{Herrero-Illana} et~al.,}{{Herrero-Illana}
  et~al.}{2017}]{Herrero17}
{Herrero-Illana} R.,  et~al., 2017, \mn@doi [\mnras] {10.1093/mnras/stx1672},
  \href {http://adsabs.harvard.edu/abs/2017MNRAS.471.1634H} {471, 1634}

\bibitem[\protect\citeauthoryear{{Ho}}{{Ho}}{2008}]{Ho08}
{Ho} L.~C.,  2008, \mn@doi [\araa] {10.1146/annurev.astro.45.051806.110546},
  \href {http://adsabs.harvard.edu/abs/2008ARA%26A..46..475H} {46, 475}

\bibitem[\protect\citeauthoryear{{Howell} et~al.,}{{Howell}
  et~al.}{2010}]{Howell10}
{Howell} J.~H.,  et~al., 2010, \mn@doi [\apj] {10.1088/0004-637X/715/1/572},
  \href {http://adsabs.harvard.edu/abs/2010ApJ...715..572H} {715, 572}

\bibitem[\protect\citeauthoryear{{Intema}, {van Weeren}, {R{\"o}ttgering}  \&
  {Lal}}{{Intema} et~al.}{2011}]{Intema11}
{Intema} H.~T.,  {van Weeren} R.~J.,  {R{\"o}ttgering} H.~J.~A.,   {Lal} D.~V.,
   2011, \mn@doi [\aap] {10.1051/0004-6361/201014253}, \href
  {http://adsabs.harvard.edu/abs/2011A%26A...535A..38I} {535, A38}

\bibitem[\protect\citeauthoryear{{Intema}, {Jagannathan}, {Mooley}  \&
  {Frail}}{{Intema} et~al.}{2017}]{Intema17}
{Intema} H.~T.,  {Jagannathan} P.,  {Mooley} K.~P.,   {Frail} D.~A.,  2017,
  \mn@doi [\aap] {10.1051/0004-6361/201628536}, \href
  {http://adsabs.harvard.edu/abs/2017A%26A...598A..78I} {598, A78}

\bibitem[\protect\citeauthoryear{{Ishwara-Chandra}, {Sirothia}, {Wadadekar},
  {Pal}  \& {Windhorst}}{{Ishwara-Chandra} et~al.}{2010}]{Ishwara-Chandra10}
{Ishwara-Chandra} C.~H.,  {Sirothia} S.~K.,  {Wadadekar} Y.,  {Pal} S.,
  {Windhorst} R.,  2010, \mn@doi [\mnras] {10.1111/j.1365-2966.2010.16452.x},
  \href {http://adsabs.harvard.edu/abs/2010MNRAS.405..436I} {405, 436}

\bibitem[\protect\citeauthoryear{{Ivezi{\'c}} et~al.,}{{Ivezi{\'c}}
  et~al.}{2002}]{Ivezic02}
{Ivezi{\'c}} {\v Z}.,  et~al., 2002, \mn@doi [\aj] {10.1086/344069}, \href
  {http://adsabs.harvard.edu/abs/2002AJ....124.2364I} {124, 2364}

\bibitem[\protect\citeauthoryear{{Ivison} et~al.,}{{Ivison}
  et~al.}{2010}]{Ivison10}
{Ivison} R.~J.,  et~al., 2010, \mn@doi [\aap] {10.1051/0004-6361/201014552},
  \href {http://adsabs.harvard.edu/abs/2010A%26A...518L..31I} {518, L31}

\bibitem[\protect\citeauthoryear{{Jester} et~al.,}{{Jester}
  et~al.}{2005}]{Jester05}
{Jester} S.,  et~al., 2005, \mn@doi [\aj] {10.1086/432466}, \href
  {http://adsabs.harvard.edu/abs/2005AJ....130..873J} {130, 873}

\bibitem[\protect\citeauthoryear{{Jeyakumar}}{{Jeyakumar}}{2016}]{Jeyakumar16}
{Jeyakumar} S.,  2016, \mn@doi [\mnras] {10.1093/mnras/stw181}, \href
  {http://adsabs.harvard.edu/abs/2016MNRAS.458.3786J} {458, 3786}

\bibitem[\protect\citeauthoryear{{Kawakatu}, {Nagai}  \& {Kino}}{{Kawakatu}
  et~al.}{2008}]{Kawakatu08}
{Kawakatu} N.,  {Nagai} H.,   {Kino} M.,  2008, \mn@doi [\apj]
  {10.1086/591900}, \href {http://adsabs.harvard.edu/abs/2008ApJ...687..141K}
  {687, 141}

\bibitem[\protect\citeauthoryear{{Kellermann}, {Sramek}, {Schmidt}, {Shaffer}
  \& {Green}}{{Kellermann} et~al.}{1989}]{Kellermann89}
{Kellermann} K.~I.,  {Sramek} R.,  {Schmidt} M.,  {Shaffer} D.~B.,   {Green}
  R.,  1989, \mn@doi [\aj] {10.1086/115207}, \href
  {http://adsabs.harvard.edu/abs/1989AJ.....98.1195K} {98, 1195}

\bibitem[\protect\citeauthoryear{{Kellermann}, {Sramek}, {Schmidt}, {Green}  \&
  {Shaffer}}{{Kellermann} et~al.}{1994}]{Kellermann94}
{Kellermann} K.~I.,  {Sramek} R.~A.,  {Schmidt} M.,  {Green} R.~F.,   {Shaffer}
  D.~B.,  1994, \mn@doi [\aj] {10.1086/117145}, \href
  {http://adsabs.harvard.edu/abs/1994AJ....108.1163K} {108, 1163}

\bibitem[\protect\citeauthoryear{{Kellermann}, {Condon}, {Kimball}, {Perley}
  \& {Ivezi{\'c}}}{{Kellermann} et~al.}{2016}]{Kellermann16}
{Kellermann} K.~I.,  {Condon} J.~J.,  {Kimball} A.~E.,  {Perley} R.~A.,
  {Ivezi{\'c}} {\v Z}.,  2016, \mn@doi [\apj] {10.3847/0004-637X/831/2/168},
  \href {http://adsabs.harvard.edu/abs/2016ApJ...831..168K} {831, 168}

\bibitem[\protect\citeauthoryear{{Kimball} \& {Ivezi{\'c}}}{{Kimball} \&
  {Ivezi{\'c}}}{2008}]{Kimball08}
{Kimball} A.~E.,  {Ivezi{\'c}} {\v Z}.,  2008, \mn@doi [\aj]
  {10.1088/0004-6256/136/2/684}, \href
  {http://adsabs.harvard.edu/abs/2008AJ....136..684K} {136, 684}

\bibitem[\protect\citeauthoryear{{Komossa}}{{Komossa}}{2008}]{Komossa08}
{Komossa} S.,  2008, in Revista Mexicana de Astronomia y Astrofisica Conference
  Series. pp 86--92 (\mn@eprint {arXiv} {0710.3326})

\bibitem[\protect\citeauthoryear{{Komossa}, {Voges}, {Xu}, {Mathur}, {Adorf},
  {Lemson}, {Duschl}  \& {Grupe}}{{Komossa} et~al.}{2006}]{Komossa06}
{Komossa} S.,  {Voges} W.,  {Xu} D.,  {Mathur} S.,  {Adorf} H.-M.,  {Lemson}
  G.,  {Duschl} W.~J.,   {Grupe} D.,  2006, \mn@doi [\aj] {10.1086/505043},
  \href {http://adsabs.harvard.edu/abs/2006AJ....132..531K} {132, 531}

\bibitem[\protect\citeauthoryear{{Kunert-Bajraszewska}, {Gawro{\'n}ski},
  {Labiano}  \& {Siemiginowska}}{{Kunert-Bajraszewska} et~al.}{2010}]{Kunert10}
{Kunert-Bajraszewska} M.,  {Gawro{\'n}ski} M.~P.,  {Labiano} A.,
  {Siemiginowska} A.,  2010, \mn@doi [\mnras]
  {10.1111/j.1365-2966.2010.17271.x}, \href
  {http://adsabs.harvard.edu/abs/2010MNRAS.408.2261K} {408, 2261}

\bibitem[\protect\citeauthoryear{{Lacy}, {Laurent-Muehleisen}, {Ridgway},
  {Becker}  \& {White}}{{Lacy} et~al.}{2001}]{Lacy01}
{Lacy} M.,  {Laurent-Muehleisen} S.~A.,  {Ridgway} S.~E.,  {Becker} R.~H.,
  {White} R.~L.,  2001, \mn@doi [\apjl] {10.1086/319836}, \href
  {http://adsabs.harvard.edu/abs/2001ApJ...551L..17L} {551, L17}

\bibitem[\protect\citeauthoryear{{Laor}}{{Laor}}{2000}]{Laor2000}
{Laor} A.,  2000, \mn@doi [\apjl] {10.1086/317280}, \href
  {http://adsabs.harvard.edu/abs/2000ApJ...543L.111L} {543, L111}

\bibitem[\protect\citeauthoryear{{Mahony} et~al.,}{{Mahony}
  et~al.}{2016}]{Mahony16}
{Mahony} E.~K.,  et~al., 2016, \mn@doi [\mnras] {10.1093/mnras/stw2225}, \href
  {http://adsabs.harvard.edu/abs/2016MNRAS.463.2997M} {463, 2997}

\bibitem[\protect\citeauthoryear{{Marecki}, {Spencer}  \& {Kunert}}{{Marecki}
  et~al.}{2003}]{Marecki03}
{Marecki} A.,  {Spencer} R.~E.,   {Kunert} M.,  2003, \mn@doi [\pasa]
  {10.1071/AS02051}, \href {http://adsabs.harvard.edu/abs/2003PASA...20...46M}
  {20, 46}

\bibitem[\protect\citeauthoryear{{Marscher}}{{Marscher}}{2016}]{Marscher16}
{Marscher} A.,  2016, \mn@doi [Galaxies] {10.3390/galaxies4040037}, \href
  {http://adsabs.harvard.edu/abs/2016Galax...4...37M} {4, 37}

\bibitem[\protect\citeauthoryear{{Massaro}, {Giommi}, {Leto}, {Marchegiani},
  {Maselli}, {Perri}, {Piranomonte}  \& {Sclavi}}{{Massaro}
  et~al.}{2009}]{Massaro09}
{Massaro} E.,  {Giommi} P.,  {Leto} C.,  {Marchegiani} P.,  {Maselli} A.,
  {Perri} M.,  {Piranomonte} S.,   {Sclavi} S.,  2009, \mn@doi [\aap]
  {10.1051/0004-6361:200810161}, \href
  {http://adsabs.harvard.edu/abs/2009A%26A...495..691M} {495, 691}

\bibitem[\protect\citeauthoryear{{Mateos} et~al.,}{{Mateos}
  et~al.}{2012}]{Mateos12}
{Mateos} S.,  et~al., 2012, \mn@doi [\mnras]
  {10.1111/j.1365-2966.2012.21843.x}, \href
  {http://adsabs.harvard.edu/abs/2012MNRAS.426.3271M} {426, 3271}

\bibitem[\protect\citeauthoryear{{Mori{\'c}}, {Smol{\v c}i{\'c}}, {Kimball},
  {Riechers}, {Ivezi{\'c}}  \& {Scoville}}{{Mori{\'c}} et~al.}{2010}]{Moric10}
{Mori{\'c}} I.,  {Smol{\v c}i{\'c}} V.,  {Kimball} A.,  {Riechers} D.~A.,
  {Ivezi{\'c}} {\v Z}.,   {Scoville} N.,  2010, \mn@doi [\apj]
  {10.1088/0004-637X/724/1/779}, \href
  {http://adsabs.harvard.edu/abs/2010ApJ...724..779M} {724, 779}

\bibitem[\protect\citeauthoryear{{Moshir} \& {et al.}}{{Moshir} \& {et
  al.}}{1990}]{Moshir90}
{Moshir} M.,  {et al.} 1990, in IRAS Faint Source Catalogue, version 2.0
  (1990).

\bibitem[\protect\citeauthoryear{{Netzer} \& {Trakhtenbrot}}{{Netzer} \&
  {Trakhtenbrot}}{2007}]{Netzer07}
{Netzer} H.,  {Trakhtenbrot} B.,  2007, \mn@doi [\apj] {10.1086/509650}, \href
  {http://adsabs.harvard.edu/abs/2007ApJ...654..754N} {654, 754}

\bibitem[\protect\citeauthoryear{{O'Dea} \& {Baum}}{{O'Dea} \&
  {Baum}}{1997}]{ODea97}
{O'Dea} C.~P.,  {Baum} S.~A.,  1997, \mn@doi [\aj] {10.1086/118241}, \href
  {http://adsabs.harvard.edu/abs/1997AJ....113..148O} {113, 148}

\bibitem[\protect\citeauthoryear{{Orienti}}{{Orienti}}{2016}]{Orienti16}
{Orienti} M.,  2016, \mn@doi [Astronomische Nachrichten]
  {10.1002/asna.201512257}, \href
  {http://adsabs.harvard.edu/abs/2016AN....337....9O} {337, 9}

\bibitem[\protect\citeauthoryear{{Osterbrock} \& {Pogge}}{{Osterbrock} \&
  {Pogge}}{1985}]{Osterbrock85}
{Osterbrock} D.~E.,  {Pogge} R.~W.,  1985, \mn@doi [\apj] {10.1086/163513},
  \href {http://adsabs.harvard.edu/abs/1985ApJ...297..166O} {297, 166}

\bibitem[\protect\citeauthoryear{{Padovani}}{{Padovani}}{2016}]{Padovani16}
{Padovani} P.,  2016, \mn@doi [\aapr] {10.1007/s00159-016-0098-6}, \href
  {http://adsabs.harvard.edu/abs/2016A%26ARv..24...13P} {24, 13}

\bibitem[\protect\citeauthoryear{{Paliya}, {Rajput}, {Stalin}  \&
  {Pandey}}{{Paliya} et~al.}{2016}]{Paliya16}
{Paliya} V.~S.,  {Rajput} B.,  {Stalin} C.~S.,   {Pandey} S.~B.,  2016, \mn@doi
  [\apj] {10.3847/0004-637X/819/2/121}, \href
  {http://adsabs.harvard.edu/abs/2016ApJ...819..121P} {819, 121}

\bibitem[\protect\citeauthoryear{{Panessa} et~al.,}{{Panessa}
  et~al.}{2011}]{Panessa11}
{Panessa} F.,  et~al., 2011, \mn@doi [\mnras]
  {10.1111/j.1365-2966.2011.19268.x}, \href
  {http://adsabs.harvard.edu/abs/2011MNRAS.417.2426P} {417, 2426}

\bibitem[\protect\citeauthoryear{{Prandoni}, {Parma}, {Wieringa}, {de Ruiter},
  {Gregorini}, {Mignano}, {Vettolani}  \& {Ekers}}{{Prandoni}
  et~al.}{2006}]{Prandoni06}
{Prandoni} I.,  {Parma} P.,  {Wieringa} M.~H.,  {de Ruiter} H.~R.,  {Gregorini}
  L.,  {Mignano} A.,  {Vettolani} G.,   {Ekers} R.~D.,  2006, \mn@doi [\aap]
  {10.1051/0004-6361:20054273}, \href
  {http://adsabs.harvard.edu/abs/2006A%26A...457..517P} {457, 517}

\bibitem[\protect\citeauthoryear{{Rafter}, {Crenshaw}  \& {Wiita}}{{Rafter}
  et~al.}{2009}]{Rafter09}
{Rafter} S.~E.,  {Crenshaw} D.~M.,   {Wiita} P.~J.,  2009, \mn@doi [\aj]
  {10.1088/0004-6256/137/1/42}, \href
  {http://adsabs.harvard.edu/abs/2009AJ....137...42R} {137, 42}

\bibitem[\protect\citeauthoryear{{Rafter}, {Crenshaw}  \& {Wiita}}{{Rafter}
  et~al.}{2011}]{Rafter11}
{Rafter} S.~E.,  {Crenshaw} D.~M.,   {Wiita} P.~J.,  2011, \mn@doi [\aj]
  {10.1088/0004-6256/141/3/85}, \href
  {http://adsabs.harvard.edu/abs/2011AJ....141...85R} {141, 85}

\bibitem[\protect\citeauthoryear{{Rakshit}, {Stalin}, {Chand}  \&
  {Zhang}}{{Rakshit} et~al.}{2017}]{Rakshit17}
{Rakshit} S.,  {Stalin} C.~S.,  {Chand} H.,   {Zhang} X.-G.,  2017, \mn@doi
  [\apjs] {10.3847/1538-4365/aa6971}, \href
  {http://adsabs.harvard.edu/abs/2017ApJS..229...39R} {229, 39}

\bibitem[\protect\citeauthoryear{{Readhead}, {Taylor}, {Xu}, {Pearson},
  {Wilkinson}  \& {Polatidis}}{{Readhead} et~al.}{1996}]{Readhead96}
{Readhead} A.~C.~S.,  {Taylor} G.~B.,  {Xu} W.,  {Pearson} T.~J.,  {Wilkinson}
  P.~N.,   {Polatidis} A.~G.,  1996, \mn@doi [\apj] {10.1086/176996}, \href
  {http://adsabs.harvard.edu/abs/1996ApJ...460..612R} {460, 612}

\bibitem[\protect\citeauthoryear{{Rengelink}, {Tang}, {de Bruyn}, {Miley},
  {Bremer}, {Roettgering}  \& {Bremer}}{{Rengelink} et~al.}{1997}]{Rengelink97}
{Rengelink} R.~B.,  {Tang} Y.,  {de Bruyn} A.~G.,  {Miley} G.~K.,  {Bremer}
  M.~N.,  {Roettgering} H.~J.~A.,   {Bremer} M.~A.~R.,  1997, \mn@doi [\aaps]
  {10.1051/aas:1997358}, \href
  {http://adsabs.harvard.edu/abs/1997A%26AS..124..259R} {124, 259}

\bibitem[\protect\citeauthoryear{{Richards} \& {Lister}}{{Richards} \&
  {Lister}}{2015}]{Richards15}
{Richards} J.~L.,  {Lister} M.~L.,  2015, \mn@doi [\apjl]
  {10.1088/2041-8205/800/1/L8}, \href
  {http://adsabs.harvard.edu/abs/2015ApJ...800L...8R} {800, L8}

\bibitem[\protect\citeauthoryear{{Roy}, {Norris}, {Kesteven}, {Troup}  \&
  {Reynolds}}{{Roy} et~al.}{1998}]{Roy98}
{Roy} A.~L.,  {Norris} R.~P.,  {Kesteven} M.~J.,  {Troup} E.~R.,   {Reynolds}
  J.~E.,  1998, \mn@doi [\mnras] {10.1046/j.1365-8711.1998.02060.x}, \href
  {http://adsabs.harvard.edu/abs/1998MNRAS.301.1019R} {301, 1019}

\bibitem[\protect\citeauthoryear{{Saikia}}{{Saikia}}{1999}]{Saikia99}
{Saikia} D.~J.,  1999, \mn@doi [\mnras] {10.1046/j.1365-8711.1999.02295.x},
  \href {http://adsabs.harvard.edu/abs/1999MNRAS.302L..60S} {302, L60}

\bibitem[\protect\citeauthoryear{{Shemmer}, {Netzer}, {Maiolino}, {Oliva},
  {Croom}, {Corbett}  \& {di Fabrizio}}{{Shemmer} et~al.}{2004}]{Shemmer04}
{Shemmer} O.,  {Netzer} H.,  {Maiolino} R.,  {Oliva} E.,  {Croom} S.,
  {Corbett} E.,   {di Fabrizio} L.,  2004, \mn@doi [\apj] {10.1086/423607},
  \href {http://adsabs.harvard.edu/abs/2004ApJ...614..547S} {614, 547}

\bibitem[\protect\citeauthoryear{{Sikora}, {Stawarz}  \& {Lasota}}{{Sikora}
  et~al.}{2007}]{Sikora07}
{Sikora} M.,  {Stawarz} {\L}.,   {Lasota} J.-P.,  2007, \mn@doi [\apj]
  {10.1086/511972}, \href {http://adsabs.harvard.edu/abs/2007ApJ...658..815S}
  {658, 815}

\bibitem[\protect\citeauthoryear{{Singh}, {Shastri}, {Ishwara-Chandra}  \&
  {Athreya}}{{Singh} et~al.}{2013}]{Singh13}
{Singh} V.,  {Shastri} P.,  {Ishwara-Chandra} C.~H.,   {Athreya} R.,  2013,
  \mn@doi [\aap] {10.1051/0004-6361/201221003}, \href
  {http://adsabs.harvard.edu/abs/2013A%26A...554A..85S} {554, A85}

\bibitem[\protect\citeauthoryear{{Singh}, {Ishwara-Chandra}, {Wadadekar},
  {Beelen}  \& {Kharb}}{{Singh} et~al.}{2015}]{Singh15a}
{Singh} V.,  {Ishwara-Chandra} C.~H.,  {Wadadekar} Y.,  {Beelen} A.,   {Kharb}
  P.,  2015, \mn@doi [\mnras] {10.1093/mnras/stu2124}, \href
  {http://adsabs.harvard.edu/abs/2015MNRAS.446..599S} {446, 599}

\bibitem[\protect\citeauthoryear{{Sirothia}, {Dennefeld}, {Saikia}, {Dole},
  {Ricquebourg}  \& {Roland}}{{Sirothia} et~al.}{2009}]{Sirothia09}
{Sirothia} S.~K.,  {Dennefeld} M.,  {Saikia} D.~J.,  {Dole} H.,  {Ricquebourg}
  F.,   {Roland} J.,  2009, \mn@doi [\mnras]
  {10.1111/j.1365-2966.2009.14317.x}, \href
  {http://adsabs.harvard.edu/abs/2009MNRAS.395..269S} {395, 269}

\bibitem[\protect\citeauthoryear{{Snellen}, {Schilizzi}, {Miley}, {de Bruyn},
  {Bremer}  \& {R{\"o}ttgering}}{{Snellen} et~al.}{2000}]{Snellen2000}
{Snellen} I.~A.~G.,  {Schilizzi} R.~T.,  {Miley} G.~K.,  {de Bruyn} A.~G.,
  {Bremer} M.~N.,   {R{\"o}ttgering} H.~J.~A.,  2000, \mn@doi [\mnras]
  {10.1046/j.1365-8711.2000.03935.x}, \href
  {http://adsabs.harvard.edu/abs/2000MNRAS.319..445S} {319, 445}

\bibitem[\protect\citeauthoryear{{Stern} et~al.,}{{Stern}
  et~al.}{2012}]{Stern12}
{Stern} D.,  et~al., 2012, \mn@doi [\apj] {10.1088/0004-637X/753/1/30}, \href
  {http://adsabs.harvard.edu/abs/2012ApJ...753...30S} {753, 30}

\bibitem[\protect\citeauthoryear{{Taylor}}{{Taylor}}{2005}]{Taylor05}
{Taylor} M.~B.,  2005, in {Shopbell} P.,  {Britton} M.,   {Ebert} R.,  eds,
  Astronomical Society of the Pacific Conference Series Vol. 347, Astronomical
  Data Analysis Software and Systems XIV. p.~29

\bibitem[\protect\citeauthoryear{{Urry}}{{Urry}}{1996}]{Urry96}
{Urry} C.~M.,  1996, in {Miller} H.~R.,  {Webb} J.~R.,   {Noble} J.~C.,  eds,
  Astronomical Society of the Pacific Conference Series Vol. 110, Blazar
  Continuum Variability. p.~391 (\mn@eprint {} {astro-ph/9609023})

\bibitem[\protect\citeauthoryear{{Urry} \& {Padovani}}{{Urry} \&
  {Padovani}}{1995}]{Urry95}
{Urry} C.~M.,  {Padovani} P.,  1995, \mn@doi [\pasp] {10.1086/133630}, \href
  {http://adsabs.harvard.edu/abs/1995PASP..107..803U} {107, 803}

\bibitem[\protect\citeauthoryear{{Vardoulaki} et~al.,}{{Vardoulaki}
  et~al.}{2015}]{Vardoulaki15}
{Vardoulaki} E.,  et~al., 2015, \mn@doi [\aap] {10.1051/0004-6361/201424125},
  \href {http://adsabs.harvard.edu/abs/2015A%26A...574A...4V} {574, A4}

\bibitem[\protect\citeauthoryear{{V{\'e}ron-Cetty}, {V{\'e}ron}  \& {Gon{\c
  c}alves}}{{V{\'e}ron-Cetty} et~al.}{2001}]{Veron-Cetty01}
{V{\'e}ron-Cetty} M.-P.,  {V{\'e}ron} P.,   {Gon{\c c}alves} A.~C.,  2001,
  \mn@doi [\aap] {10.1051/0004-6361:20010489}, \href
  {http://adsabs.harvard.edu/abs/2001A%26A...372..730V} {372, 730}

\bibitem[\protect\citeauthoryear{{Wadadekar}}{{Wadadekar}}{2004}]{Wadadekar04}
{Wadadekar} Y.,  2004, \mn@doi [\aap] {10.1051/0004-6361:20034244}, \href
  {http://adsabs.harvard.edu/abs/2004A%26A...416...35W} {416, 35}

\bibitem[\protect\citeauthoryear{{Wajima}, {Fujisawa}, {Hayashida}, {Isobe},
  {Ishida}  \& {Yonekura}}{{Wajima} et~al.}{2014}]{Wajima14}
{Wajima} K.,  {Fujisawa} K.,  {Hayashida} M.,  {Isobe} N.,  {Ishida} T.,
  {Yonekura} Y.,  2014, \mn@doi [\apj] {10.1088/0004-637X/781/2/75}, \href
  {http://adsabs.harvard.edu/abs/2014ApJ...781...75W} {781, 75}

\bibitem[\protect\citeauthoryear{{Wang}, {Zhou}, {Wang}, {Lu}  \& {Lu}}{{Wang}
  et~al.}{2006}]{Wang06}
{Wang} T.-G.,  {Zhou} H.-Y.,  {Wang} J.-X.,  {Lu} Y.-J.,   {Lu} Y.,  2006,
  \mn@doi [\apj] {10.1086/504397}, \href
  {http://adsabs.harvard.edu/abs/2006ApJ...645..856W} {645, 856}

\bibitem[\protect\citeauthoryear{{Williams}, {Intema}  \&
  {R{\"o}ttgering}}{{Williams} et~al.}{2013}]{Williams13}
{Williams} W.~L.,  {Intema} H.~T.,   {R{\"o}ttgering} H.~J.~A.,  2013, \mn@doi
  [\aap] {10.1051/0004-6361/201220235}, \href
  {http://adsabs.harvard.edu/abs/2013A%26A...549A..55W} {549, A55}

\bibitem[\protect\citeauthoryear{{Willott}, {Rawlings}, {Blundell}, {Lacy}  \&
  {Eales}}{{Willott} et~al.}{2001}]{Willott01}
{Willott} C.~J.,  {Rawlings} S.,  {Blundell} K.~M.,  {Lacy} M.,   {Eales}
  S.~A.,  2001, \mn@doi [\mnras] {10.1046/j.1365-8711.2001.04101.x}, \href
  {http://adsabs.harvard.edu/abs/2001MNRAS.322..536W} {322, 536}

\bibitem[\protect\citeauthoryear{{Wright} et~al.,}{{Wright}
  et~al.}{2010}]{Wright10}
{Wright} E.~L.,  et~al., 2010, \mn@doi [\aj] {10.1088/0004-6256/140/6/1868},
  \href {http://adsabs.harvard.edu/abs/2010AJ....140.1868W} {140, 1868}

\bibitem[\protect\citeauthoryear{{Yuan}, {Zhou}, {Komossa}, {Dong}, {Wang},
  {Lu}  \& {Bai}}{{Yuan} et~al.}{2008}]{Yuan08}
{Yuan} W.,  {Zhou} H.~Y.,  {Komossa} S.,  {Dong} X.~B.,  {Wang} T.~G.,  {Lu}
  H.~L.,   {Bai} J.~M.,  2008, \mn@doi [\apj] {10.1086/591046}, \href
  {http://adsabs.harvard.edu/abs/2008ApJ...685..801Y} {685, 801}

\bibitem[\protect\citeauthoryear{{Zhou}, {Wang}, {Yuan}, {Lu}, {Dong}, {Wang}
  \& {Lu}}{{Zhou} et~al.}{2006}]{Zhou06}
{Zhou} H.,  {Wang} T.,  {Yuan} W.,  {Lu} H.,  {Dong} X.,  {Wang} J.,   {Lu} Y.,
   2006, \mn@doi [\apjs] {10.1086/504869}, \href
  {http://adsabs.harvard.edu/abs/2006ApJS..166..128Z} {166, 128}

\makeatother
\end{thebibliography}
\appendix
\section{A representative sub-sample of radio-detected NLS1s}
\begin{table*}
\centering
\rotatebox{90}{
\begin{minipage}{240mm}
\caption{A representative sub-sample of our radio-detected NLS1s.}
\resizebox{18cm}{!}{ 
\begin{tabular}{@{}ccccccccccccccc@{}}
\hline
RA          &        DEC    &     $z$      & FWHM (H${\beta}$) & log(L$_{\rm 4400{\AA}}$) & S$_{\rm 1.4 GHz}^{\rm FIRST}$  &  S$_{\rm 1.4 GHz}^{\rm NVSS}$ & S$_{\rm pol}^{\rm NVSS}$ &  ${\sigma}_{\rm var}$ & S$_{\rm 327 MHz}^{\rm WENSS}$ & S$_{\rm 150 MHz}^{\rm TGSS}$ & log(L$_{\rm 1.4 GHz}$) & ${\alpha}_{\rm 150~MHz}^{\rm 1.4~GHz}$ &    Size     & log(R$_{\rm 1.4 GHz}$)   \\   
(h m s)     &    (d m s)    &              &    (km s$^{-1}$)  & (erg s$^{-1}$)           &     (mJy)                      &        (mJy)                  &       (mJy)              &                       &        (mJy)                  &       (mJy)                  &      (W~Hz$^{-1}$)     &                                        &    (kpc)    &                           \\  \hline   
01 46 45    &  $-$00 40 43  &    0.0824    &     1234$\pm$20   &   43.18                  &    4.54$\pm$0.69               &        6.9$\pm$0.5            &     ...                  &    $-$11.2            &         ...                   &       ...                    &    22.88$\pm$0.07      &        $<$ $-$0.75                     &    16.2     &   1.59                   \\
03 38 11    &  $-$06 20 41  &    0.5474    &     784$\pm$47    &   44.46                  &   2.73$\pm$0.22                &        4.0$\pm$0.5            &     ...                  &    $-$3.3             &         ...                   &       ...                    &    24.52$\pm$0.03      &        $<$ $-$0.98                     &  $<$ 18.3   &   1.95                    \\
07 51 15    &  +25 16 32    &    0.3364    &     1314$\pm$134  &   43.43                  &   42.95$\pm$0.13               &        47.5$\pm$1.5           &     ...                  &    $-$1.9             &         ...                   &       ...                    &    25.21$\pm$0.01      &        $<$ 0.25                        &  $<$ 3.9    &   3.68                    \\
08 20 08    &  +37 28 40    &    0.0819    &     1694$\pm$121  &   43.01                  &    1.64$\pm$0.17               &         ...                   &     ...                  &     ...               &         ...                   &       ...                    &    22.43$\pm$0.04      &        $<$ $-$1.21                     &  $<$ 5.9    &   1.31                    \\
09 32 41    &  +53 06 34    &    0.5970    &     1897$\pm$181  &   44.36                  &  411.43$\pm$0.15               &        481.6$\pm$14.4         &     ...                  &    $-$3.9             &      434.0$\pm$4.1            &       566.7$\pm$57           &    26.79$\pm$0.01      &          $-$0.14$\pm$0.03              &  $<$ 12.5   &   4.33                    \\
09 53 22    &  +29 11 51    &    0.1448    &     1283$\pm$210  &   43.28                  &   1.75$\pm$0.17                &         ...                   &     ...                  &     ...               &         ...                   &        ...                   &    22.99$\pm$0.04      &        $<$ $-$1.18                     &  $<$ 9.9    &   1.61                    \\
10 04 11    &  +52 30 25    &    0.2987    &     1970$\pm$212  &   43.33                  &   6.64$\pm$0.25                &         7.6$\pm$0.5           &     ...                  &    $-$6.9             &         ...                   &       26.7$\pm$43            &    24.28$\pm$0.02      &          $-$0.62$\pm$0.02              &    17.7     &   2.85                     \\
10 46 35    &  +13 45 03    &    0.0098    &     1850$\pm$11   &   42.19                  &   40.95$\pm$0.14               &        117.6$\pm$4.2          &     ...                  &    $-$23.5            &         ...                   &        ...                   &    21.94$\pm$0.01      &        $<$ 0.23                        &     7.3     &   1.63                    \\
11 57 01    &  +32 44 58    &    0.4862    &     2049$\pm$72   &   44.28                  &  150.68$\pm$0.18               &        159.9$\pm$4.8          &    7.89$\pm$0.44         &    $-$10.3            &        518$\pm$3.1            &       880.2$\pm$88.2         &    26.14$\pm$0.01      &           $-$0.79$\pm$0.01             &    25.6     &   3.74                    \\
13 38 17    &  +48 16 41    &    0.0270    &     484$\pm$31    &   42.69                  &   47.20$\pm$0.23               &        126.3$\pm$4.5          &     ...                  &    $-$20.7            &         ...                   &       619.0$\pm$62.3         &    22.91$\pm$0.01      &           $-$1.15$\pm$0.01             &    2.0      &   2.11                   \\ \hline
\end{tabular}}
\label{table:Sample} 
\end{minipage}
}
\end{table*}
\bsp	
\label{lastpage}
\end{document}